\RequirePackage{graphicx}

\documentclass[12pt]{iopart}

\usepackage{cite}
\usepackage{xspace}
\usepackage{bm}
\usepackage{color}
\usepackage{slashed}
\usepackage{hyperref}
\usepackage{array}
\usepackage[utf8]{inputenc}
\usepackage{graphicx}
\usepackage{comment}

\usepackage{lineno}

\usepackage{upgreek}

\def\beq{\begin{equation}}
\def\eeq{\end{equation}}
\def\bea{\begin{eqnarray}}
\def\eea{\end{eqnarray}}
\def\D0{D\O }

\newcommand{\qsq}{\ensuremath{q^{2}}\xspace}

\newcommand{\Dssm}{\ensuremath{D^{*}}\xspace}

\newcommand{\Bs}{\ensuremath{B}\xspace}

\newcommand{\FourS}{\ensuremath{\Upsilon(4S)}\xspace}

\newcommand{\mev}{\ensuremath{\mathrm{\,Me\kern -0.1em V}}\xspace}
\newcommand{\gev}{\ensuremath{\mathrm{\,Ge\kern -0.1em V}}\xspace}

\newcommand{\BBb}{\ensuremath{B\overline{B}}\xspace}

\def\mmsq {\ensuremath{M_{miss}^{2}}\xspace}

\newcommand{\dzb}{\ensuremath{{\overline D}^0}\xspace}

\newcommand{\btodslnu}{\ensuremath{B\to D^{*}\ell\nu_\ell}\xspace}
\newcommand{\bztodslnu}{\ensuremath{B^0\to D^{*-}\ell^+\nu_\ell}\xspace}
\newcommand{\bptodslnu}{\ensuremath{B^+\to D^{*0}\ell^+\nu_\ell}\xspace}
\newcommand{\btodlnu}{\ensuremath{B\to D\ell\nu_\ell}\xspace}
\newcommand{\bztodlnu}{\ensuremath{B^0\to D^{-}\ell^+\nu_\ell}\xspace}
\newcommand{\bptodlnu}{\ensuremath{B^+\to \dzb\ell^+\nu_\ell}\xspace}

\newcommand*{\eqref}[1]{(\ref{#1})}

\begin{document}

\title[J. Phys. G: Nucl. Part. Phys]{Determination of the   Cabibbo-Kobayashi-Maskawa matrix element $|V_{cb}|$  }

\author{Giulia Ricciardi }

\address{Dipartimento di Fisica E. Pancini, Universit\`{a} di Napoli Federico II and \\
I.N.F.N. Sezione di Napoli,
\\
Complesso Universitario di Monte Sant'Angelo, Ed. 6, Via Cintia, 80126 Napoli, Italy}
\ead{giulia.ricciardi@na.infn.it}

\author{ Marcello Rotondo}

\address{Laboratori Nazionali dell'INFN di Frascati,
Via Enrico Fermi 40, 00040 Frascati (Roma), Italy}
\ead{marcello.rotondo@lnf.infn.it}

\vspace{10pt}
\begin{indented}
\item[]Nov 2019
\end{indented}

\clearpage

\begin{abstract}

In this review we present and discuss the determination of the magnitude of the Cabibbo-Kobayashi-Maskawa (CKM) matrix parameter $V_{cb}$.  The CKM matrix   parametrizes  the  weak  charged  current  interactions  of  quarks in the Standard Model (SM), and a precise determination of its elements has always been one  of  the most important targets of particle physics. The precise knowledge of the $|V_{cb}|$ value plays a pivotal role in testing the flavour sector of the SM and in the analyses of the unitarity  of the CKM matrix. 

The SM does not predict the values of the CKM matrix elements, which have to be extracted by  experimental data. 
Given the variety of channels that allow the extraction of $|V_{cb}|$, different theoretical and experimental techniques are mustered for the $|V_{cb}|$ determination. The exertion toward precision  represents not only a significant test of our theoretical procedures but a stimulus towards better  detection performances. 

The most precise measurements of $|V_{cb}|$ come from semileptonic decays, that being tree level at the lowest order in the SM are generally considered unaffected by  physics beyond the SM. After summarizing the characteristics  of the
SM that set the frame for the determination of $|V_{cb}|$, we discuss
 inclusive and exclusive semileptonic $B$ decays. We analyze the  $|V_{cb}|$ extraction methods and recent results, detailing both the theoretical and experimental techniques, and, finally, outline future prospects. We also comment on exclusive decays into heavy leptons, on the observables $R(D)$ and  $R(D^\ast)$, on decays to excited $D$ meson states and on baryon decays.

\end{abstract}

%
%
%
%
%

\clearpage
\tableofcontents

\clearpage

\section{Introduction}

Nowadays accuracy in  measurements and theoretical calculations of physical observables is  indispensable to  check the  Standard Model (SM) and explore the small region of parameters space left to its extensions, at our energies.
The increase in precision  demands an accurate knowledge of the parameters of the Cabibbo-Kobayashi-Maskawa (CKM) matrix,  which are not
 predictable within the SM, and must be extracted by data.
 In the last decades, a large effort has gone towards their determination, mostly driven by increasingly higher statistics at new and improved facilities, accompanied by more complex and sophisticated theoretical computations.
 
 Among the    CKM matrix elements, $V_{cb}$ takes central stage.
 Its role is pivotal in the unitarity analyses of the CKM matrix.
 The  so-called  unitarity  clock,  the  circle  around  the origin  in  the  $\bar \rho -\bar \eta$ plane, is proportional to the ratio $|V_{ub}/V_{cb}|$, and $|V_{cb}|$ normalizes the whole unitarity triangle. Relations  between $|V_{cb}|$ and  other observables can be exploited to estimate their values, within or beyond the SM, and an accurate determination of $|V_{cb}|$ is necessary for their correct assessment.
One example are $B$ decays  originated  by flavour changing neutral currents, such as rare radiative $B \to X_s \gamma$  or semileptonic  $B \to X_s \ell^+ \ell^-$ decays, where $X_s$ are hadronic states with strangeness different from zero. In the SM, the $b \to s $ quark transitions cannot occur at tree level, but start at one loop, mediated by the so-called penguin diagrams, with an up-type quark running in the loop. Top and charm quark contributions  are proportional to $V_{tb} V^\star_{ts}$ and $V_{cb}V^\star_{cs}$ respectively (unitarity can be used to cancel $V_{ub}V^\star_{us}$  in the rate).
Other examples are in the kaon sector, where $\epsilon_K$, $\epsilon^\prime /\epsilon$ and branching ratios of rare kaon decays  depend sensitively on values of $|V_{cb}|$ (and $|V_{ub}|$)~\cite{Buras:2014sba}.
 
The semileptonic decays of beauty hadrons, dominated at the quark level by the weak transition $b\to c \ell \nu_\ell$, are used to determine with high precision the magnitude of the  matrix element $V_{cb}$. 
The heavy mass of the $B$ meson allows to exploit simplifications in the limit of infinite quark mass and to better separate perturbative and non-perturbative  regimes.  
Another advantage is that semileptonic decays are mediated at leading order in perturbation theory tree-level processes. 
The exchange of a new physics (NP) particle is strongly constrained at tree level. A clean determination of CKM parameters from tree level processes is therefore a valuable input for other NP  more  sensitive  estimates. 
Past, present and future $B$ factories have provided and will provide an unparalleled level of precision in branching ratios and related observables, and  LHCb is following suit.

 There are two approaches to determine $|V_{cb}|$, which allow almost equally precise measurements: 
the inclusive and the exclusive approach. In the inclusive approach, the $ B \rightarrow X_c \ell \nu_\ell$ decays, where $X_c$, the hadronic state originated by the charm quark, is not reconstructed in any
specific final state. 
Sufficiently inclusive quantities can be expressed as a double series in $\alpha_s$ and $\Lambda_{QCD}/m_b$, in the framework of the Heavy Quark Expansion (HQE). 
In the exclusive approach, one consider decays where a specific hadronic final state is reconstructed, as  $B\to D \ell \bar \nu_\ell$ and $B\to D^{\ast} \ell \bar \nu_\ell$ decays.
The inclusive and exclusive semileptonic determinations rely on different theoretical calculations  and on different experimental techniques which have, to a large extent, uncorrelated statistical and systematic uncertainties. 
This independence makes their expected agreement a useful test of our understanding of both experiments and theory.
 Since at least three decades, there is a tension among the  $|V_{cb}|$ values, depending on whether they are extracted using exclusive or inclusive semileptonic channels. In  the present  general scenario  of data in optimal  agreement within  the SM, this tension is  intriguing, and alone motivates, in our view, more and more precise theoretical and experimental investigations.

In this paper we  review the theoretical  background and the experimental techniques relevant for the  $|V_{cb}|$ determination. In section~\ref{Scenary} we introduce the flavour sector of the SM Lagrangian and the CKM matrix. In section~\ref{Semileptonicmesondecays} we discuss exclusive and inclusive semileptonic decays (into light and heavy leptons) and the theoretical tools necessary for their analyses. In section~\ref{Experimentaltechniques} we review the experimental techniques used at the $B$-Factories and LHCb to study semileptonic decays, pointing out  the various sources of systematic uncertainties. Sections~\ref{InclusiveVcbdetermination} and \ref{ExclusiveVcbdetermination} are devoted to inclusive and exclusive $|V_{cb}|$ determinations, respectively.
Finally, in section~\ref{futureprospects}, we examine future prospects at Belle-II and LHCb facilities, and future theoretical directions of development.

\section{The flavour scenary}
\label{Scenary}

\subsection{The Yukawa terms in the SM Lagrangian}
The SM  is a gauge field theory describing the electromagnetic, weak interactions and strong interactions of  quarks and leptons.  It has supported 
 calculations of physical quantities with unflinching
precision for the past 50 years. Although there are  challenges that the SM does not address,  a complete, coherent framework, in agreement with data, which encompasses and extends the SM, has still to emerge. 

The SM Lagrangian is invariant under  $SU(2)_L \otimes U(1)_Y \otimes  SU(3)_c$ gauge transformations.
 Fields in the SM Lagrangian are classified according irriducible representations of this gauge group.
Gauge invariance in the SM Lagrangian leads one to expect massless vector bosons, in contrast with the experimental evidence that the weak interactions are short ranged. Such impasse is surmounted by the so-called  Higgs mechanism.
According to the Higgs mechanism, the vector  bosons $
W^\pm$ and $Z^0$ couple through the EW covariant derivative   to a complex scalar $\phi$, the
Higgs (or Brout-Englert-Higgs) field, which
  behaves as  a doublet under the $SU(2)_L$ symmetry  and  has  hypercharge 1/2. 
When
$\phi$
gets a vacuum expectation value different from zero (spontaneously symmetry breaking), the SM
Lagrangian acquires extra terms which are precisely mass terms  for the Higgs and the $
W^\pm$ and $Z^0$ bosons.

In order to give mass to quarks and charged leptons, and additional gauge invariant Lagrangian, the  Yukawa Lagrangian  $  {\cal  L}_{Y}$,  is added to the SM  Lagrangian
%
\beq\label{yuk3higgs}
  {\cal  L}_{Y} =  -\sum_{i,j=1}^{3} \left(Y^{(d)}_{ij} \,  \overline{q^i}_L \, \phi \, d^j_R +Y^{(u)}_{ij} \,  \overline{q^i}_L \, \phi_C
    \, u^j_R + Y^{(\ell)}_{ij} \,  \overline{l^i}_L \, \phi \, e^j_R+ h. c. \right) \nonumber
\eeq
where  $h.c.$
stands for Hermitian conjugate. The fields $\phi$  and  its charge conjugate $\phi_C  \equiv  i \tau_2 \phi^\star$   are  Higgs doublets of  hypercharge $Y=1/2$ and  $Y=-1/2$, 
 $q_L^i$ and $\ell_L^i$ are the $SU(2)_L$ left-handed fondamental doublets for three generations, $u_R^i$, $d_R^i$ and $e_R^i$ are  right-handed up-type, down-type quarks and charged leptons, respectively.
 The gauge symmetry does not constrain the boson-fermion $Y^{(u,d,\ell)}$
couplings, referred as Yukawa
couplings, which are complex number completely arbitrary.

%


After spontaneous symmetry breaking,
the Yukawa Lagrangian  in the quark sector can be written  as
\begin{equation}\label{yuk2}
   {\cal  L}_{Y^q}=-  \overline{\hat{d}}_L M^{(d)}
   \hat{d}_R -\overline{\hat{u}}_L M^{(u)}
    \hat{u}_R + h. c.
\end{equation}
where $M_{ij}$ are three by three complex matrices, connected to the Yukawa couplings and equally arbitrary. The up-type quarks have been indicated with $\hat{u} \equiv (u,c,t)$ and the down-type quark with
$\hat{u} \equiv (u,c,t)$ and $ \hat{d}\equiv  (d,s,b)$.
These are flavour eigenstates, that is states participating in gauge interactions, but not yet mass eigenstates. Indeed,
the $M^{(u)}$  and $M^{(d)}$  matrices are not necessarily Hermitian, nor there is an a priori theoretical reason that they should be diagonal in the generation index. By what is known in mathematics as  a singular value decomposition, they can be both made hermitian and diagonal by a
 bi-unitary transformation
\beq\label{diag1}
U^{u\dagger}_L M^{(u)} U^u_R = M^u_D
\qquad \qquad
U^{d\dagger}_L M^{(d)} U^d_R = M^d_D
\eeq
where $M^u_D$ and $M^d_D$ are diagonal with positive eigenvalues and  $U^{u(d)}_{L(R)} $ are unitary matrices.
It corresponds to the
transformations of the quark states
\begin{eqnarray}\label{diag}
\hat{u}_L &\rightarrow& U^u_L \, \hat{u}_L
\qquad \qquad
\hat{u}_R \rightarrow U^u_R \, \hat{u}_R
\\ \nonumber
\hat{d}_L &\rightarrow& U^d_L \, \hat{d}_L \qquad \qquad
\hat{d}_R \rightarrow U^d_R \, \hat{d}_R
\end{eqnarray}
The new states are the physical ones, since the mass matrix
is diagonal in that basis.

%


The change from flavour to mass quark eigenstates \eqref{diag} in the Yukawa sector has to be registered by other
sectors of the Lagrangian.
One can easily observe that the neutral  and
electromagnetic currents remain invariant, since they couple separately  up-type and down-type quarks. On the contrary, the charged current
interactions are affected by this change of basis and
 the part of the Lagrangian describing the hadronic exchanges of charged bosons $W^\pm$ becomes
\begin{equation}\label{LCC}
    {\cal L}_{CC}=  \frac{g}{\sqrt{2}}
(W^+_\mu \overline{\hat{u}_L}\gamma^\mu V \hat{d}_L + W^-_\mu
  \overline{\hat{d}_L}\gamma^\mu  V^\dagger \hat{u}_L )
\end{equation}
in terms of the quark mass eigenstates.
A new unitary matrix,
the Cabibbo-Kobayashi-Maskawa (CKM) matrix, defined as
\beq V \equiv U^{u\dagger}_L \, U^d_L \eeq
 has appeared in the SM. 
It is a unitary matrix, being the product of  unitary matrices, and it parameterizes the change of basis \eqref{diag}, but its
 elements are otherwise completely arbitrary and has to be determined  experimentally.

\subsection{The Cabibbo-Kobayashi-Maskawa matrix}
In the SM, the  CKM  matrix is a key element in describing the flavour dynamics. As  seen above, it is  unitary, but this is its only theoretical constraint. 
The parameters of the CKM, which can be complex, have to be determined experimentally, and there is no a priori theoretical way to determine their values within the SM framework. 
The CKM matrix $V$ induces flavour-changing transitions inside and between generations in the charged  sector  at  tree  level.   By  contrast,  there  are  no  flavour-changing transitions in the neutral sector at tree level. We can write 
\begin{equation}
V= \left(%
\begin{array}{ccc}
  V_{ud} & V_{us} & V_{ub} \\
  V_{cd} & V_{cs} & V_{cb} \\
  V_{td} & V_{ts} & V_{tb} \\
\end{array}%
\right)
\end{equation}
Due to the unitarity, not all the entries of the CKM matrix are independent.
The independent parameters are four in the case of three generations, and can be interpreted as  three
rotation angles and one phase. There are several equivalent parameterization of the  CKM matrix.
A common one is
\begin{eqnarray}
V&=& \left(%
\begin{array}{ccc}
c_{12} c_{13} & s_{12} c_{13}  & s_{13} e^{-i \delta} \\
  -s_{12} c_{23}-c_{12} s_{23} s_{13} e^{i \delta} &
   c_{12} c_{23}-s_{12} s_{23} s_{13} e^{i \delta} & s_{23} c_{13}  \\
s_{12} s_{23}-c_{12} c_{23} s_{13} e^{i \delta } &
 -c_{12} s_{23}-s_{12} c_{23} s_{13} e^{i \delta} & c_{23} c_{13} \\
\end{array}%
\right),  \label{chaupar}
\end{eqnarray}
where $c_{ij}=\cos \theta_{ij}$ and $s_{ij}=\sin \theta_{ij}$, with
$i$ and $j$ labeling families that are coupled through that angle ($i, j = 1, 2, 3$). This CKM parameterization can be seen as the product of three rotations, with the phase put on the smallest element. The rotation
angles may be restricted to lie in the first quadrant, provided
one allows the phase $\delta$ to be free. As a consequence,
$c_{ij}$ and $s_{ij}$ can all be chosen to be positive. 
The angle $\theta_{12}$ is generally called the Cabibbo angle ($\theta_C$), and $\sin \theta_C  \simeq 0.22$,  corresponding to a value   $\theta_C \simeq 13^0$. The angle of mixing between the second and the third
family is $\theta_{23} \simeq 2^0$, and between the first and the third is $\theta_{13} \simeq 0.2^0$.  The phase $\delta$ is constrained by measurements of the CP violation in $K$ decays  to be in the range $ 0 < \delta < \pi $. Its value is approximately $\delta \simeq 1.2$.
In this parameterization, 
the $s_{ij}$ are simply related to directly measurable quantities
\bea s_{13} &=& |V_{ub}| \nonumber \\ s_{12} &=& |V_{us}|/ \sqrt{1-|V_{ub}|^2} \sim |V_{us}| \nonumber \\
 s_{23}&=& |V_{cb}|/ \sqrt{1-|V_{ub}|^2} \sim |V_{cb}|
 \eea
where we have set  $|V_{ub} |\ll 1$, as indicated by  data.

According to experimental evidence, the CKM matrix has a hierarchical structure.  Transitions within the same generation are characterized by  matrix elements of order $O(1)$.  Transitions between the first and second generations are suppressed by a factor of $O(10^{-1})$,   between the second and third generations by a factor of $O(10^{-2})$ and  between the first and third generations by a factor of $O(10^{-3})$.  This hierarchy has prompted another useful parameterization, the so-called Wolfenstein parameterization~\cite{Wolfenstein:1983yz},
based on a series expansion in the small parameter $\lambda=|V_{us}|$.
At  order $\lambda^3$ we have 
\begin{eqnarray}
V
& = & \left(%
\begin{array}{ccc}
  1 -\lambda^2/2 & \lambda & A \lambda^3 (\rho -i \eta) \\
  -\lambda & 1 -\lambda^2/2 & A \lambda^2 \\
 A \lambda^3 (1-\rho -i \eta) & -A \lambda^2 & 1 \\
\end{array}%
\right)  + O(\lambda^4).  \label{wolfpar}
\end{eqnarray}
This parameterization corresponds to a particular choice of phase
convention which eliminates as many phases as possible and puts
the one remaining complex phase in the matrix elements $V_{ub}$
and $V_{td}$. In  this parameterization the unitarity of the matrix is explicit, up to
$\lambda^3$  corrections. The real, independent, parameters $A$, $\rho$ and $\eta$  are
known to be roughly of order unity, while $\lambda$,  that is essentially the sine of the Cabibbo angle, $s_{12} $, is a small number, of order 0.2.
Relative sizes of amplitudes depending on  CKM parameters can be
roughly estimated by counting powers of $\lambda$ in the
Wolfenstein parameterization. 

It is convenient to express the Wolfenstein parameters through phase convention-independent quantities
\bea
s_{12}^2&=&\lambda^2= \frac{|V_{us}|^2}{|V_{ud}|^2+|V_{us}|^2} \nonumber \\
s_{23}^2&=& A^2 \lambda^4= \frac{|V_{cb}|^2}{|V_{ud}|^2+|V_{us}|^2}
\nonumber \\
\bar \rho+i \bar \eta &=&-\frac{V_{ud} V_{ub}^\star}{V_{cd} V_{cb}^\star}
\eea
where $\bar \rho$ and $\bar \eta$ are two new parameters that substitute $\rho$ and $\eta$. These relations ensure that   the CKM matrix written in terms of
$\lambda$, A, $\bar\rho$, and $\bar \eta$ is unitary to all orders in $\lambda$ \cite{Buras:1994ec}. %
When terms of the ${\cal O}(\lambda ^6)$ are neglected, we  have
\begin{eqnarray}
\label{CKM6}
 \hspace*{-1cm}
{\bf {\rm V}}_{\rm CKM} \simeq
\left(\footnotesize
\begin{array}{ccc}
 1 - \frac{1}{2}\lambda ^2 - \frac{1}{8} \lambda ^4 & \lambda  &  A\lambda^3(\bar \rho - i\bar \eta)  \\
-\lambda +\frac{1}{2}A^2 \lambda^5[1-2(\bar \rho +i\bar \eta)]        & 1 -  \frac{1}{2}\lambda ^2 - \frac{1}{8}\lambda ^4(1+4A^2)&  A\lambda^2
\\
A\lambda^3[1- (\bar \rho + i\bar \eta)]     & -A\lambda^2 + \frac{1}{2}A\lambda^4[1-2(\bar \rho +i \bar \eta)]       &1-\frac{1}{2}A^2\lambda^4
\end{array} \right) 
\end{eqnarray}
%
Since we have defined 
\beq
s_{13} e^{i \delta} = V_{ub}^\star = A \lambda^3 (\rho -i \eta)
\eeq
 the following relation holds
\beq 
 \rho+i  \eta= \left( 1+ \frac{\lambda^2}{2} \right) \bar \rho+i \bar \eta + O(\lambda^4)
\eeq
Thus one can reproduce  the CKM matrix (\ref{CKM6}) at the same order in $\rho$ and $\eta$ by the substitutions $\bar \rho \to \rho$ and $\bar \eta \to \eta$ in all entries, except $V_{td}$ where the substitution is
$ (\bar \rho + i\bar \eta) \to (1-\frac{1}{2}\lambda^2)   (\rho + i\eta)$.

\subsection{The unitarity triangles}
 
\begin{figure}[tb!]
  \begin{center}
    \includegraphics[width=0.7\linewidth]{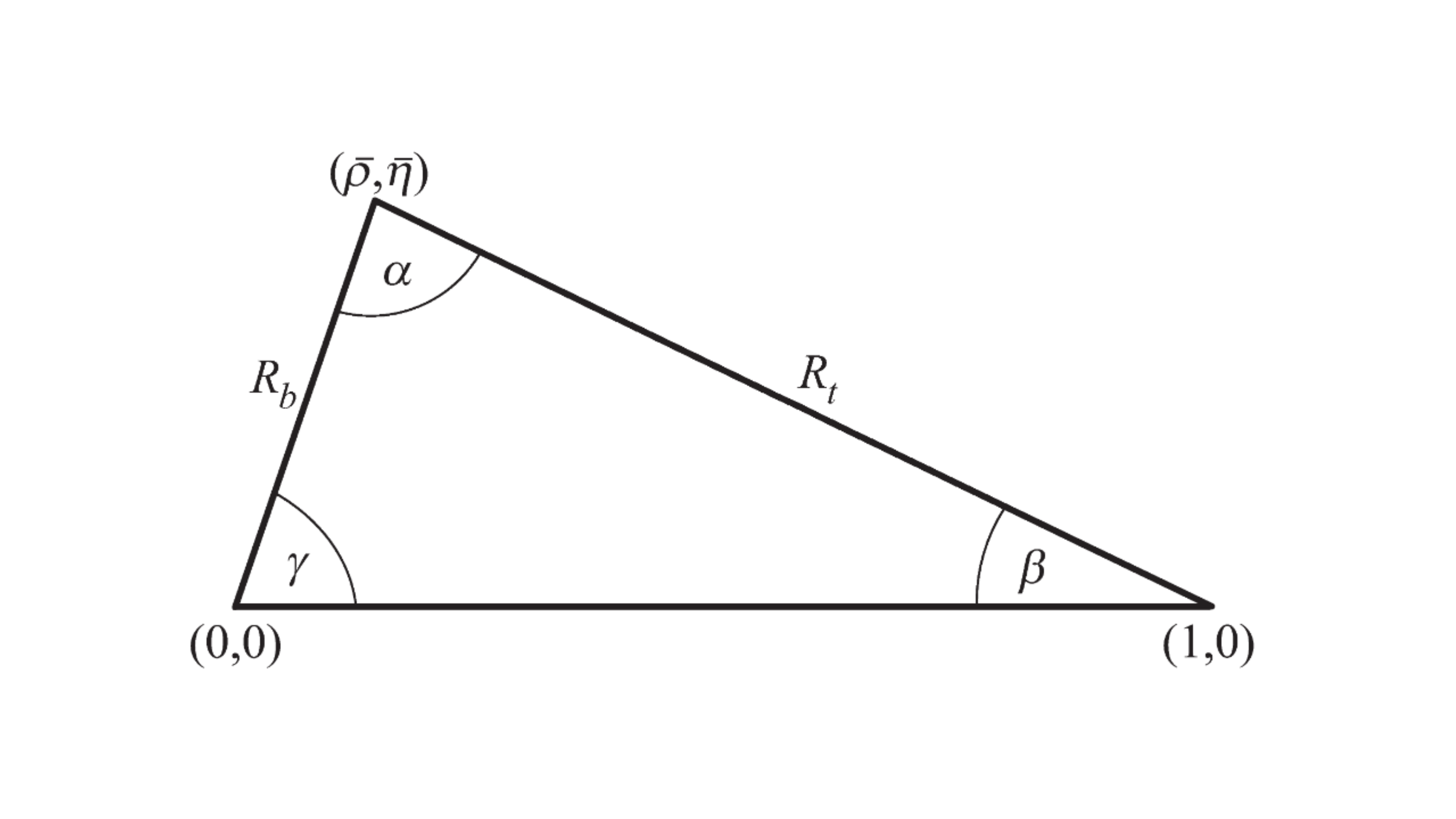}
  \caption{The unitarity triangle in the complex plane.}
   \label{fig:UTtriangle}
  \end{center}
\end{figure}

The unitarity of the CKM matrix  implies
\bea
\sum_{i=1}^3 |V_{ij}|^2 &=& 1 \quad j=1,2,3 \nonumber \\
\sum_{i=1}^3 V_{ji}V^*_{ki} &=& \sum_{i=1}^3 V_{ij}V^*_{ik} =0     \quad  j,k=1,2,3 \quad j\neq k
\label{UNI}
\eea
The   equalities for the off-diagonal terms are sums of three complex numbers, depending on the four CKM parameters.   They are
\bea
&&V_{ud}V^*_{us} \;   [{\cal O}(\lambda )] + V_{cd}V^*_{cs} \;
[{\cal O}(\lambda )] +
 V_{td}V^*_{ts} \; [{\cal O}(\lambda ^{5} )] = 0
 \label{OLD11}\\
&& V^*_{ud}V_{cd} \;  [{\cal O}(\lambda )] + V^*_{us}V_{cs} \;   [{\cal O}(\lambda )] +
V^*_{ub}V_{cb} \; [{\cal O}(\lambda ^{5} )] = 0
\label{OLD12} \\
&& V_{us}V^*_{ub} \;   [{\cal O}(\lambda ^4)] + V_{cs}V^*_{cb} \;  [{\cal O}(\lambda ^{2} )] +
V_{ts}V^*_{tb} \;  [{\cal O}(\lambda ^2  )] = 0
\label{OLD21}\\
&& V^*_{cd}V_{td} \;   [{\cal O}(\lambda ^4 )] + V^*_{cs}V_{ts} \;  [{\cal O}(\lambda ^{2})] +
V^*_{cb}V_{tb} \;  [{\cal O}(\lambda ^{2} )] = 0
\label{OLD22}\\
&& V_{ud}V^*_{ub} \;  [{\cal O}(\lambda ^3)] + V_{cd}V^*_{cb} \;    [{\cal O}(\lambda ^{3} )] +
V_{td}V^*_{tb} \;  [{\cal O}(\lambda ^3  )] = 0
\label{OLD31}
\\
&& V^*_{ud}V_{td} \; [{\cal O}(\lambda ^3 )] + V^*_{us}V_{ts} \;
[{\cal O}(\lambda ^{3})] +
V^*_{ub}V_{tb} \;  [{\cal O}(\lambda ^3 )] = 0
\label{OLD32}
\eea
In these relations it is  indicated in parenthesis the order of each term  in the expansion parameter $\lambda$. 
These equalities  give way to a geometric representation in terms of $\bar \rho, \bar \eta, A$ and  $\lambda$, since
in the complex plane  they can be geometrically represented by triangles, all characterized by the same area~\cite{Jarlskog:1985ht}. 
Only the last two of the six triangles corresponding to these equalities have sides of the same order of magnitude, $ {\cal O}(\lambda ^3 )$ (i.e., the triangles are not squashed).  In particular, the triangle defined by  (\ref{OLD31}), rescaled by a factor $V_{cd} V^\star_{cb}$ is commonly referred to as the unitarity triangle (UT) (see figure~\ref{fig:UTtriangle}).
Because it involves the term $V_{cd} V^\star_{cb}$ and $V_{ud} V^\star_{ub}$, the UT arises naturally in analyses involving $B$ mesons.
With the bases of the UT normalized to unity, the coordinates of the UT apex are $(\bar \rho, \bar \eta)$. The sides $R_b$ and $R_t$ are given by the magnitudes of 
\bea  
R_b &=& =  \frac{V_{ud} V^\star_{ub}}{V_{cd} V^\star_{cb}}  = \left( 1- \frac{\lambda^2}{2} \right) \frac{1}{\lambda} \frac{V_{ub}^\star}{|V_{cb}|} \nonumber \\
R_t &=& =  \frac{V_{td} V^\star_{tb}}{V_{cd} V^\star_{cb}}  = \frac{1}{\lambda} \frac{V_{td}}{|V_{cb}|}
\eea 
As can be seen,  a special role is played by $|V_{cb}|$, which normalizes the UT triangle.
 Due to its economical structure in terms of only four parameters, the CKM matrix can be determined experimentally by exploiting several  different  flavour changing decays or  processes related to neutral-meson mixing. 
 One tries to measure as many observable as possible, in function of the UT triangle parameters, over-constraining the shape of the triangle and testing that it closes. The consistency of the various measurements probes the consequences of unitarity in the three generations SM and discrepancies with the SM expectations  signal the possibility of NP in some observable.
 An extensive program of measurements of the UT parameters has  been  carried through  at  different  experiments since the nineties.
Due to the complexity of non-perturbative strong interactions, it is convenient to analyze processes with a limited number of hadrons in the initial or final state, as semileptonic $B$ decays into one hadron, or  observables (typically ratios) for which uncertainties due to such QCD effects reduce or cancel.
Besides, since the potential sensitivity to NP is limited for tree-level processes, they are often preferred to fix the CKM parameters. Tree level processes are e.g. the semileptonic  $B$ decays into charmed states,   mediated  by the  quark decay $b\to c \ell \nu_\ell$ at the lowest order in the SM. The results from tree-level processes can be used as input for precise SM predictions of rare, loop-induced  processes. Since the start of the analyses on the UT triangle, there has always been an intensive strain to  combine all available measurements (global analysis)  in order to obtain statistically meaningful constraints on the CKM parameters,  in the framework of the SM and some of its extensions\footnote{A systematic program in this direction is carried on by the CKMfitter \cite{CKMfitter} and UTfit collaborations\cite{UTfit}.}.


\section{
Semileptonic $B$ meson decays}
\label{Semileptonicmesondecays}
\begin{figure}[t!]
  \begin{center}
    \includegraphics[width=0.80\linewidth]{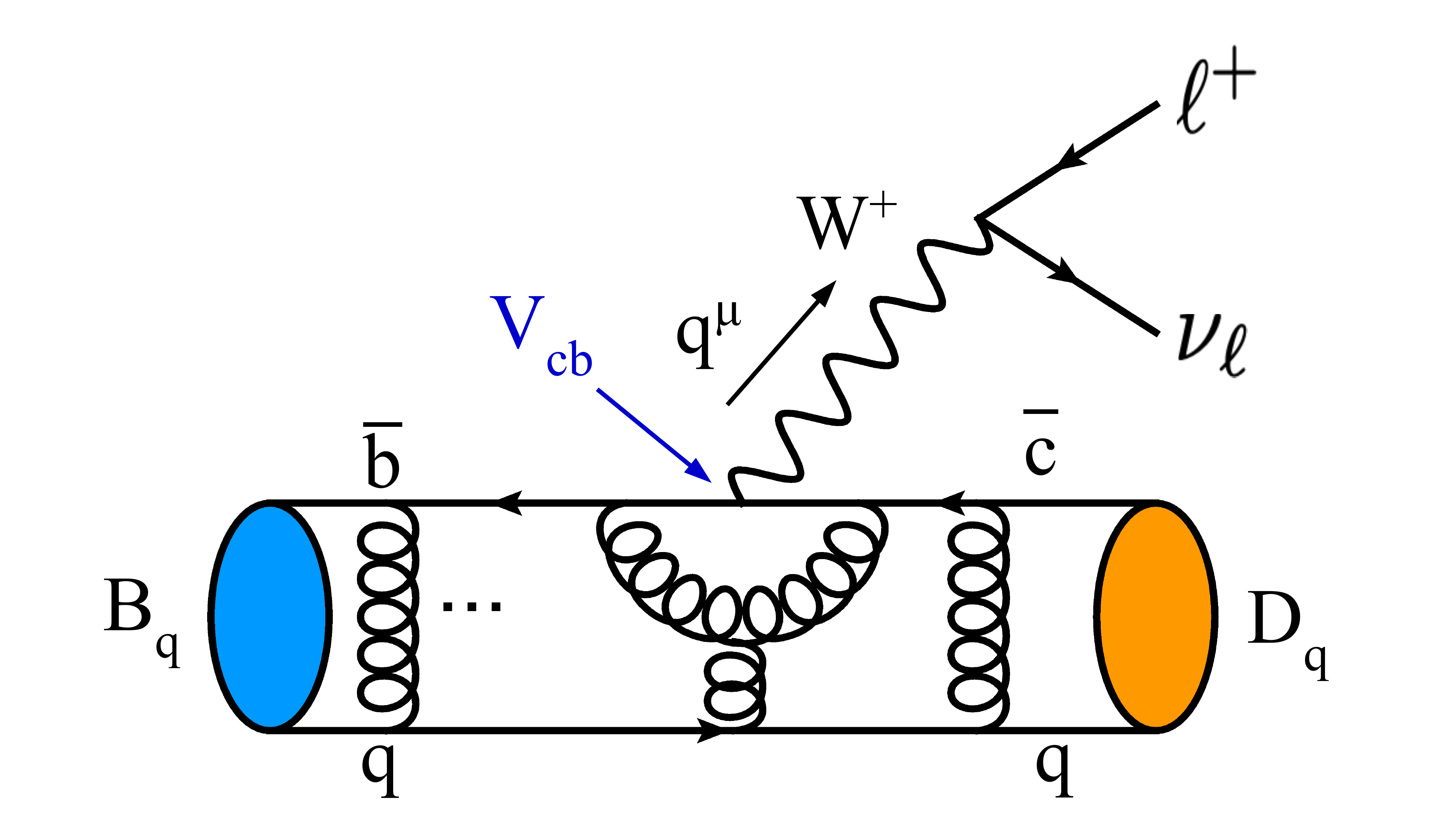}
  \caption{Diagram of the $B_q\to D_q \ell^+\nu_\ell$ decay. The amplitude of this process is proportional to $V_{cb}$. The transfer four-momentum $q^\mu$ is given by $q^\mu\equiv p^\mu_B-p^\mu_D = p^\mu_\ell-p^\mu_\nu$. The diagram is at the lowest order in the weak interactions and the gluon configuration depicted is merely indicative.}
  \label{fig:semilep_diagram}
  \end{center}
\end{figure}

Semileptonic $B$ decays are the processes of election when it comes to a precise determination of the magnitude of the CKM matrix element $V_{cb}$. 
At the lowest order in the SM, semileptonic $B$ decays are mediated  by the a tree level quark decay, the $b\to c \ell \nu_\ell$ decay, whose  amplitude is proportional to  $V_{cb}$, as illustrated in figure \ref{fig:semilep_diagram}. The presence of leptons in the final states simplifies the QCD analyses, since  hadronic and  leptonic currents factorize.  

There are two methods  for  $|V_{cb}|$ determination with semileptonic $B$ decays, taking the name from the hadronic processes involved.
In the so-called exclusive method, $|V_{cb}|$ is extracted by studying  exclusive decays, in particular $B \to D^{(\ast)} \ell \nu_\ell $. Having  only one hadron in the final state   facilitates the analysis (e.g. no final state rescattering). The inclusive method refers to the investigation of the inclusive semileptonic decay $ B \rightarrow X_c \, \ell \, \nu_\ell$ decays,  where the final state
$X_c$ is an hadronic state originated by the charm  quark.  The inclusive and exclusive determinations rely on different theoretical calculations and make use of different techniques which, to a large extent, have uncorrelated experimental uncertainties. 
Comparing the results of these two largely independent approaches represents  also a powerful test of our understanding of  hadron dynamics. We detail both approaches in the following. 
	
\subsection{Inclusive decays}
\label{inclusivedecays}

In  inclusive $ B \rightarrow X_c \, \ell \, \nu_l$ decays,  the final state
$X_c$ is an hadronic state originated by the charm  quark. 
Inclusive decays can be interpreted as a sum over all possible hadronic final states;   the details of the hadronic final states are lost, and   transition amplitudes are expected  to be sensitive only to the  dynamics of the initial $B$ meson.
Quark-hadron duality is generally assumed, which means, loosely speaking, that  the inclusive hadronic observables, when integrated over large enough portions of phase space, are described in terms of the underlying parton-level processes, provided all possible sources of corrections  stemming from QCD  are properly accounted for \footnote{For reviews on quark-hadron duality see for instance~\cite{Shifman:2000jv, Bigi:2001ys} }. 

Both perturbative and non-perturbative  QCD interactions affect  decay processes in an essential way. A basic tool to disentangle their contributions to the decay amplitude in a systematic fashion is provided by the operator product expansion (OPE). The OPE formalism allows us to express the non-perturbative physics in terms of $B$ meson
matrix elements of local, gauge invariant, operators, and  the perturbative physics in terms of   Wilson coefficients, which can be computed as a  series in a perturbative QCD coupling $\alpha_s$. In other terms, the OPE separates the physics associated with soft scales (parametrized by the matrix elements of the local operators) from that associated with hard scales, which determine the Wilson coefficients.
Semileptonic $B$ decays have an intrinsic large 'dynamic' scale of energy release of the order of  the $b$-quark mass, while the soft scale is of the order of the hadronic scale $\Lambda_{QCD}$. The large hierarchy between these two scales leads naturally to $\Lambda_{QCD}/m_b$ as an expansion parameter of non-perturbative effects and to a description of the heavy $b$-quark
in  the framework of   the Heavy Quark Effective Theory  (HQET) (for a  review see for instance~\cite{Manohar:2000dt}).

Jumping to the conclusions, sufficiently inclusive quantities (typically the 
 total semileptonic width and the moments
of the kinematic distributions)  can be expressed as a double series in $\alpha_s$ and $\Lambda_{QCD}/m_b$.  This expansion is referred to as Heavy Quark Expansion (HQE). The expansion for the total semileptonic width takes the form
\bea
\label{HQE}
\Gamma(B\rightarrow X_c l \nu) &=&\frac{G_F^2m_b^5}{192 \pi^3}
|V_{cb}|^2 [ c_3 \langle O_3 \rangle + \nonumber  \\  
&+& \left. c_5\frac{ \langle O_5 \rangle }{m_b^2}+c_6\frac{ \langle O_6 \rangle }{m_b^3}+O\left(\frac{\Lambda^4_{QCD}}{m_b^4},\; \frac{\Lambda^5_{QCD}}{m_b^3\, m_c^2}, \dots \right) \right] 
\eea
Here  $c_d$ ($d=3,5,6 \dots$) are short distance coefficients, calculable  in perturbation theory as a series in the strong coupling $\alpha_s$, and
$O_d$ denote local operators of (scale) dimension $d$. 
The hadronic expectation values of the operators
encode the nonperturbative corrections and can be parametrized in terms of  HQE  parameters, whose number grows with powers of $\Lambda_{QCD}/m_b$.
The leading term  is given by the free $b$-quark decay (parton model). A  remarkable feature  of (\ref{HQE})
is the absence of a contribution of order $1/m_b$, due to the
absence of an independent gauge invariant operator of
dimension four. The power corrections start at $O(\Lambda_{QCD}^2/m_b^2)$, and are comparatively suppressed.
The fact that nonperturbative, bound state
effects in inclusive decays are strongly suppressed (at least two
powers of the heavy quark mass) explains a posteriori the success
of parton model in describing such processes.
Due to the relative sizes of the $b$ and $c$ quarks, at
higher orders in the expansion, terms suppressed by powers of $m_c$ also appear, starting
with $O(\Lambda_{QCD}^5/ m_b^3\, m_c^2)$. 

Similar expansions give the moments of distributions of
charged-lepton energy, hadronic invariant mass and hadronic energy.
As most experiments can detect the
leptons only above a certain threshold in energy, the charged-lepton energy moments are
defined as
\beq \langle E^n_\ell \rangle= \frac{1}{\Gamma_{E_\ell>E_{cut} }} \,\int_{E_\ell>E_{cut} }  E^n_\ell \, \frac{d \Gamma}{dE_\ell} \, dE_\ell
\label{eq:lepton_moments}
\eeq
where $E_\ell$ is the charged lepton energy in the $B\rightarrow X_c \ell \nu_\ell$ decays, $n$ is the order of the moment, $\Gamma_{E_\ell>E_{cut} }$ is the semileptonic width
above the energy threshold $E_{cut}$ and $d \Gamma/dE_\ell$ is the differential semileptonic width
as a function of $E_l$.  The hadronic mass moments are
\beq \langle m^{2n}_X \rangle=\frac{1}{\Gamma_{E_\ell>E_{cut} }} \,\int_{E_\ell>E_{cut} }  m^{2 n}_X \, \frac{d \Gamma}{dm^2_X} \, dm^2_X
\label{eq:hadron_moments}
\eeq
Other moments (and cuts on other observables)  can be defined in a similar way.
It is sometimes  convenient  to  employ central moments,  computed  relative to the averages $\langle E_l \rangle$ and $\langle m^2_X \rangle$, that is
\beq 
l_n(E_{cut}) \equiv \langle (E_l-\langle E_l \rangle )^n \rangle \qquad 
h_n(E_{cut}) \equiv \langle (m^2_X- \langle m^2_X \rangle)^n \rangle 
\eeq 

Let us stress that the HQE  is valid only for sufficiently inclusive measurements
and away from perturbative singularities, therefore the relevant quantities to be measured are
global shape parameters (moments of various kinematic distributions)
and the total rate.
While
the general structure of the expansion is the same for all the above mentioned  observables,  the perturbative  coefficients  are in general different.

Details on HQE will be given in  section~\ref{HeavyQuarkExpansion} and  the sensitivity of rates and momenta  to the definition of quark masses   briefly discussed in section~\ref{massschemes}. In section ~\ref{InclusiveVcbdetermination} we will draw conclusions on the inclusive $|V_{cb}|$ extraction.
%
%

\subsubsection{Heavy Quark Expansion}
\label{HeavyQuarkExpansion}

In order to discuss the characteristics and the status of the HQE in 
$B\rightarrow X_c \ell \nu_\ell$ decays,
let us go back to the  expansion for the total semileptonic width in \eqref{HQE}.
The hadronic
expectation values of the local operators  $ O_d  $ are the (normalized) forward matrix elements,  written in the short-hand notation as
\begin{equation}
\langle O_d \rangle \equiv \frac{\langle B|O_d|B \rangle}{2 m_B}
\label{mb-norm}
\end{equation}
where $m_B$ is the $B$ meson mass,  included in the definition for the normalization and dimensional counting. This set of operators, built with dimensional criteria using HQET $b$ quarks fields,  is basically the same set of operators, albeit with
different weights, that appears in other $B$ decay rates as well as distributions.
 While we can easily identify these operators
and their dimensions, we cannot  compute their hadronic expectation
values from first principles, and we have to express them in function of a   number of  HQET  parameters, which  increases with powers of  $1/m_b$. 


The lowest-order terms of HQE are the dimension-three operators. 
In the HQET formalism,  $v_\mu$ is the $B$ meson velocity ($v^2=1$, $v_0 > 0$) and $b_v(x) = e^{-i m_b v\cdot x} \, b(x) $  is the $b$ field whose space time dependence  is determined by the residual momentum $k_\mu = p_\mu-m_b v_\mu$, which is due to binding effects of the heavy quark inside the heavy $B$ meson, and it is  of order $\Lambda_{QCD}$.
Owing to Lorentz invariance and parity there are
only two combinations which can appear, namely $O_3 =
\bar{b_v} \slashed v b_v$ and $O_3^\prime =
\bar{b_v}b_v$. Since the operators $b_v$ differ from the full QCD operators only by a phase redefinition,
the equalities $\bar{b_v} \slashed v b_v= \bar{b} \slashed v b $   and $\bar{b_v} b_v= \bar{b} b  $ hold. The matrix element of the former is
\begin{equation}
  \langle B|\bar{b} \slashed v b | B \rangle    = v^\mu  \langle B| \bar{b} \gamma_\mu b  | B \rangle =  v^\mu (2 m_B v_\mu) = 2 m_B 
\end{equation}
The penultimate equality follows because $\bar{b} \gamma_\mu b $ is the conserved $b$ quark
number current.
%
The hadronic expectation value of the operator $\bar{b_v} b_v  $ between the heavy meson states can be expanded in $1/m_b$, finding that it differs from the hadronic expectation value of the operator $\bar{b_v} \slashed v b_v  $ by terms of order $1/m_b^2$. Thus the matrix elements of the dimension-three contribution are known; they incorporate the parton model result which dominates asymptotically, i.e. for $m_b\rightarrow \infty$.

At order $1/m_b^0$ in the HQE, that is at the parton level,  the  perturbative corrections up to order $\alpha_s^2$ to the width and to the moments of the lepton energy and hadronic mass
distributions are known completely \cite{Melnikov:2008qs, Gambino:2011cq, Trott:2004xc, Aquila:2005hq, Pak:2008qt, Pak:2008cp, Biswas:2009rb}. The terms of order $\alpha_s^{n+1} \beta_0^n$, where $\beta_0$ is the first coefficient of the QCD $\beta$ function, $\beta_0= (33-2 n_f)/3$, have also been computed following  the
 Brodsky-Lepage-Mackenzie (BLM) procedure \cite{Aquila:2005hq, Benson:2003kp}.
 
By using the equation of motion in HQET, one can check that 
there are no matrix elements of dimension four
operators that occur in the HQE. This means that there are no corrections suppressed by a single power
of $\Lambda_{QCD}/m_b$.

The next order  is $ \Lambda_{QCD}^2/m_b^2$, and at this order the HQE includes
two  operators, called the kinetic energy  and the chromomagnetic operator. Their matrix elements,   $\mu^2_{\pi}$ and  $\mu^2_{G}$, respectively, are defined as 
\begin{eqnarray}
\mu^2_\pi &\equiv & \frac{1}{2 m_B}  \langle B|\bar{b_v} \vec \pi^2 b_v|B\rangle \nonumber \\
\mu^2_G &\equiv & \frac{1}{2 m_B}  \langle B|\bar{b_v} \frac{i}{2} \sigma_{\mu\nu} G^{\mu\nu} b_v|B\rangle
\label{mb-norm1}
\end{eqnarray}
where $\vec \pi= -i \vec D$, $D^\mu$ is the covariant derivative and $G^{\mu\nu}$ is the gluon field tensor. The matrix element $\mu^2_\pi $ is naturally associated with the average kinetic energy of the $b$ quark inside the $B$ meson while the matrix element $\mu^2_G$  is connected to  the $B^{*}-B$ hyperfine mass splitting. Both matrix elements generally depend on a cut-off $\mu$ chosen to separate soft and hard physics, which can be implemented in different ways, or schemes.
Perturbative corrections to the coefficients   of the kinetic operator  \cite{Becher:2007tk,Alberti:2012dn}
and  the chromomagnetic operator
\cite{Alberti:2013kxa, Mannel:2014xza, Mannel:2015jka}   have been
 evaluated   at order $\alpha_s$.

Two independent parameters, $\rho^3_{D,LS}$, are also needed to describe matrix elements of operators of dimension six, that is 
at order $1/m_b^3$. Their coefficients
have long been known at tree level, i.e.
neglecting  perturbative corrections \cite{Gremm:1996df}. Very recently  an analytical calculation of the $\alpha_s$ corrections for the coefficient $\rho^3_{D}$ has been presented~\cite{Mannel:2019qel}.

Starting at  order  $\Lambda_{QCD}^3/m_b^3$,   terms with an infrared sensitivity to the charm mass appear, at this order as a $\log m_c$ contribution \cite{Bigi:2005bh,Breidenbach:2008ua, Bigi:2009ym}. At higher orders these  contributions, sometimes dubbed intrinsic charm  contribution, in form of powers of $\Lambda_{QCD}/m_c$, have to be considered as well.  Indeed, roughly speaking, since $m^2_c \sim O( m_b \Lambda_{\mathrm{QCD}})$ and $\alpha_s(m_c) \sim O(\Lambda_{\mathrm{QCD}})$, contributions of order $\Lambda^5_{\mathrm{QCD}}/m^3_b \, m^2_c$ and $\alpha_s(m_c) \Lambda^4_{\mathrm{QCD}}/m^2_b\, m^2_c$  are expected comparable in size to  contributions of order $\Lambda^4_{\mathrm{QCD}}/m^4_b$.

Presently, the matrix elements have been  identified and estimated up to the order  $1/m_b^{4}$ and $1/m_b^{5}$ ~\cite{Dassinger:2006md, Mannel:2010wj,Heinonen:2014dxa}.
In HQE the number of  independent parameters needed to describe the nonperturbative physics of matrix elements grows with the order in $1/m_b$. 
At dimension seven  and eight, nine and  eighteen independent matrix elements appear, respectively, and for higher orders one has an almost
factorial increase in the number of independent parameters.

\subsubsection{Mass schemes}
\label{massschemes}

In QED, the location of the divergence in the  propagator of the electron can be taken as a physical definition of the electron mass, and it is indicated as on-shell or pole mass. 
This definition is not naturally extended to quarks, which are confined and  can never be seen as asymptotic states. 
 While
not measurable per se due to confinement, 
one can still define a pole mass for quarks in a formally consistent way within perturbation theory. However, this mass will be plagued  by ambiguities related to
non-perturbative effects in QCD, the so-called renormalon ambiguities (for a review see for instance~\cite{Beneke:1998ui}),  when related to observable quantities.
Alternative definitions of  mass for a quark can be used, each with its own advantages and disadvantages, but all requiring a careful description of the adopted framework (prescription or scheme).

The HQE nonpertubative parameters depend on the heavy quark mass, although sometimes the infinite mass
limit of these parameters is taken.
They are affected by the
 particular theoretical scheme that is
used to define the quark masses.

A commonly used definition of the mass of the quark is the minimal subtraction (MS) mass, which corresponds to the running renormalized mass in perturbative QCD, when, in dimensional regularization, the finite parts of the relevant counterterms are set to zero. In the $\overline{\rm MS}$ subtraction scheme, also  $\ln(4\pi)$ and $\gamma_E$ factors are subtracted off. The $\overline{\rm MS}$  prescription has the advantage of computational simplicity.
 The mass $m_b^{\overline{\rm MS}}(\mu)$ depends on a scale $\mu$ and 
 it is not affected by renormalon ambiguities. It is sometimes referred as a short-distance mass, since it is well
defined in the infrared regime. 
The $\overline{\rm MS}$ is quite appropriate for describing heavy flavour production,  but  not  for treating heavy meson decays, where the dynamics is characterized by  scales lower than the heavy scale $m_b$. 
 
Alternative scheme have been proposed, sometimes referred as threshold schemes; we list the most commonly used to describe heavy quarks in heavy mesons.
In the kinetic scheme\cite{Bigi:1994ga,Bigi:1996si,Bigi:1997fj}, the so-called “kinetic mass” $m^{kin}_b(\mu)$
is the mass entering the non-relativistic expression for the kinetic energy of a heavy
quark. It is defined  by introducing an explicit factorization  scale,   and subtracting the physics at scales below this scale from the quark-mass definition. More technically, its definition requires using heavy-quark sum rules for semileptonic $b\to c$ decays in the small velocity (SV) limit. %

Other examples of threshold schemes are the PS (Potential subtracted) scheme \cite{Beneke:1998rk} and the 1S scheme\cite{Hoang:1998ng, Hoang:1998hm, Hoang:1999zc}.
The PS mass  and   the  kinetic  mass  are similar, in the sense that they  both  subtract out the troublesome infrared part by introducing an explicit factorization scale.  The PS scheme is based on the properties of nonrelativistic
quark-antiquark systems, whose dynamics depends on the total static energy. The contribution to the  potential from the region of small momenta, identified as the source of the
leading renormalon, is subtracted from the PS mass.
The 1S mass is defined as
half the energy of the 1S state
 $\Upsilon$ state calculated in perturbation theory.
 In the 1S scheme there is a mismatch with the usual perturbation theory, overcome by
a working tool, the so-called `$\Upsilon$  expansion', whose validity has been questioned \cite{Uraltsev:2004ra}.
%

%



\subsection{Exclusive decays into light leptons}
\label{subsectionExclusive decays}

In this section we discuss the exclusive semileptonic CKM favoured  $B \to D^{(\ast)} \ell  \nu_\ell$ decays, when  $\ell=e, \mu$. Neglecting lepton masses, their  SM differential ratios 
can be 
written as
\begin{eqnarray}
\frac{d\Gamma}{dw}(\btodslnu)
&=&  \frac{G_F^2}{48 \pi^3}  (m_B-m_{D^\ast})^2 m_{D^\ast}^3 \chi (w)  (w^2-1)^{\frac{1}{2}}
 |V_{cb}|^2  |\eta_{EW}|^2 |{\cal F}(w)|^2
 \nonumber \\
 \frac{d\Gamma}{d w} (\btodlnu)  &=&
 \frac{G_F^2}{48 \pi^3}\,   (m_B+m_D)^2  m_D^3 \,
(w^2-1)^{\frac{3}{2}}\,
 |V_{cb}|^2  |\eta_{EW}|^2 | {\cal G}(w)|^2
 \label{diffrat0}
\end{eqnarray}
where  $m_X$ is the mass of the $X$ meson, $p_X$ its 4-momentum and $w$ is 
the  recoil parameter, defined as $w = p_B \cdot p_{D^{(\ast)}}/(m_B \, m_{D^{(\ast)}})=v_B\cdot v_{D^{(\ast)}}$;  $v_B$ and $v_{D^{(\ast)}}$ are the 4-velocities of the initial and final-state mesons. The recoil parameter is related  to the energy transferred to the leptonic pair $q^2=(p_B-p_{D^{(\ast)}})^2= (p_\ell+p_{\nu_\ell})^2$, namely $w=(m_B^2+m_{D^{(\ast)}}^2-q^2)/(2 m_B m_{D^{(\ast)}})$. In the $B$ meson rest frame the expression for $w$ reduces to the Lorentz boost $w= \gamma_{D^{(\ast)}}=E_{D^{(\ast)}}/m_{D^{(\ast)}}$. The values of the recoil parameter are limited by kinematics.  The superior limit occurs when $ q^2=q^2_{min}=m_\ell^2$, that is at $w=(m_B^2+m_{D^{(\ast)}}^2)/(2 m_B m_{D^{(\ast)}})$, assuming massless leptons.  The inferior limit  (the zero recoil point) is at $w=1$, and corresponds at $ q_{max}^2=(m_B-m_{D^{(\ast)}})^2\simeq 11 $ GeV$^2$. Intuitively an higher $q^2$ (lower $w$) corresponds to an higher mass of the virtual $W$ boson, which, at the two-body decay level, implies a lower “kick” to the ${D^{(\ast)}}$.
\begin{figure}[t!]
  \begin{center}
    \includegraphics[width=0.95\linewidth]{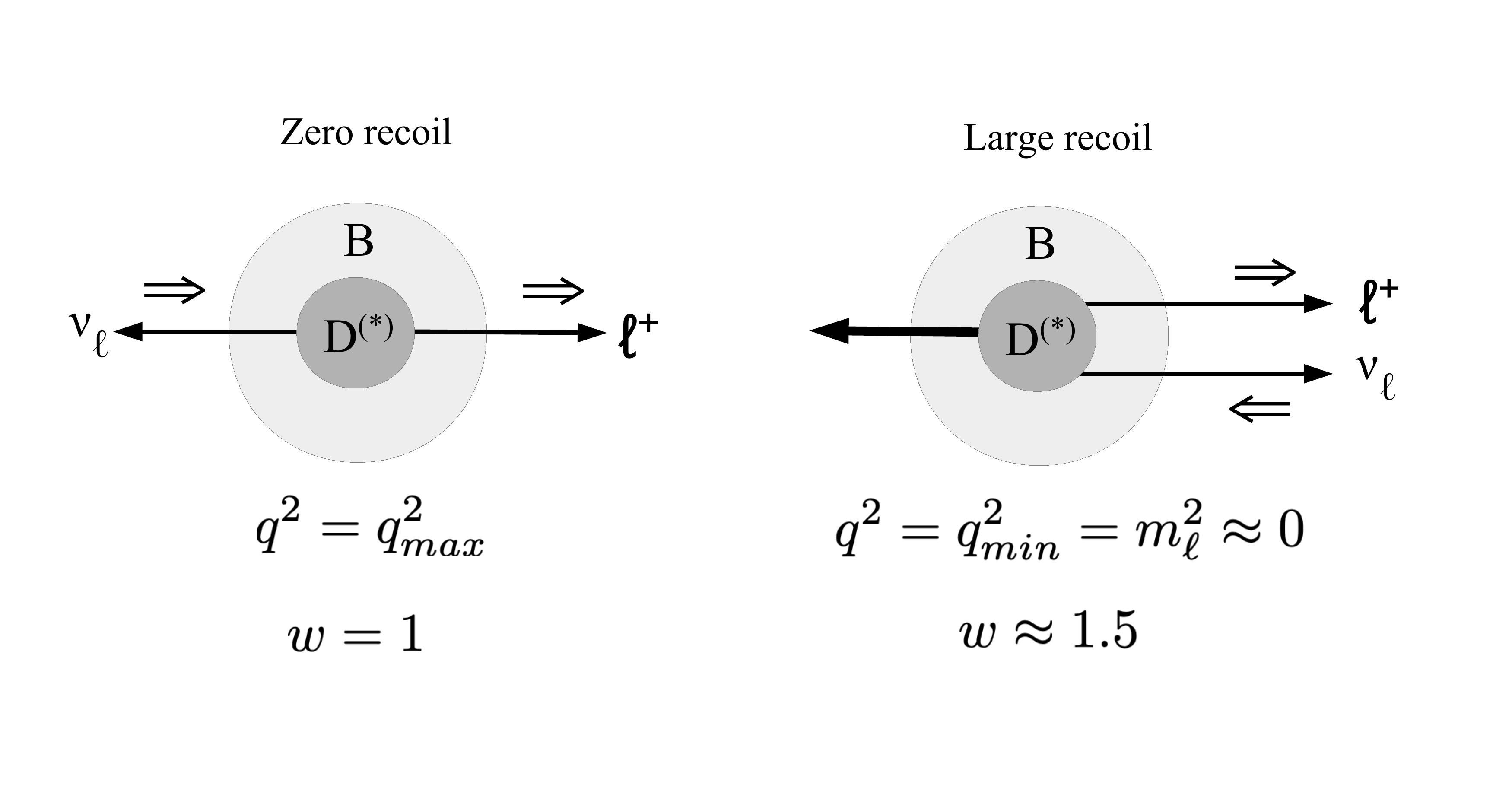}
  \vspace*{-15mm}     
  \caption{In the "zero recoil" kinematic configuration the hadron is at the rest in the $B$ meson rest frame, so the two leptons are produced back to back. In the "large recoil" limit the leptons are produced parallel and opposite to the hadron that acquires the largest momentum. The spin of the right-handed neutrino and the left-handed positron are also depicted.}
  \label{fig:semilep_limits}
  \end{center}
\end{figure}

In figure \ref{fig:semilep_limits}, we give an illustration of the kinematics of the decays at low and high $q^2$  in the $B$ meson rest frame. These simple pictures can be useful to gain some intuition about semileptonic $B$ decays. For example the large {\it helicity} suppression at zero recoil of \btodlnu decay compared to \btodslnu decay, can be easily understood:
the lepton and the neutrino are back to back, this means that the component of the total angular momentum of the leptons along their line of flight is unity and cannot be compensated by the pseudoscalar $D$ meson. At the other extreme, $q^2\approx 0$, where the hadron recoil velocity is maximum, the lepton and the neutrino are parallel and their combined spin along this direction is null. For the \btodslnu decays this means that the $D^{\ast}$ is fully polarized having null spin projection along the lepton direction.

 As seen in  (\ref{diffrat0}),  the differential cross sections are proportional to \begin{itemize} \item  the squared modulus of the CKM matrix element: $|V_{cb}|^2$ \item  a single form factor, ${\cal F}(w)$ and  ${\cal G}(w)$, for \btodslnu and \btodlnu, respectively
\item  $\eta_{EW}$,  a structure-independent   correction  factor  that  accounts for electroweak effects \cite{Sirlin:1981ie}. In the literature, a long-distance EM radiation effect (Coulomb correction) is sometimes added to this factor\cite{Bailey:2014tva}
\item a phase space factor, $(w^2-1)^{1/2}$  and  $(w^2-1)^{3/2}$ for  \btodslnu decays and \btodlnu decays, respectively, that vanish at the zero recoil point. For \btodslnu there is an additional  phase space factor $\chi (w)$  
 \beq \chi (w)= (w+1)^2 \left( 1 + \frac{ 4 w}{w + 1} \frac{ m_B^2 - 2 w \, m_B m_{D^\ast} + m_{D^\ast}^2}{(m_B-m_{D^\ast})^2} \right)\eeq
 \end{itemize}
 
The hardship of the extraction of $|V_{cb}|$ is due to the presence of the form factors, which cannot be computed in the framework of perturbation theory. In the
heavy quark limit ($m_{b/c} \to \infty $), that is to lowest  order in heavy quark effective theory, heavy quark symmetries predict that
both form factors equal a single universal Isgur-Wise
function,  ${\cal F}(w) = {\cal G}(w)  = {\cal  \xi} (w)   $, which is
absolutely normalized to unity at zero recoil,  that is  ${\cal \xi }(w=1) =1 $.
This property has  an intuitive reason. The no-recoil point corresponds to the kinematic situation where the $D$ meson stays at rest in the rest frame of the decaying $B$ ($v=v^\prime$); the decaying $b$-quark, at rest, is transformed into a $c$-quark, also at rest. The light hadronic cloud does not notice the flavour change $b \to c$ and it is transferred from the $B$ to the $D$ meson with probability one. 
The form factor function is identical for $B \to D$ and $B \to D^{\ast}$ transitions, because these are related by the heavy-quark spin symmetry.
For a realistic analysis, corrections to the heavy-quark limit have to be considered. 
At zero recoil, the heavy quark symmetries also provide the structure of the symmetry breaking non-perturbative corrections at finite heavy quark mass $m$, which start at order $1/m^2 $ and  $ 1/m$ for the  ${\cal F}(w=1)$ and ${\cal G}(w=1)$ form factors, respectively. 

In order to extract $|V_{cb}|$, we  need  not only to compute the form factors, but also to measure experimental decay rates. The advantage in the computation of the form factors provided by the heavy quark symmetries at $w=1$  has the hindrance that the differential rates in  \eqref{diffrat0} vanish
 at zero-recoil.
Thus
 one  needs to extrapolate  the experimental points taken at $w \neq 1$ to the zero recoil point $w=1$, using a parameterization of the dependence on $w$ of the form factors, which introduces additional uncertainties.
%
In other words, the $|V_{cb}|$ determination may proceed according to the following steps:
\begin{enumerate}
    \item[1)]
    theoretical determination of the form factors $ {\cal F}/{\cal G}$ at zero recoil $w = 1$;
    \item[2)]
        theoretical parameterization of the $w$ dependence;
    \item[3)]
 experimental measurements of the exclusive decays rates at non-zero recoil points, yielding the  products $ |\eta_{EW}|^2 \: |{\cal F}(w)|^2\; |V_{cb}|^2$  or  $ |\eta_{EW}|^2 \: |{\cal G}(w)|^2\; |V_{cb}|^2$;
    \item[4)]
    extrapolation of the experimental points to zero recoil and $|V_{cb}|$ extraction.
\end{enumerate}
\noindent Since  a few years there is an endeavor to amend this strategy, by calculating form factors directly at  non-zero recoil points, with evident advantages on the extraction of $|V_{cb}|$. Some results are already available in  the \btodlnu channel. 

Several parameterizations for the momentum dependence of the form factors are on the market.
Traditionally, the form factors are  parameterized with an explicit pole and a sum of effective poles, see e.g. Ball and Zwicky \cite{Ball:2001fp, Ball:2004ye} and 
Becirevic and Kaidalov\cite{Becirevic:1999kt}. Although these parameterizations capture some known properties of form factors, in general they do not allow  an easy quantification of systematic uncertainties.
Recent determinations adopt  a  more systematic approach that aims at exploiting the positivity and analyticity properties of two-point functions of vector currents. In these parameterizations 
 $w$ is mapped onto a complex variable $z$ via the conformal transformation
$
z= (\sqrt{w+1}-\sqrt{2})/(\sqrt{w+1}+\sqrt{2})
$. The form factors may be written in form of an
expansion in $z$, which converges rapidly in the kinematic
region of heavy hadron decays. The coefficients of the expansions  are subject to unitarity  bounds  based  on  analyticity \cite{Meiman:1963,Okubo:1972ih,Singh:1977et,Bourrely:1980gp}. 
To this type belong the
 CLN
(Caprini-Lellouch-Neubert) \cite{Caprini:1997mu},
 the BGL
(Boyd-Grinstein-Lebed)  \cite{Boyd:1994tt} and the  BCL (Bourrely-Caprini-Lellouch) \cite{Bourrely:2008za} parameterizations.
Further details are given in section~\ref{parameterizations}.
The experimental measurements of the form factors are described in sections~\ref{BDstarchannel} and~\ref{BDchannel} for the \btodslnu and \btodlnu decays, respectively.

\subsubsection{Form factors}
\label{Formfactors}

 From the field theory point of view, it is  convenient to define form factors as coefficients of independent Lorentz structures appearing in the hadronic transition matrix elements. 
In the framework of HQET, the independent Lorentz 4-vectors are the velocities of the two  mesons, rather than their momenta.
This can be intuitively understood by considering that
in the heavy flavour limit,
$m_{b,c} \to \infty$ ($m_b/m_c$ fixed),
when the weak current changes the flavour $b \to c$,
the light degrees of freedom inside the meson become  aware of the change in the  heavy quark velocities, $v_B  \to  v_{D^{(\ast)}} $ ($v_B \equiv p_B/m_B$, $v_{D^{(\ast)}} \equiv p_{D^{(\ast)}}/m_{D^{(\ast)}}$),
 rather than of the change in momenta.
Since the only scalar formed from the velocities ($v_B^2= v_{D^{(\ast)}}^2=1$ by definition) is  $ w= v_B \cdot v_{D^{(\ast)}}$, 
we can set \cite{Manohar:2000dt}
\bea
\frac{\langle D|V^\mu|B \rangle }{\sqrt{m_B m_D}}  &=&  h_+ (w) (v_B+v_D)^\mu+ h_-(w) (v_B-v_D)^\mu \nonumber \\
 \frac{\langle D^\ast|V^\mu|B \rangle}{\sqrt{m_B m_{D^\ast}}}  &=&  h_V (w) \varepsilon^{\mu \nu\rho \sigma} {v_B}_\nu {{v_{D^*}}_{\rho}} {\epsilon^{\ast}}_\sigma \nonumber \\
\frac{\langle D^\ast|A^\mu|B \rangle}{\sqrt{m_B m_{D^\ast}}}  &=& i h_{A_1} (w) (1+w) \epsilon^{\ast ^\mu}-i \left[  h_{A_2}(w) v_B^\mu + h_{A_3} (w) v_{D^\ast}^\mu \right] \epsilon^\star \cdot v_B
\label{FF11}
\eea
where $\epsilon^{\ast \mu}$ is the $D^\ast$  polarization vector,  which respects the equality $\sum_{\alpha=1}^3 \epsilon^{\ast \mu}_\alpha  \epsilon^{\ast \nu}_\alpha= - g^{\mu \nu} + v_{D^\ast}^\mu  v_{D^\ast}^\nu$.  In the conventional, relativistic normalization  of the meson states  $ |B(D^{(\ast)})\rangle $,  the  factor $1/ \sqrt{m_{B(D^{(\ast)})}}$ on the left side of Eqs.~(\ref{FF11}) is omitted; its addition pertains to  a mass independent renormalization \cite{Manohar:2000dt}.

In the heavy flavour limit
there is only one form factor, the Isgur-Wise function $\xi (w)$ \cite{Isgur:1989vq, Isgur:1989ed}. In that limit, the form factors become
\beq
h_+ (w)= h_V (w)= h_{A_1} (w)  = h_{A_3} (w)= \xi (w) \qquad
h_-(w)= h_{A_2}(w)=0
\label{clnrelhq}
\eeq
The form factor ${\cal G}(w)$ in ~\eqref{diffrat0} can be expressed as a combination of $h_+(w)$ and $h_-(w)$ \cite{Manohar:2000dt}
\beq
 {\cal G}(w)= h_+(w) -\frac{m_B-m_D}{m_B+m_D} h_-(w)
\eeq
Similarly, the form factor ${\cal F}(w)$ can be written as \cite{Manohar:2000dt}
\bea
{\cal F}(w) &=& \left\{ 2 (1-2 w r + r^2) \left[ h_{A_1}^2 + \left( \frac{w-1}{w+1} \right) h_V^2 \right] +\right. \nonumber \\
&+& \left[(1-r) h_{A_1} + (w-1) (h_{A_1}-h_{A_3}-r  h_{A_2}) \right]^2 \} \times \nonumber \\
& \times & \left\{ (1-r)^2 + \frac{4 w}{w+1} ( 1- 2 w r + r^2) \right\}^{-1}
\eea
where $r=m_{D^{\ast}}/m_B$. The form factor ${\cal F}(w)$ is dominated by the axial vector form factor $ h_{A_1}$ as $w \to 1$. 
%
 It is sometimes convenient
to define two ratios of the form factors
\beq
R_1= \frac{h_V}{h_{A_1}} \qquad R_2= \frac{h_{A_3} + r h_{A_2} }{h_{A_1}} 
\label{R1R2}
\eeq
In the infinity mass limit, heavy quark spin symmetry implies that $R_1 = R_2 =1$, independently of the $w$ value.

With respect to comparison with experimental results, the above definition of form factors is not the most convenient, since 
 the combinations of form factors most easily obtained from data are those appearing in a sum of squares in the differential rates, namely, the helicity amplitudes. They  are particular linear combinations of the original form factors, and thus simply form a different basis for the description of the matrix elements.
In  the \btodslnu decay, one can use three helicity amplitudes, labeled $H_\pm$  and $H_0$,  corresponding to the three polarization states of the $D^\ast$, two transverse and one longitudinal.
 The form factor ${\cal F}(w) $ can be expressed in terms of the helicity amplitudes as
 \beq
\chi (w) | {\cal F}(w) |^2 = \frac{1- 2 w r +r^2}{12 m_B m_{D^\ast} (1-r)^2}
\left( H_0^2(w) + H_+^2(w)+ H_-^2(w) \right)
 \eeq
 The helicity amplitudes, in turn, depend on the $h_x(w)$ form factors
 \bea 
 H_0(w) &=& \frac{\sqrt{m_B m_{D^\ast} }}{1- 2 w r +r^2} (w+1) \left[ (w-r) h_{A_1}(w) - 
 (w-1) ( r h_{A_2}(w)+h_{A_3}(w) \right] \nonumber \\
 H_\pm(w) &=& \sqrt{m_B m_{D^\ast} }(w+1) \left[  h_{A_1}(w) \pm 
\sqrt{\frac{w-1}{w+1}}h_V(w) \right]
\label{heloamp}
 \eea
Other
details on the $w$ dependence of form factors and helicity amplitudes is given in section~\ref{parameterizations12}.

\subsubsection{Zero recoil and beyond}
\label{Formfactors2}
 
 Since more than a decade,  the lattice community performs computations of the $B \to D^{(\ast)}$ form factors.
 The difficulties related to heavy fermions on lattice can be n\"aively summarized by observing that direct simulation of high mass such $m a \geq 1$, where $a$ represent a lattice spacing, gives discretization errors out of control.   As of today
$ m_b \sim 1/a
$ and  no direct simulation is possible. The main
 way out is  the usage of
effective theories, as HQET
\cite{Isgur:1989vq} and Non-Relativistic QCD (NRQCD) \cite{Caswell:1985ui}. In broad terms, they eliminate high degrees of freedom, aided by systematic expansions in $\Lambda_{QCD}/m_b$.
The downside is the introduction of new sources of errors (matching of HQET to QCD, renormalization, control of extrapolation, etc.) to take care of.

Another common approach to non-perturbative calculations of form factors are  QCD sum rules.
The sum rules are  based on the general idea of calculating a relevant quark-current correlation function and relating it to the hadronic parameters of interest via a dispersion relation. They have reached  wide application for calculation of exclusive  amplitudes and form factors in the form of
light cone sum rules (LCSR), employing light-cone OPE of the relevant  correlation functions. Uncertainties may originate from the truncation of the expansions, the input parameter uncertainties, and
the assumption of quark-hadron duality.
Direct sum rules calculations, without extrapolations, hold in the kinematic region of large recoil (small $q^2$), where the lattice calculation are substantially more difficult, and are in this respect complementary to lattice analyses.

Let us now report recent results in literature, starting from the $B \to D^{\ast} \ell  \nu_\ell$ channel, which  is less suppressed in the phase space and whose branching fractions are  more precise (even twice) in the majority of experimental measurements.


The form factor for the $B \to D^{\ast} \ell  \nu$ channel,
in the lattice unquenched $N_f= 2+1$  approximation has been estimated at zero recoil.
The FNAL/MILC collaboration, which used Wilson fermions for both $c$ and $b$
heavy quarks, gives \cite{Bailey:2014tva}
\beq  {\cal F}(1)
=0.906\pm 0.004_{stat} \pm  0.012_{sys}  \label{VcbexpF2}  \eeq
The first error is statistical and the second one  is the sum in quadrature of all systematic errors.
The total uncertainty is around  the (1-2)\% level.
The largest error is the heavy quark discretization error related to the Fermilab action. 

A more recent value of the lattice form factor ${\cal F}(1)$ at zero recoil has been presented by the HPQCD collaboration, which used the fully relativistic
 HISQ (Highly
improved
staggered quarks) action for light, strange and charm quarks and the  NRQCD  action for the $b$ quark \cite{Harrison:2017fmw}
\beq  {\cal F}(1)
=0.895\pm 0.010_{stat} \pm  0.024_{sys}  \label{VcbexpF2HISQ}  \eeq
The dominant error arises from missing $O(\alpha^2_s)$ matching of NRQCD currents to QCD.
Both the  results in (\ref{VcbexpF2}) and (\ref{VcbexpF2HISQ}) are in good agreement.
Another recent calculation  by  HPQCD  focuses  on $B_s \to D_s^\ast \ell \nu_\ell$~\cite{McLean:2019sds}. They use  the HISQ action for all valence quarks in order to perform the normalizations  of  all  required  currents  non-perturbatively and avoid a large source of systematic uncertainty.
From their result for ${\cal F}_s(1)$ they extract ${\cal F}(1)$ by using the ${\cal F}(1)/{\cal F}_s(1)$  ratio computed in their older paper~\cite{Harrison:2017fmw}, and
 obtain~\cite{McLean:2019sds}
 \beq {\cal F}(1)= 0.914 \pm 0.024 \eeq in agreement with all previously mentioned determinations. All the above form factor values are reported in Table \ref{tab:formf}.

The  LANL/SWME collaboration is working~\cite{Bailey:2017xjk, Bhattacharya:2018gan, Bhattacharya:2020xyb} to reduce the charm discretization error, the dominant ($\sim 1\%$) error in~\cite{Bailey:2014tva},   to below the percent level~\cite{Bailey:2020uon} by using an improved version of the Fermilab action, the Oktay-Kronfeld action. 
Their calculation is carried out on the $N_f=2+1+1$  MILC  HISQ  ensembles, at two lattice spacings $a\sim 0.12$, 0.09 fm and  pion masses $m_\pi\sim 220$, 310 MeV.
Preliminary results for  $B \to D^{\ast} \ell  \nu_\ell$ decays form factor $h_{A1}$ at zero recoil are reported.
A crucial planned  step will be to improve the currents up to order $\lambda^3$, where $\lambda_{b,c} \sim \Lambda_{QCD}/2 m_{b,c}$.
They also plan to analyze two more data sets measured, to include other physical pion masses and finer lattices and to increase statistics.

At the current level of precision, it would be important to extend
form factor unquenched  calculations for   $B \to D^{\ast}$ semileptonic decays
  to non-zero recoil, in order to reduce the uncertainty due to the extrapolation to $w=1$.
Indeed, at finite momentum transfer,  only old  quenched lattice results are  available \cite{deDivitiis:2008df}.
Stimulated by this objective, theory work on lattice is rapidly progressing. 

Nearly final results at non-zero recoil, with $w \in [1,1.1]$,  are already available from the Fermilab/MILC collaboration \cite{Aviles-Casco:2017nge, Aviles-Casco:2019vin, Aviles-Casco:2019zop}.
Their latest  analysis includes 15 MILC asqtad ($a^2$, tadpole improved) ensembles with $N_f=2+1$ flavors of sea quarks and lattice spacings ranging from $a \sim 0.15$ fm down to 0.045 fm.  The valence light quarks employ the asqtad action, whereas the $b$ and $c$ quarks are treated using the Fermilab action.  The analysis shows a larger slope at small recoil than the experimental measurements,  and the source of this behavior is currently under investigation.

The work in progress of the JLQCD collaboration is  based on  M$\mathrm{\ddot{o}}$bius domain-wall quarks,  at zero and non-zero recoil, from $N_f= 2+1$
QCD~\cite{Kaneko:2018mcr}. The systematics of the continuum and chiral extrapolation are under investigation.
A recent update~\cite{Kaneko:2019vkx} extends the $w$ range to $w \in [1,1.1]$ and simulates $b$ quark masses up to 0.7 $a^{-1}$ (at lattice cutoffs $a^{-1}\sim 2.4$, 3.6 and 4.5 GeV) to control discretization errors. 
Their preliminary results  for $h_{A_1}(1)$ are  in reasonable agreement with the previous estimates by Fermilab/MILC~\cite{Bailey:2014tva}  and HPQCD~\cite{Harrison:2017fmw}.


For the \btodlnu decays, 
 lattice-QCD calculation of the hadronic form factors
at  non-zero  recoil have  become available since 2015 \footnote{Prior results at non-zero recoil were only available  in the quenched approximation \cite{deDivitiis:2007otp}.}.
In 2015,  the FNAL/MILC collaboration has calculated the form factors
for a range of recoil momenta and parameterized their dependence on momentum transfer
using  the BGL z-expansion. Their analysis employs ensembles at four values of the lattice spacing ranging between approximately 0.045~fm and 0.12~fm. The $z$ expansion fit to lattice-only data gives \cite{Lattice:2015rga}
\beq
{\cal G}(1)= 1.054 \pm 0.004_{stat} \pm  0.008_{sys}   
\eeq
Two months later,  new results on
\btodlnu
form factors at non-zero recoil were announced by the HPQCD Collaboration~\cite{Na:2015kha}. Their results are based on NRQCD action
for  $b$ quarks  and  the  
HISQ action for $c$ quarks, together with $N_f=2+1$ MILC gauge configuration.
By using the CLN parameterization they obtain at zero recoil
\beq
{\cal G}(1)= 1.035\pm 0.040  
\eeq
Both the FNAL/MILC~\cite{Lattice:2015rga} and HPQCD~\cite{Na:2015kha}  estimates for the form factors at zero recoil are reported in Table \ref{tab:formf}. They are in good agreement, although the HPQCD one has larger errors coming mainly from discretization effects and 
 the systematic uncertainty associated with the perturbative matching, as in  the $B \to D^\ast$ case.

\begin{table}[bt!]
\footnotesize
\begin{center}
\caption{Latest lattice form factor estimates at zero recoil}
\vspace*{4mm}
\label{tab:formf}
\begin{tabular}{l l c || l c }
\hline 
Collaboration	& Refs. & ${\cal F}(1)$ 	&	 Refs.	&	 ${\cal G}(1) $  \\
\hline
FNAL/MILC  & \cite{Bailey:2014tva} 			& $0.906\pm 0.004 \pm  0.012 $			&	   \cite{Lattice:2015rga} & 
$ 1.054 \pm 0.004 \pm  0.008 $    \\
HPQCD  & \cite{Harrison:2017fmw} &	
$0.895\pm 0.010 \pm  0.024 $			&	  \cite{Na:2015kha}  & $ 1.035\pm 0.040 $       \\
HPQCD  & \cite{McLean:2019sds}	&
$0.914 \pm 0.024 $ &  	&  \\
\hline
 & & ${\cal F}^{ B_s \to D^\ast_s}$(1) &  & ${\cal G}^{ B_s \to D_s}$(1) \\
 \hline
HPQCD & \cite{McLean:2019sds} &  $0.9020 \pm 0.0096 \pm 0.0090$ & \cite{Monahan:2017uby} 
 & $1.068\pm 0.004$
\\
 \hline 
Atoui et al. &  &   & \cite{Atoui:2013zza}
 & $1.052 \pm 0.046$
\\
 \hline 
\end{tabular}
\end{center}
\end{table}

Until the very recent  LHCb measurement \cite{Aaij:2020hsi}, the lattice QCD results for $ B_s \rightarrow D_s^{(\ast)} $ form factors could not be compared with experiment.
Now the  $ B_s \rightarrow D_s^{(\ast)}  \ell  \nu_\ell $  decays supply a new method for precisely determining $|V_{cb}|$. 
These decays are more advantageous from the the point of view of lattice, since the larger mass of the valence $s$ quark compared to $u$ or $d$ quarks makes the calculations of the form factors less computationally expensive.

There are two analyses of the $B_s \to D_s^\ast$ zero-recoil form factors \cite{Harrison:2017fmw,McLean:2019sds}, both from the HPQCD collaboration using $N_f$=2+1+1 MILC HISQ ensembles. These analyses differ in the treatment of the $b$ quark.
The calculation of \cite{Harrison:2017fmw} uses an NRQCD $b$-quark, while \cite{Harrison:2017fmw} uses the relativistic ‘heavy-HISQ’ approach on fine ensembles down to $a \sim 0.45$~fm to avoid the main systematic uncertainty, which comes from the perturbative current matching known to $O(\alpha_s)$.
The  results are in agreement, and in Table \ref{tab:formf} we have reported
the more recent value
 ${\cal F}^{ B_s \to D^\ast_s}(1)= 0.9020 \pm 0.0096_{stat} \pm 0.0090_{sys}$ \cite{McLean:2019sds}.

 Lattice  QCD  calculations  of $B_s \rightarrow D_s$ form factors have already been performed at high $q^2$, close to zero recoil, where statistical errors are smaller.
 The signal/noise degrades exponentially as the spatial momentum of the meson in the final state grows.  Systematic errors from missing discretization (and relativistic) corrections also grow away from zero recoil. 
%

A recent published result for the zero-recoil vector form factors  
${\cal G}^{B_s \to D_s}(1)=1.068\pm 0.004$ was provided by the HPQCD Collaboration \cite{Monahan:2017uby} and it is reported in Table \ref{tab:formf}. The dominant source of uncertainty is due to discretization  effects, followed by perturbative matching uncertainties.  In Table \ref{tab:formf} we also report the zero recoil  value given by a $N_f=2$ determination which uses twisted Wilson quarks  
${\cal G}^{B_s \to D_s}(1)=  1.052 \pm 0.046$ \cite{Atoui:2013zza}.
The MILC collaboration has determined the ratios  between the semileptonic decay $\bar B^0 \to D^+ \ell^- \bar \nu_\ell$ and
$ \bar B^0_s  \rightarrow D_s^{+}  \ell^- \bar  \nu_\ell $ \cite{Bailey:2012rr}, and, more recently,
 the ratios of the scalar and vector form factors for the decays $B_s \to K \ell \nu_\ell$ and
$ B_s \rightarrow D_s  \ell  \nu_\ell $ \cite{Bazavov:2019aom}.
They have used $N_f=2+1$ asqtad ensembles, and the clover action with Fermilab interpretation
for $b$ and $c$ valence quarks.
Preliminary results on  semileptonic $ B_s \rightarrow D_s $ form factors have also been presented by the RBC/UKQCD Collaboration 
 \cite{Flynn:2016vej, Flynn:2019any, Flynn:2019jbg}.
 In the valence sector they have used domain wall fermions for $u$/$d$,  $s$ and $c$ quarks,  whereas $b$ quarks have been simulated with the relativistic heavy quark action.

Very recently, the  HPQCD Collaboration has  presented a lattice QCD determination of the $ B_s \rightarrow D_s  \ell  \nu_\ell $  scalar and vector form factors over the  full  physical  range  of  momentum  transfer \cite{McLean:2019qcx}. They work  with  a  highly  improved  quark  action and cover a range of values of the lattice spacing that includes very fine lattices and results from lighter than physical $b$ quarks. 

In alternative to lattice,   form factor  estimates are available via zero recoil sum rules, giving
 \cite{Gambino:2010bp, Gambino:2012rd}
${\cal F}(1) = 0.86 \pm 0.02 \label{gmu} $,
in good agreement with the lattice value in \eqref{VcbexpF2}, but  slightly lower in the central value.
Recently, information on all form factors parameterizing matrix elements of the basis of dimension-six operators, including those appearing only in connection of new physics effects, has become available in the framework of QCD LCSR  ~\cite{Gubernari:2018wyi}, and exploited for $|V_{cb}|$ determinations from
 $ B \rightarrow D^{(\ast)} \, \ell \, \nu_\ell$ decays~\cite{Bordone:2019vic}.

\subsubsection{Unitarity bounds}
\label{parameterizations}

As mentioned above,
the extraction of $|V_{cb}|$ 
involves an extrapolation to the zero-recoil point, for which a parameterization of the form factors in terms of $w$ is needed.
In this section we describe briefly parameterizations built on the basis of dispersion relations
and unitarity  bounds.
Since  more than 50 years,  it has been known that nontrivial constraints on an hadronic form factor can be derived starting from a given inequality on a suitable integral of the  square modulus of the form factor,  along the unitarity cut.
Let $F(t)$ denote a generic form factor, depending on a variable $t$, which is real analytic in the complex $t$-plane cut along the positive real axis from the lowest unitarity branch point $t_+$ to $\infty$. The essential inequality just mentioned is expressed as 
\beq
\int_{t_+}^\infty dt \rho(t) |F(t)|^2 < I
\label{ineq}
\eeq
where both the function $\rho(t) \ge 0$  and the quantity $I$ are known.
Such integral condition can be provided by  an observable  or, alternatively, by the dispersion relation satisfied by a suitable correlator.
The positive spectral function of the correlator has, by unitarity, a lower bound involving the modulus squared of the relevant form factor. Therefore, the constraints derived in this framework are often referred to as “unitarity bounds”.
Through complex analysis, this condition leads to constraints on the values at interior points or on the expansion parameters. 

Many  applications  of  this  approach  to the heavy-to-heavy and   heavy-to-light form factors,   the  light-meson form factors,  the electro-magnetic form factor of the pion, the strangeness changing $K \pi$ form factors, and so on, can be found in literature (for a  review see  e.g.~\cite{Abbas:2010jc}).
Here we sketch the   application  to $B \to D^{(\ast)}$ decays; details and demonstrations can be found elsewhere (e.g. in~\cite{Bourrely:1980gp,Boyd:1997kz, Boyd:1997qw, Grinstein:2015wqa} and  therein).
%
The two-point QCD  function $\Pi^{2P}$ of a flavor-changing current $J$ is rendered finite by making one or two subtractions, leading to  dispersion relations. For one subtraction one can write
\beq
\chi \equiv \frac{\partial  }{\partial q^2}  \Pi^L(q^2) = \frac{1}{\pi} \int_{0}^{\infty} dt \frac{{\rm{Im}}\, \Pi^L(t)}{(t-q^2)^2} 
\eeq
where $\Pi^{2P}(q^2) = 1/q^2 (q^\mu q^\nu -q^2 g^{\mu\nu}) \, \Pi^T(q^2) + q^\mu q^\nu /q^2 \, \Pi^L(q^2)$. Similarly for   $\Pi^T(q^2)$. The functions $\chi$ may be computed reliably in perturbative QCD for values of $q^2$ far from the kinematic region where the current  can  produce  manifestly  nonperturbative  effects, like pairs of hadrons. 
For heavy quarks    a reasonable choice is  $ q^2 = 0 \ll (m_b+m_c)^2$.
The spectral functions ${\rm{Im}}\, \Pi$ are evaluated by unitarity,  inserting  into  the  unitarity  sum  a  complete  set  of states $X$ that couple the current to the vacuum
\beq
{\rm{Im}}\, \Pi^L = \frac{1}{2}\sum_X (2 \pi)^4 \delta^4(q-p_X) | \langle 0|J|X \rangle|^2
\eeq 
Since the sum is semi-positive definite, by taking a subset of hadronic states,  namely the states with only the two heavy mesons, one can obtain a strict inequality. We recover an upper bound of the form of ~\eqref{ineq} in the pair-production region, that is
\beq 
\frac{1}{\pi \chi} \int_{t_+}^\infty dt \frac{W(t) |F(t)|^2}{(t-q^2)^2} \le 1
\label{inequalityrough}\eeq
where $W(t)$ is a computable  function, expressed as a product of phase-space factors, and $t_+=(m_B+m_{D^{(\ast)}})^2$ is  the  unitarity  threshold. 
A similar result holds for $\Pi^T$. In  the  case  of semileptonic $B$ decays,
$q^2$ ranges  from approximately zero to  $t_-=(m_B-m_{D^{(\ast)}})^2$, but  the  form  factors  can  be continued  analytically  in  the complex  plane. 

The  inequality \eqref{inequalityrough} makes clear how the perturbative calculation constrains the magnitude of the form factor in the pair-production region, but to  turn it into a constraint in the semileptonic region requires that the integrand is analytic below the pair-production threshold $t \le t_+$.
The form factor $F(t)$  may have poles arising from the contribution of bound states, the $B_c$ resonances  with the appropriate quantum numbers.

Let us consider a conformal variable transformation as
\beq
z(t;t_0) \equiv \frac{\sqrt{t_+-t_0} -\sqrt{t_+-t}}{\sqrt{t_+-t}+\sqrt{t_+-t_0}}
= \frac{t-t_0}{(\sqrt{t_+-t}+\sqrt{t_+-t_0})^2}
\label{confoprmalv}
\eeq 
 This transformation maps the  complex $t$-plane, which contains a branch cut extending from $t_+$ to $\infty$,  onto the unit disc $|z| < 1$  in the $z(t)$ plane.
The branch point $t_+$ is  mapped onto $z=  1$  and  the  two  edges  of  the  unitarity  cut $t \ge t_+$ map  to  the  boundary $|z|=  1$.  
We can see that $z$ is real for $t \le t_+$ and a pure phase for $t \ge t_+$;
  $t_0$ is a free parameter that  represents the $t$-point mapped onto the origin of the $z$ plane.
 Let us observe that a simple pole in $t_0$ can be eliminated by multiplying by $z(t;t_0)$. The change of variable \eqref{confoprmalv} simplifies the next step, aimed at
 isolating  factors  that encode the nonanalytic behavior of the form factor $F(t)$, so that the  inequality ~\eqref{inequalityrough} becomes
 \beq
 \frac{1}{2 \pi i} \int_C \frac{dz}{z} |\phi(z) P(z) F(z)|^2 \le 1
 \label{expF00}
 \eeq 
 where $C$ is the unit circle in the complex $z$ plane. Here $\phi(z) $ is an outer function, defined in  complex  analysis   as  an  analytic  function  lacking zeros in $|z|<1$, and  $ P(z)$  is known as a Blaschke factor (or inner function), a products of suitable chosen $z(t;t_0)$ removing singularities due to the resonances below the  pair-production threshold. Since $\phi(z) P(z) F(z)$ is analytic on the whole unit disc, we have managed to isolate the analytic structure of the form factor and can write an expansion as
 \beq 
 F(t) = \frac{1}{\phi(t;t_0) |P(t)|} \sum_{n=0}^\infty a_n z^n(t;t_0)
 \label{expF}
 \eeq 
with  unknown coefficients $a_n$. This  coefficients are different for each form factor, and must be determined by experiment. 
Inserting \eqref{expF} back into \eqref{expF00} gives the constraint
\beq
\sum_{n=0}^\infty a_n^2 \le 1,\\
 \label{expF10}
\eeq 
which is known  as  the  weak  unitarity  constraint, and holds for each set of form factors sharing parity and spin quantum numbers. All possible functional dependence of the form factor $F(t)$ consistent with the analyticity, unitarity, and explicit QCD information discussed before are now encoded into the coefficients $a_n$, which are highly constrained by  ~\eqref{expF10}. A randomly chosen shape for a form factor would almost inevitably have some $a_n>1$, disallowing the bond given by ~\eqref{expF10}. 
In case the allowed kinematic range for $z$ has $|z| \ll 1$, as for semileptonic $B\to D^{(\ast)}$ decays, the convergence of the series is geometrically fast, and only the first few   $a_n$ coefficients are  relevant to the shape of the form factor.
In that case the sum in ~\eqref{expF10} is well approximated by a sum limited  by a finite number, depending on the form factor analysed, rather than by $\infty$.

One  can  further  constrain  the  coefficients  of  the $z$ expansion by considering several decays related by crossing symmetry; these additional constraints are known as the strong unitarity constraints.

We conclude this section by observing that in case of semileptonic $B \to D^{(\ast)}$ decay the above formalism is generally expressed in terms of parent and daughter velocity 4-vectors, and the parameter $w=(m^2_B+m^2_{ D^{(\ast)}}-t)/2 m_B m_{ D^{(\ast)}}$. 
The latter kinematic variable turns out to be more convenient than the momentum transfer variable $t= (p_B-p_{D^{(\ast)}})^2$ in the framework of  heavy quark symmetries.
The conformal transformation $t \to z$ in  \eqref{confoprmalv} becomes $w \to z$, and we have
\cite{Boyd:1997kz}
\beq
z(w;{\cal N}) \equiv \frac{\sqrt{1+w} -\sqrt{2 {\cal N}}}{\sqrt{1+w}+\sqrt{2 {\cal N}}} \qquad \qquad {\cal N}\equiv \frac{t_+-t_0}{t_+-t_-}
\label{confoprmalv2}
\eeq 
where $z(w;{\cal N})$  maps the physical region $1< w <1.5$ onto $0< z <0.056$ and vanishes at $w= 2{\cal N}-1$. There are several parameterizations of the form factors for semileptonic $B \to D^{(\ast)}$ decays based on the approach outlined in this section; we  discuss two examples in the next section.

\subsubsection{BGL and CLN parameterizations}
\label{parameterizations12}




The  unitarity and dispersion relations outlined in section ~\ref{parameterizations} are at the basis of several different parameterization for the exclusive semileptonic $B \to D^{(\ast)}$ decays.

Let us consider the $B \to D^\star$ channel.
%
In the so-called Boyd, Grinstein and Lebed (BGL) parameterization~\cite{Boyd:1994tt,Bourrely:2008za,Boyd:1995sq}, it is convenient to 
set
\bea
H_0(w) &= F_{1}(w)/\sqrt{\qsq} \ , \nonumber\\
H_{\pm}(w) &=  f(w) \mp m_{\Bs}m_{\Dssm}\sqrt{w^2 -1}g(w) 
\label{eq:Hel_amp_BGL}
\eea
These equalities define
the form factors $F_{1}(z)$, $f(z)$, and $g(z)$ in terms of the helicity amplitudes; looking at \eqref{heloamp}, we observe that $F_{1}(z)$ and $f(z)$ are connected to axial form factors, and $g(z)$  to the vector one.  
These new form factors can be expressed by  series in the variable $z$, as seen in section ~\ref{parameterizations}
\bea
f(z) &=& \frac{1}{P_{1^{+}}(z)\phi_{f}(z)}\sum_{n=0}^{\infty}a_{n}^{f}z^{n} \ ,\nonumber\\
F_{1}(z) &=& \frac{1}{P_{1^{+}}(z)\phi_{F_{1}}(z)}\sum_{n=0}^{\infty}a_{n}^{F_{1}}z^{n}  \ ,\nonumber \\ 
g(z) &=& \frac{1}{P_{1^{-}}(z)\phi_{g}(z)}\sum_{n=0}^{\infty}a_{n}^{g}z^{n}
\label{eq:Form_factors_BGL}
\eea
The $\phi$ functions are the outer functions~\cite{Boyd:1997kz}.
The $P_{1^{\pm}}$  factors are the Blaschke factors,
which take into account the sub-threshold $B_c$ resonances with the same quantum numbers as the current involved in the definition of the form factor, and depend on the masses of such resonances. Recent determinations can be found  in Refs.~\cite{Bigi:2017jbd,Gambino:2019sif}. The coefficients $a_n$ are the parameters that need to be fitted on data,  subject to unitary constraints
\beq 
\sum_{n=0}^{n_g} (a_n^g)^2 < 1, \qquad \qquad
\sum_{n=0}^{n_f} (a_n^f)^2 +\sum_{n=0}^{n_{F_1}} (a_n^{F_1})^2 < 1
 \label{expF13}
\eeq 
They ensure the convergence of the series over the whole physical region $0 < z < 0.056$~\cite{Gambino:2019sif}. The series are truncated at different $n_i$. A similar analysis can be done for the $B \to D$ channel.


Another common parameterization is the so-called  Caprini, Lellouch and Neubert  (CLN)  parameterization \cite{Caprini:1997mu}. This parameterization is based on the same unitarity bounds as  the  BGL parameterization, but it employs  strong unitarity constraints to reduce the number of parameters of the more general expansion. It makes  use of the relations among the form factors due to heavy quark symmetries (HQS), in particular of the connection, at the leading order in the $1/m_b$ expansion, of  all  the  form  factors  to  the  single  Isgur-Wise  function $\xi(w)$. 
In the heavy-quark limit, all form factors become identical  and equal to $\xi(w)$
(see ~(\ref{clnrelhq})). In order to incorporate corrections to that limit, one form factor, $F_{ref}(w)$,  is chosen  a reference form factor and  expanded around $w=1$. Its derivatives
are bounded by unitarity  relations of the kind of  \eqref{expF10}.
The first derivative, the slope,   is defined as
\beq
\rho^2=- \left.\frac{\partial F_{ref}(w)}{\partial w}\right|_{w=1}
\eeq
The ratio of all  other form factors with the reference one are obtained by including the leading short-distance and $1/m_b$ corrections, and  expressed in terms of the reference parameters as $\rho$. Roughly speaking we have, for each form factor $F(w)$
\beq 
F(w) = \left(\frac{F}{F_{ref}}\right)_{HQS} F_{ref}(w)
\eeq

%
%
For $B\to D^*\ell\nu$ decays, in the CLN parameterization, the more convenient variables are the leading form factor $h_{A_1}(w)$ and the ratios of form factors $R_{1}(w)$, and $R_{2}(w)$ 
defined in  Eqs.~\eqref{R1R2}. 
The  form factor $h_{A_1}(w)$  up to symmetry-breaking corrections coincides with the Isgur-Wise function, while the two form-factor ratios are equal to 1 in the heavy flavour limit, independently of $w$. The reference form factor is taken to be the axial vector form factor, see formula (35) in \cite{Caprini:1997mu}.
These parameters are expanded for
$w \to 1$, fixing the series coefficients using dispersive bounds. They
are given by \cite{Caprini:1997mu}
\bea
h_{A_{1}}(w) &= h_{A_{1}}(1)[1 - 8\rho^{2}z +(53\rho^{2} - 15)z^{2} - (231\rho^{2} - 91)z^{3}] \ , \nonumber\\
R_{1}(w) &= R_{1}(1) - 0.12(w - 1) + 0.05(w - 1)^{2} \ , \nonumber\\
R_{2}(w) &= R_{2}(1) + 0.11(w - 1) - 0.06(w - 1)^{2} 
\label{eq:CLN:param}
\eea
where $z=(\sqrt{w+1}- \sqrt{2})/ (\sqrt{w+1}+\sqrt{2})$. In the $B\to D\ell\nu$ decays, the reference function is taken to be ${\cal G} (w)$, yielding, in the $z$ variable \cite{Caprini:1997mu}
\beq 
{\cal G} (z) = {\cal G}(1)[1 - 8\rho_D^{2}z +(51\rho_D^{2} - 10)z^{2} - (252\rho_D^{2} - 84)z^{3}]. 
\eeq 

In this section, we have restricted our discussion  to  BGL and CLN  parameterizations, whose comparison has excited  lively discussions since a couple of years.  Indeed, in 2017 the reliability of the CLN approach has been questioned in both $B \to D\ell\nu_\ell  $ \cite{Bigi:2016mdz} and $B\to D^*\ell\nu_\ell$ channels \cite{Bigi:2017njr, Grinstein:2017nlq}.
Details and updates on the current situation are given in section \ref{result1}.

\subsection{Decays to excited $D$-meson states}
\label{DecaystoexcitedD-Mesonstates}

The  interest in
semileptonic $B$ decays to excited states of the
charm meson spectrum  derives mostly  by the fact that they
contribute
as a background to the direct decay $ B \to D^{(\ast )} \ell  \nu_\ell$ at the B factories, and, as a consequence, as
 a source of systematic error in the $|V_{cb}|$ measurements.
 Precise knowledge of the properties of the excited D meson states is important to reduce uncertainties in the measurements of semileptonic decays.

The spectrum of mesons consisting of a charm and a $\overline u$
or $\overline d$ (open charm mesons) is poorly known.
A QCD framework for their analysis can be set
up by using  HQET.
In the limit  of infinite heavy quark mass,   the spin  of the heavy quark $\vec{s_h}$
 is conserved
and  decouples from the total angular momentum of the light
degrees of freedom $\vec{j_l}$, which  becomes  a conserved quantity as well.
The
separate conservation in strong interaction processes of $\vec{s_h}$ and $\vec{j_l}$  permits a classification of heavy mesons of given radial (principal)  quantum
number according to the
value of $\vec{j_l}$.
Mesons can be collected in doublets: the two states in each doublet (spin partners)
have total angular momentum $\vec{J} = {\vec{j_l}} + 1/2 \, \hat{s_h}$ and parity
$P = (-1)^{L+1}$,  since ${\vec{j_l}} \equiv   {\vec{L}+{\vec{s_l}}}$, where $\vec{L}$ is the orbital angular momentum  and $\vec{s_l}$  the spin of the light degrees of freedom.
Within each doublet
the two states are degenerate in the limit of infinite heavy quark mass.

\begin{figure}[tb!]
  \begin{center}
  \includegraphics[width=0.9\linewidth]{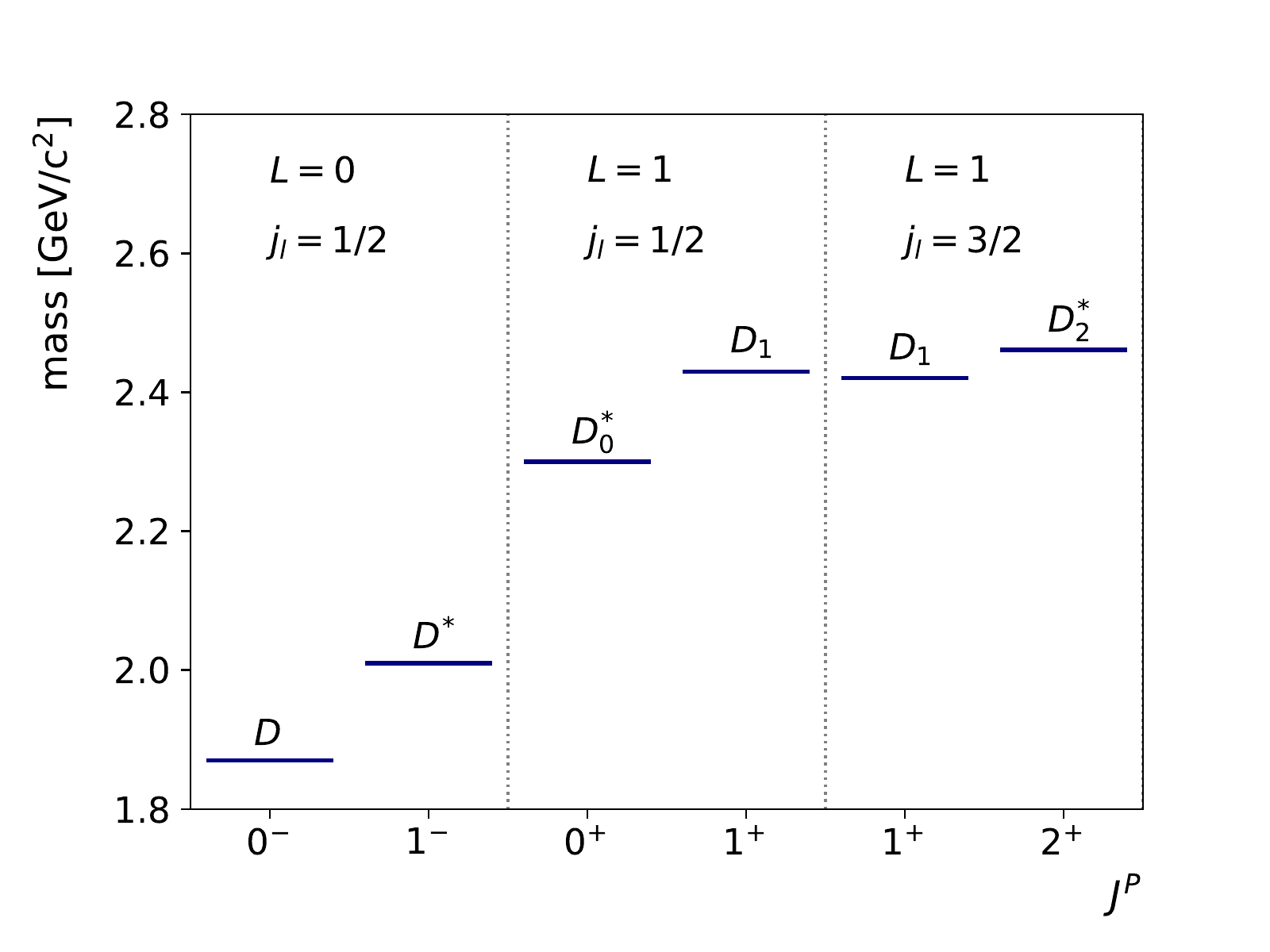}
  \caption{Low-mass $D$-meson spectrum. 
  For the interpretation of the  states see the text.}
    \label{fig:spectrum}
  \end{center}
\end{figure}


The low-mass spectrum includes the ground states, with principal (radial) quantum number $n=1$ and  $L = 0$ (1S, in the spectroscopic notation), which implies ${j_l}^P={\frac{1}{2}}^-$. The ground state doublet consists of two
states with $J^P = (0^-, 1^-)$,
that is $D$ and $D^\ast$ mesons~\footnote{The naming convention followed is to use $D^\ast(mass)$ to denote the states having $P=(-1)^J$, that is $J^P=0^+,1^-,2^+, \dots$ (natural spin-parity) and with $D(mass)$ all the others (unnatural spin-parity).}.


When $L=1$, there are four states ($1P$ states), which are generically referred to as $D^{\ast \ast}$ \footnote{Sometimes in literature this term is extended to include all particles in the low-mass spectrum
except the ground states.}. The doublet having ${j_l}^P={\frac{1}{2}}^+$ is named ($D_0^\ast, D_1$) and corresponds to $J^P=(0^+, 1^+)$. These states are identified with $D_0^\ast(2300)$  (it was $D_0^\ast(2400)$, see~\cite{Tanabashi:2018oca}) and $D_1(2430)$.  The doublet having ${j_l}^P={\frac{3}{2}}^+$ is named ($D_1, D_2^\ast$) and corresponds to $J^P=(1^+, 2^+)$.
These states are identified with 
$D_1(2420)$ and $D_2^\ast (2460)$.
%
%
For the states with $j_l=\frac{3}{2}$,  the two-body decay  $ D^{\ast\ast} \rightarrow D^{(\ast)} \pi $ must be in the D-wave to conserve $j_l$. Therefore, the width should be narrow and relatively easy to observe.
$D_1(2420)$ and $D^\ast_2(2460)$
 have relatively narrow widths, about 30~MeV, and have been observed and studied  by a number of experiments
since the nineties. In contrast, for the state with $j_l=\frac{1}{2}$  the same two-body decay should proceed in S-wave, and widths should be wide. Therefore,   $D^\ast_0(2300)$ and $D_1(2430)$  are more difficult to detect due to the large width, about 200-400~MeV,  and were not observed prior to the $B$-Factory era. The state $D^\ast_0(2300)$  has  been  studied  by Belle, BaBar and LHCb collaborations in exclusive $B$ decays~\cite{Abe:2003zm, Aubert:2009wg, Aaij:2015vea, Aaij:2015sqa, Aaij:2016fma}, while the state $D_1(2430)$   has been observed by Belle collaboration~\cite{Abe:2003zm}, but its production in semileptonic $B$ decays, studied by BaBar~\cite{Aubert:2008ea} and Belle~\cite{Liventsev:2007rb} gives contradictory results.
We have reported the above mentioned states in Fig.~\ref{fig:spectrum}.

When a new state is observed, the concept of a heavy quark spin doublet is the guiding principle to understand the nature of the observed state.
However, the spectroscopic identification for heavier states is
not very clear.
In 2010 BaBar has observed,  for the first time, candidates for the radial excitation (2S) of the $D^0$, $D^{\ast 0}$ and $D^{\ast +}$, as well as the $L=2$ excited states of the $D^0$ and $D^+$ \cite{delAmoSanchez:2010vq}. Resonances in the $2.4$-$2.8$  ${\mathrm{GeV/c}}^2$ region  of hadronic masses have also  been identified at LHCb
\cite{Aaij:2013sza, Aaij:2015vea, Aaij:2015sqa, Aaij:2016fma}.
%

%
%

Limits in the experimental scenario concerning $B$ decays into excited states are mirrored by  theoretical ambiguities. The analyses from Belle~\cite{Liventsev:2007rb} and BaBar~\cite{Aubert:2007qw}, which combined one additional pion to the ground and first excited states, revealed a couple of interesting anomalies. 

The first is the fact the $ B \to D^{\ast \ast} \to D^{(\ast)} \pi l \nu_l$ branching fraction is composed of approximately equal contributions from the $j_l=1/2$  and $j_l=3/2$ states. This is unexpected as most  theoretical calculations, using sum rules \cite{LeYaouanc:1996bd,Uraltsev:2000ce}, quark models \cite{Morenas:1996yq, Morenas:1997nk,  Ebert:1998km, Ebert:1999ga} (but not   constituent quark models, see e.g. \cite{Segovia:2011dg}),  OPE \cite{Leibovich:1997em, Bigi:2007qp} , indicate that  the narrow width states dominate over
the broad  $D^{\ast\ast}$ states (the ``1/2 vs 3/2" puzzle).

The other puzzle is that the sum of the measured semileptonic exclusive rates having $D^{(\ast)}$ in the final state is less than the inclusive one (``gap" problem)~\cite{Liventsev:2007rb, Aubert:2007qw}.
Indeed, decays into $D^{(\ast)}$ make up $\sim$ 70\% of the total inclusive $ B \to X_c l \bar \nu$ rate and decays into $D^{(*)} \pi$  make up another $\sim $ 15\%, leaving a gap of about 15\%. This  is  in  contrast  to  the  situation  with  the  tauonic  channels,  where  the branching fractions of the $B \to D^{(\ast)} \tau \nu_\tau$ saturate the inclusive $B \to X_c \tau \nu_\tau$ rate measured at LEP~\cite{Tanabashi:2018oca}. 
BaBar used the full dataset to improve the precision on decays involving $D^{(*)} \pi\, l \, \nu$  and to search for 
$D^{(*)} \pi\,\pi  l \, \nu$ decays \cite{Lees:2015eya}. These result have assigned about 0.7\% to the $D^{(*)} \pi\,\pi  l \, \nu$ branching ratio, reducing the significance of the gap from $7\sigma$ to $3 \sigma$.

One possible  weakness common to most theoretical approaches is that they are derived in the heavy quark limit and
corrections might be  large. For instance, it is expected that $1/m_c$ corrections induce a significant mixing
 between the two $D_1$ states, which could soften the 1/2-3/2 puzzle~\cite{Klein:2015doa}. The possibility of a larger than expected contribution of the first radial excitation of the $D^{*}$ to the $B$ semileptonic decay into charmed mesons has also been advanced~\cite{Bernlochner:2012bc, Becirevic:2017vxe}.
However,  no firm conclusion can be drawn until more high quality data on the masses and the widths of the orbitally excited $D$
meson states become available.

\subsection{Decays into heavy leptons}
\label{Exclusivedecaysintoheavyleptons}

Exclusive  $B$ decays into $\tau$ leptons were
first observed by the Belle Collaboration in 2007 \cite{Matyja:2007kt}.
Subsequent  measurements  by  BaBar  and  Belle    reported  branching  fractions  above-yet  consistent  with-the  SM  predictions until 2012, when a significant  excess  over  the  SM  expectation was reported by BaBar~\cite{Lees:2012xj}.
The discrepancy with the SM persists today,  triggering a relevant amount of theoretical  analyses. No extraction of $|V_{cb}|$  performed so far makes use of semitauonic $B$ meson decays.

Measurements  and  predictions  are  usually  quoted  as branching fraction ratio
\begin{equation}
R(D^{(\ast)}) \equiv  \frac{{\cal{B}}( B \to D^{(\ast)} \tau \nu_\tau)}{{\cal{B}}( B \to D^{(\ast)} \ell  \nu_\ell)}
\label{ratiotau0}
\end{equation}
where the denominator is the average for $\ell \in \{e, \mu\}$. This ratio
 is typically used instead of the absolute branching fraction
of $ B \to D^{(\ast)} \tau  \nu_\tau$ decays, in order to cancel  uncertainties common to the numerator and the denominator. These include $|V_{cb}|$ and several theoretical uncertainties on hadronic form factors and experimental reconstruction effects.
%
%
The ratio (\ref{ratiotau0}) tests the couplings of the charged gauge bosons to the different lepton families.
A discrepancy with the SM predictions challenges the universality of the SM couplings, and indicates physics beyond the SM.
Although  this ratio cannot be  used to determine $|V_{cb}|$ directly, its knowledge is still useful, indirectly, since possible new physics couplings would affect high precision semileptonic analyses aimed at $|V_{cb}|$ extraction, which motivates us to briefly outline the current experimental situation.

%
 In 2012-2013 the BaBar collaboration measured $R(D^{(\ast)})$ by using its full data sample \cite{Lees:2012xj, Lees:2013uzd}, and reported a significant excess over the SM expectation.
 In 2015 the Belle collaboration reported a measurement of $R(D)$ and $R(D^{\ast})$ \cite{Huschle:2015rga}, using the hadronic $B$-tagging in an analysis similar to the BaBar one. In the same year, LHCb collaboration reported the first measurement of $R(D^\ast)$ in $pp$ collisions \cite{Aaij:2015yra}. Both these measurements were above the SM expectations. Since then other measurements have been performed; here we report the full list:
 \begin{enumerate}
  \item  $R(D)$ and $R(D^\ast)$ with $\tau$ reconstructed in the $\tau\to \ell$ mode ($\ell \in \{e, \mu\}$), and using the hadronic $B$-tagging approach: BaBar 2012 \cite{Lees:2012xj,Lees:2013uzd}, Belle 2015 \cite{Huschle:2015rga};
 \item  $R(D^\ast)$ with the $\tau$ reconstructed in $\tau\to \mu$ mode: LHCb 2015 \cite{Aaij:2015yra};
 \item $R(D^\ast)$ with $\tau\to \ell$, using the semileptonic $B$-tagging: Belle \cite{Sato:2016svk} (this measurement has been superseded by the more recent combined $R(D)$ and $R(D^\ast)$ measurement \cite{Belle:2019rba} using the same tagging approach); 
 \item $R(D^\ast)$ and $\tau$ polarization with the $\tau$ reconstructed in hadronic $\tau\to\pi(\pi^0)\nu_\tau$ decay mode, and using the hadronic $B$-tagging: Belle 2016 \cite{Hirose:2016wfn};
 \item $R(D^\ast)$ with $\tau$ reconstructed in the hadronic $\tau\to 3\pi(\pi^0)\nu_\tau$ mode: LHCb 2017 \cite{Aaij:2017uff};
 \item $R(D)$ and $R(D^\ast)$  with $\tau\to \ell$, using the semileptonic $B$-tagging: Belle 2019 \cite{Belle:2019rba}.
 \end{enumerate}
 
\begin{figure}[tb!]
  \begin{center}
    \includegraphics[width=1\linewidth]{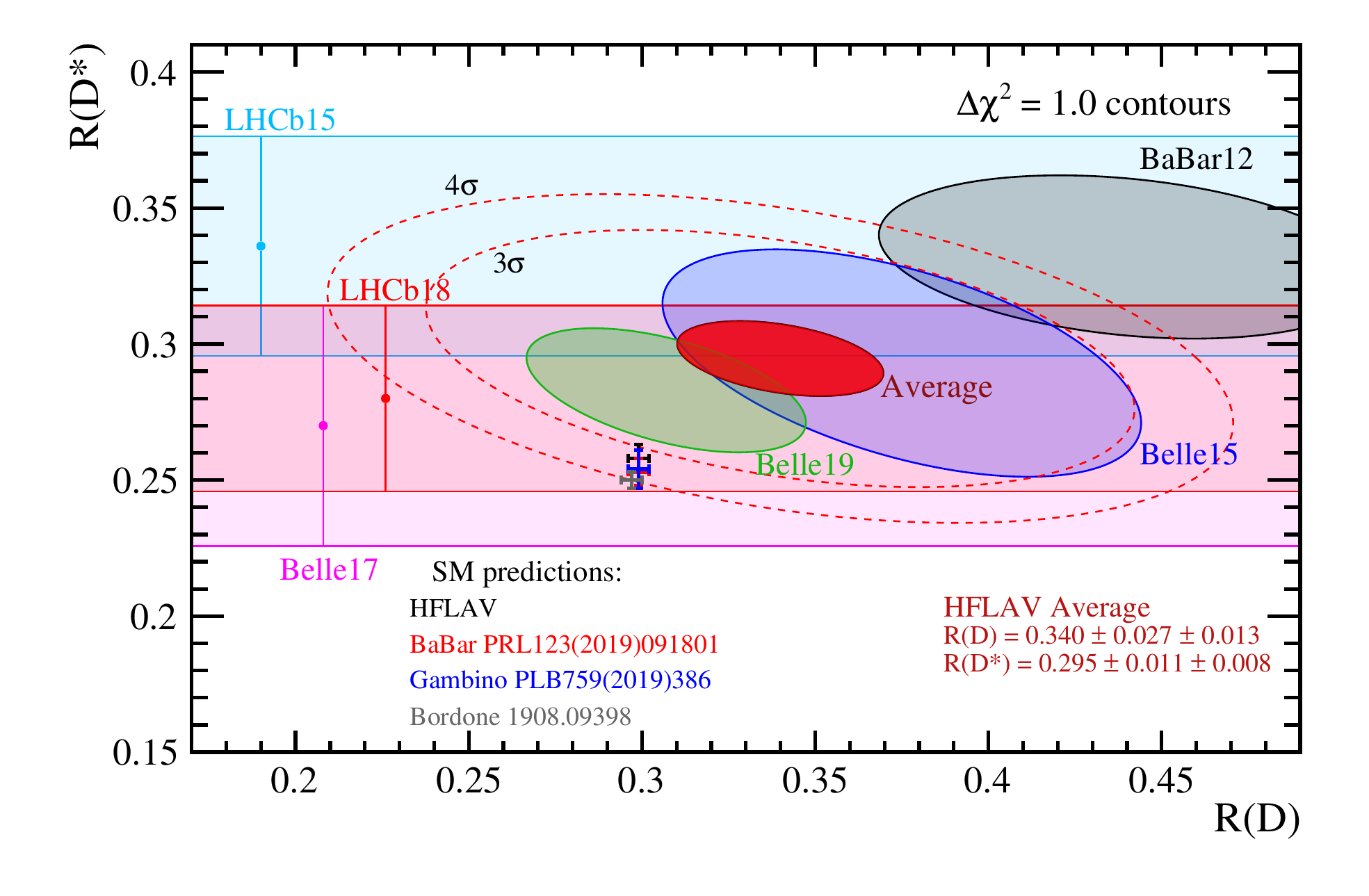}
  \caption{Measurements of $R(D)$ and $R(D^{\ast})$ and their two-dimensional HFLAV average \cite{Amhis:2019ckw}. Contours correspond to $\Delta\chi^2= 1$, i.e., 68\% CL for the bands and 39\% CL for the ellipses. The black point with errors is the average of the SM predictions used by HFLAV and obtained from \cite{Bigi:2016mdz,Bernlochner:2017jka,Bigi:2017jbd,Jaiswal:2017rve}. This prediction and the experimental average, deviate from each other by $3.1\sigma$. The dashed ellipses correspond to a $3\sigma$ (99.73\%) and $4\sigma$ contours. Also very recent predictions are reported.}
    \label{fig:rdsrds_summary}
  \end{center}
\end{figure}

\noindent By averaging the measurements \cite{Lees:2012xj,Huschle:2015rga, Aaij:2015yra, Hirose:2016wfn, Aaij:2017uff, Belle:2019rba}, the HFLAV Collaboration has found~\cite{Amhis:2019ckw}
\bea
R(D^\ast) &=& 0.295 \pm 0.011 \pm 0.008 \nonumber \\
R(D)  &=& 0.340 \pm 0.027 \pm 0.013  \qquad \qquad
\label{ratiotau}
\eea
where the first uncertainty is statistical and the second one is systematic. 
The average and the individual measurements included are shown in figure \ref{fig:rdsrds_summary}. 

Several theoretical  predictions for $R(D^\ast)$ in the SM have been performed, starting from 2012~\cite{Fajfer:2012vx}.
%
Some of them use the data presented by the Belle collaboration in 2017\cite{Abdesselam:2017kjf} and the BGL parameterization~\footnote{They were prompted by the debate on the different parameterizations outlined in section ~\ref{results633}.}~\cite{Bigi:2017jbd, Jaiswal:2017rve, Bernlochner:2017jka}. 
Their results are generally consistent with the older predictions, and 
their  arithmetic average, as given by the  HFLAV collaboration~\cite{Amhis:2019ckw}, is
\begin{equation}
R(D^\ast) = 0.258 \pm 0.005
\label{RDDSTAR}
\end{equation}
%

In the case of $R(D)$, lattice  SM predictions 
by  FNAL/MILC~\cite{Lattice:2015rga} and HPQCD~\cite{Na:2015kha} collaborations  have been averaged by the FLAG collaboration, yielding
$
R(D) = 0.300 \pm 0.008
$~\cite{Aoki:2019cca}.
Like for $ R(D^{\ast})$,  there are more recent  calculations~\cite{Bigi:2016mdz,Jaiswal:2017rve, Bernlochner:2017jka} that have performed analyses combining experimental data on \btodlnu decays from Belle and BaBar, and theory calculation; their arithmetic HFLAV average is~\cite{Amhis:2019ckw} 
\begin{equation}
R(D) = 0.299 \pm 0.003
\label{RDD333}
\end{equation}

The HFLAV predictions (\ref{RDDSTAR}) and (\ref{RDD333})  
are reported in figure \ref{fig:rdsrds_summary}.
The  averages for $R(D)$ and $R(D^\ast)$ in (\ref{ratiotau}) exceed the SM values by about 1.4$\sigma$ and 2.5$\sigma$, respectively. If one considers both deviations, the tension rises to about 3.1$\sigma$. 

More recent SM predictions \cite{Dey:2019bgc, Gambino:2019sif, Bordone:2019vic},  while compatible with the previous calculations, are slightly on the lower side, resulting in discrepancies with the HFLAV average between 3.3 and 3.9$\sigma$. We show in figure \ref{fig:rdsrds_summary} also these more recent predictions. 


\subsection{Comparison with baryon decays}
\label{Comparisonwithbaryondecays}

A significant sample of $\Lambda^0_b$ baryons is available at the LHCb experiment, opening the possibility to study  their  semileptonic decays and to
an interesting  comparison  with semileptonic  $B$ meson decays.
A $\Lambda$-type baryon consists of a heavy quark, and of a spin and isospin zero light di-quark.
 As in the $B$ meson case,  it can be viewed as a state containing a single heavy quark $Q$, dressed by light degrees of freedom to make up a color singlet hadron, and its decay can be discussed in the framework of the HQET.
 The eigenstates of the Lagrangian in HQET differ from those of the full theory in the baryon sector in the same way as in the meson sector. For the spin-1/2 $\Lambda_Q$ baryon the situation is in fact simpler, because the light degrees of freedom carry no angular momentum and hence there is no spin symmetry violating mass splitting. 
 
 Let us consider the semileptonic decay of a spin-1/2 baryon $\Lambda_Q$  to a  spin-1/2 baryon $\Lambda_{Q^\prime}$.  This transition is governed by the hadronic matrix elements of the flavor changing vector and axial vector currents.  They are conventionally parameterized in terms of six form factors $F_i$ and $G_i$, defined by
\setlength{\mathindent}{35pt}
\bea
  \langle \Lambda_{Q^\prime}(v^\prime,s^\prime) | \bar Q^\prime \gamma_\mu Q | \Lambda_{Q} (v,s) \rangle & =& \bar  u_{\Lambda^\prime} (v^\prime,s^\prime) \left[F_1 \gamma_\mu + F_2 v_\mu +F_3 v^\prime_\mu \right] u_{\Lambda} (v,s) \nonumber \\
\langle \Lambda_{Q^\prime}(v^\prime,s^\prime) | \bar Q^\prime \gamma_\mu \gamma_5Q | \Lambda_{Q} (v,s) \rangle &= & \bar u_{\Lambda^\prime} (v^\prime,s^\prime) \left[G_1 \gamma_\mu + G_2 v_\mu +G_3 v^\prime_\mu \right] \gamma_5 u_{\Lambda} (v,s) 
\eea
\setlength{\mathindent}{65pt}
where $v$ and $v^\prime$ are the velocities of the initial and final baryon.  The form factors depend on $w =v \cdot v^\prime =(m^2_Q+m^2_{Q^\prime} -q^2)/2 m_Q m_{Q^\prime}$, where $m_Q$ and $m_{Q^\prime}$ are the masses of the initial and final baryon, and $q^2$ is the squared invariant mass of the lepton pair.
 In the infinite quark mass limit,   $F_1=G_1= \zeta(\omega)$, a universal Isgur-Wise function,  and $F_2=F_3=G_2=G_3= 0$. 
 An alternate, helicity-based, definition of the form factors 
 was introduced in \cite{Feldmann:2011xf}.

 The leading power corrections to the decay rate at zero recoil  are of order $1/m^2_Q$.
 The semileptonic decay $\Lambda_b\to \Lambda_c\ell \nu_l$ is particularly simple to analyze near the zero recoil point $w= 1$, where  $q^2$  takes on its maximum value $q^2_{max}= (m_{\Lambda_b}-m_{\Lambda_c})^2$. In the limit of vanishing lepton mass, angular momentum conservation requires that the weak matrix element $ \langle \Lambda_c(v,s^\prime)|V_\mu-A_\mu|\Lambda_b(v,s)\rangle$
 depends only on the function $G_1(1)$. 
In semileptonic decay $\Lambda_b\to \Lambda_c\ell \nu_l$ a partial cancellation of $1/m_Q^2$ corrections at zero recoil was found, with the conclusion that large deviations from the infinite quark mass limit are unlikely, and the heavy quark expansion is well under control \cite{Falk:1992ws}.

 Form factors for  the   baryon decays  $\Lambda^0_b\to \Lambda_c^+\mu\bar \nu_\mu$ and $\Lambda^0_b\to p\mu^-\bar\nu_\mu$ are already available in lattice QCD.
They have been computed 
 using  RBC/UKQCD $N_f= 2 + 1$  flavors of dynamical domain-wall fermions,   six  different  pion  masses  and  two  different  lattice  spacings \cite{Detmold:2015aaa}. The importance of this computation is that, combined to a recent measurement by LHCb \cite{Aaij:2015bfa},  allows for an independent exclusive determination of the ratio $|V_{ub}|/|V_{cb}|$, as we will discuss in 
 section \ref{sec:lbtopmunu}.
 
Due to the possibility of new physics  in the ratio $
R(D^{(\ast)}) $, discussed in section \ref{Exclusivedecaysintoheavyleptons}, an  analogous ratio  for baryon decays, $R(\Lambda_c)= {\cal{B}} (\Lambda_b \to \Lambda_c \tau \bar \nu )/{\cal{B}} (\Lambda_b \to \Lambda_c \mu \bar \nu)$, has been identified  and analyzed\cite{Bernlochner:2018bfn}. 

\section{Experimental techniques}
\label{Experimentaltechniques}

\subsection{$B$-hadron production} 

The $b$-hadrons can be produced in different experimental environments:
from $e^+e^-$ annihilation, 
 collisions of protons or proton-antiproton collisions. 
The most recent results on $b$-hadron semileptonic decays come  from $e^+e^-$ experiments operating at the 
energy of the \FourS
and from $pp$ collisions at LHC. 

Understanding the features of the $b$-hadron production in various environments is crucial to understand the experimental setup and analysis techniques developed to study semileptonic decays. In the following we focus on the $b$-hadron production mechanism at the $B$-Factories and $pp$ colliders.

\subsubsection{$B$-Factories}

Studies of $B$ meson decays have been performed at $e^+e^-$ collisors working at the center-of-mass energy of $\sqrt{s}=10.58$~GeV, which corresponds to the mass of the \FourS resonance. The first two experiments working at this resonance were ARGUS (at DORIS accelerator, DESY, Germany) and CLEO (at CESR, USA).

The next generations of $e^+e^-$ collisors  have been the modern $B$-Factories, BaBar  and Belle,  designed to collect data produced in the collisions at PEP-II (at SLAC, USA) and KEKB (at KEK, Japan), respectively. A detailed description of both BaBar and Belle, their performances and their  analysis methods can be found in \cite{Bevan:2014iga}.

The main characteristic of the $B$-Factories was the very high luminosity (2 order of magnitude higher than older $e^+e^-$ collisors) 
achieved by the machines PEP-II  and KEKB. The  BaBar and Belle experiments stopped their operations in 2008 and 2010, respectively. Nowadays, a decade later, many analyses are still ongoing to exploit the full dataset collected by these two experiments. The present measurements of $|V_{cb}|$ are dominated by the $B$-Factories. 

\begin{figure}[tb!]
  \begin{center}
    \includegraphics[width=0.5\linewidth]{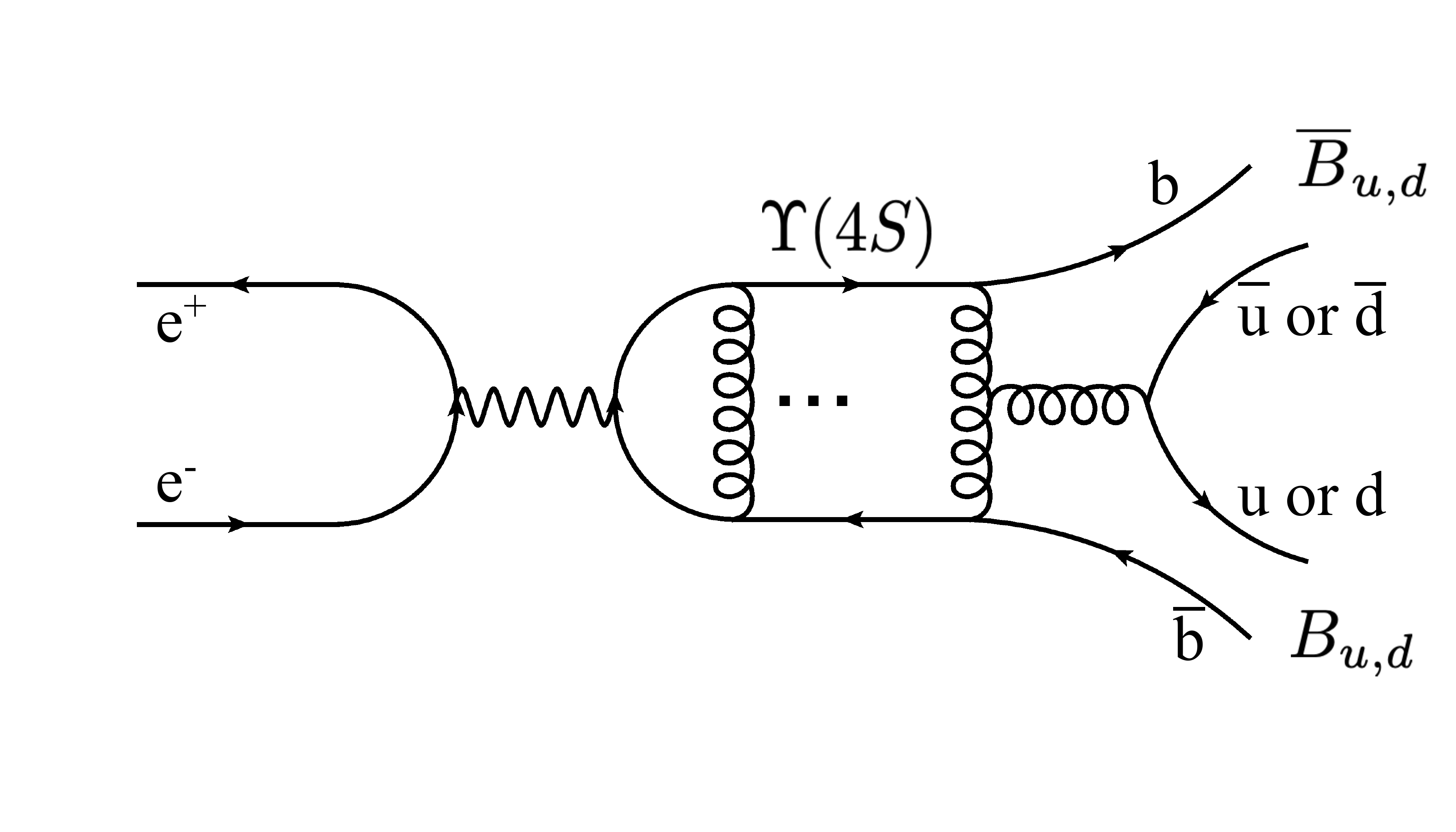} 
  \end{center}
  \vspace*{-10mm} 
  \caption{$B$ mesons at $B$-Factories are produced from the decays of the \FourS.}
  \label{fig:y4s_bb}
\end{figure}

At $B$-Factories, the $B$ mesons are produced through the decay of the \FourS. An illustration of the process involved is shown in figure~\ref{fig:y4s_bb}.
The \FourS is the lightest $b\overline{b}$ resonance with mass above the \BBb pair production threshold~\footnote{The \FourS mass is above the \BBb pair mass, so the decays proceed through strong decays which dominate over radiative or weak decays. The \FourS is accessible at  $e^+e^-$ colliders because the process $e^+e^-\to \gamma^* \to b\overline{b}$ allows only states with $J^{PC}=1^{--}$ quantum numbers.}.
This resonance decays almost exclusively in a couple of $B$ meson pairs. The probabilities to produce $B^0 \overline{B}^0$ and $B^+B^-$ from \FourS decays are about the same. The ratio of the branching fraction decays $f_{+-}/f_{00}$ differs slightly from unity because of the small difference due to Coulomb effect, which increase the rate when oppositely charged states are present in the final state. The current average value is $f_{+-}/f_{00} = 1.058\pm0.024$ \cite{Tanabashi:2018oca}. 

Because of the small mass difference between the \FourS state and \BBb pairs, the $B$ mesons are produced with very small momentum in the \FourS center of mass. In particular the $B$ meson momentum is $|\vec p_B| \simeq 320\mev$. For this reason the decay products of the two $B$'s are produced almost isotropically in the \FourS rest frame. Evens like these are usually called {\it spherical}. 

The integrated luminosity collected at \FourS energy was 426~fb$^{-1}$ and 711~fb$^{-1}$ at BaBar and Belle, respectively. The integrated luminosity collected by ARGUS and CLEO was only 0.2~fb$^{-1}$ and 16~fb$^{-1}$, respectively. The high luminosity has been paramount to study CP violation in $B$ mesons decay, because it allows the study of rare processes, with branching ratios of the order of $10^{-4}\div 10^{-6}$. The need to measure time-dependent properties of the $B$ meson decays has driven the design of the $B$-Factories. 
A unique characteristic of the $B$-Factories was the asymmetric energies of the colliding beams, so the \FourS was produced boosted. The boost allowed a better spatial separation of the two $B$ meson decay vertices. For example, in BaBar the boost was $\beta\gamma\approx 0.55$, resulting in an average distance between the two $B$ meson decay vertex of 250~$\upmu$m, 
which was in the capability of the vertex detector. 
To maximize the acceptance of the decay products of the boosted $\Upsilon(4S)$, BaBar and Belle detectors were offset from the interaction point by about 30~cm to keep high the acceptance in the direction of the higher energy beam, resulting in slightly asymmetric detectors.

At the energy $\sqrt{s}$ corresponding to the \FourS mass, the  
cross section of $e^+e^-\to \FourS$ is about 1.06$~\rm {nb}$, resulting in about $1.1\times 10^6/\rm{fb}^{-1}$ \BBb pairs. But at this energy, only about one forth of all the hadronic events produced are \FourS, the rest being non-\BBb events. The cross sections for some important processes at $\sqrt{s}=10.58$~GeV are reported in table~\ref{tab:y4s_cs}. These events are a background to the study of $B$ meson decays, called {\it continuum background}. In general they are rejected exploiting the differences between decays of the \FourS and the decays of the $e^+e^-\to q\overline q$. 
As said before, the $B$ mesons are produced almost at rest in the \FourS frame, so the decay products have a spherical topology, while in $e^+e^-\to q\overline q$ processes the tracks coming from the fragmentation of the two quarks produce a topology with two opposite jets. Furthermore, the average number of particles produced in the quark hadronization in $e^+e^-\to q\overline q$ processes is smaller than in $\FourS\to \BBb$ processes. The suppression of the continuum background is thus performed requiring a minimum number of tracks, usually three or four, and applying global event shape criteria that allow to separate jet-like events from more spherical events. 

Even with these requirements, the continuum, mainly the contribution from $e^+e^-\to c\overline{c}$, remains an important background in many semileptonic analyses. Therefore part of the data (about one tenth) are taken at a center of mass about 50~MeV below the \FourS mass, in order to have pure continuum events, needed for the study of the background. The study of these continuum events, corrected for the luminosity and for the small energy difference, can be used to predict both the absolute scale and the correct kinematics of the continuum background events.

As we will see in section \ref{susubec:btagging}, many semileptonic analyses gain a lot by an approach called {\it B-tagging} where the signal $B$ meson is reconstructed together with the second $B$ meson present in the event. The $B$-tagging is very effective in suppressing the continuum background, and more generally, to clean the event reconstruction. 

\begin{table}[bt!]
\footnotesize
\begin{center}
\caption{The cross section for some relevant processes at different colliders. The collected integrated luminosity for some of the experiments is reported in parenthesis.}
\vspace*{4mm} 
\label{tab:y4s_cs}
\begin{tabular}{l l c || l l}
\hline
Collider	& Process 				&	cross section & experiments \\
\hline
		&$b\overline b$			&	1.06~nb	&    BaBar  (426 ${\rm fb}^{-1}$) \\
$e^+e^-\to \FourS$	&$c\overline c$			&	1.30~nb&   Belle  (711 ${\rm fb}^{-1}$)\\
		&${d\overline d},{u\overline u},{s\overline s} $			&	2.09~nb &   \\
\hline
$e^+e^-\to Z$ & $b\overline b$	& 6.6~nb   &  ALEPH, DELPHI ($0.14~{\rm fb}^{-1}$), OPAL, L3\\				
\hline
$pp$ 7, 8 TeV	&  $b\overline b$ $2<\eta<5$	         &    72~$\upmu$b     	 &	LHCb (3 ${\rm fb}^{-1}$)\\
		        &  $b\overline b$ total                            &     ~295~$\upmu$b  &	CMS, ATLAS (25 ${\rm fb}^{-1}$ each)\\
\hline
$pp$ 13 TeV	& 	$b\overline b$ $2<\eta<5$	&	144~$\upmu$b  	&      LHCb (6 ${\rm fb}^{-1}$)\\
		& 	 $b\overline b$ total &       ~600~$\upmu$b  	&  CMS, ATLAS (150 ${\rm fb}^{-1}$ each)\\	
\hline
$p\overline p$  1.96 TeV & $b\overline b$ $|\eta|<1$	&     ~30~$\upmu$b  	&  CDF, D0 (10 ${\rm fb}^{-1}$ each)\\	
\hline
\end{tabular}
\end{center}
\end{table}

\subsubsection{Hadron Colliders}
\label{subsub:had_prod}

\begin{figure}[tb!]
  \begin{center}
    \includegraphics[width=0.90\linewidth]{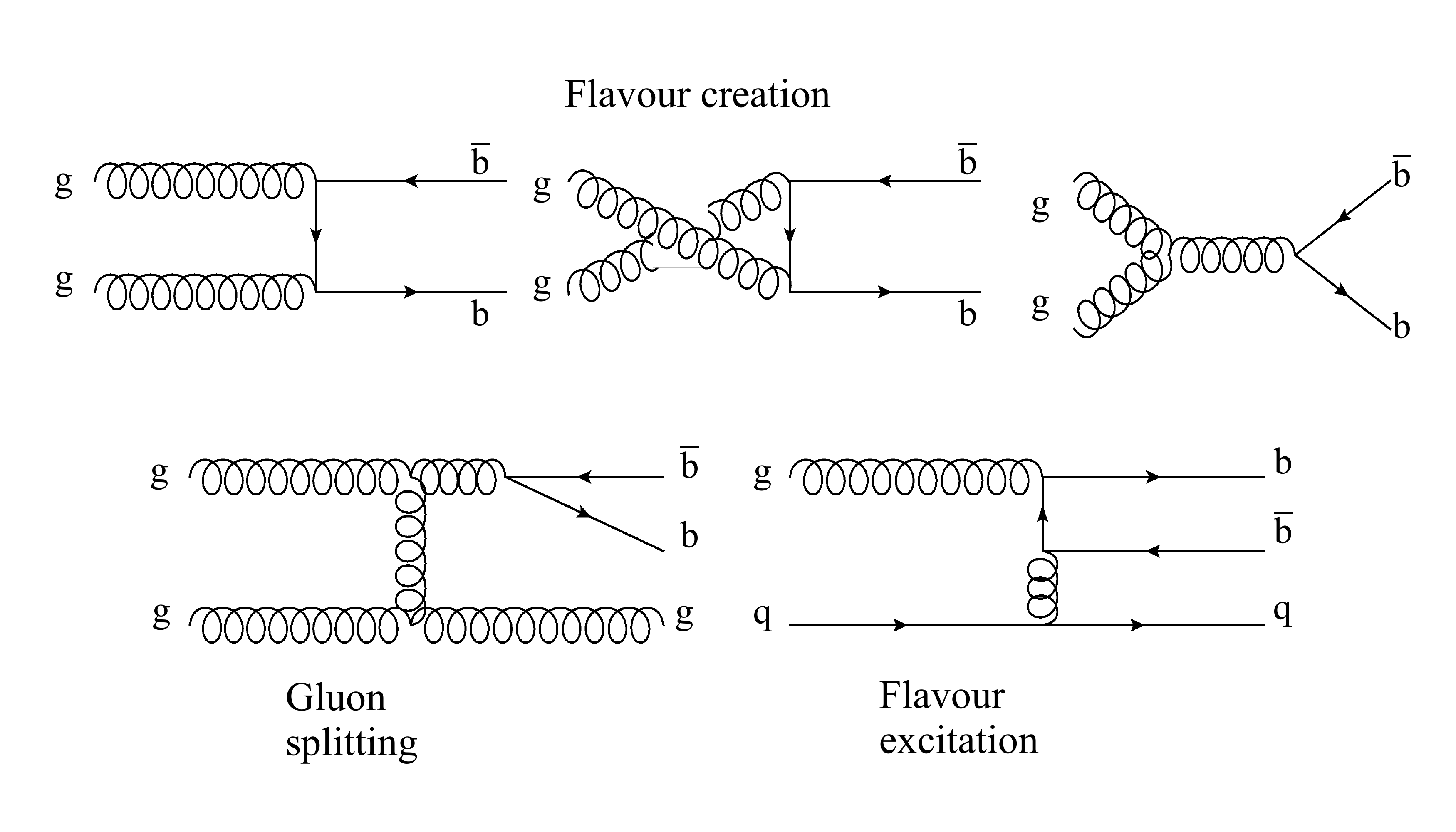} 
  \end{center}
   \vspace*{-10mm} 
  \caption{The leading order heavy flavour production processes dominant at LHC.}
  \label{fig:bb_prod}
\end{figure}

At LHC the production mechanism of $b$-quarks are the quark annihilation ${q \bar q}\to {b \bar b}$ and gluon fusion processes ${q \bar q},{gg}\to {b \bar b}$, with the latter ones largely dominating \cite{Nason:1999ta}. 
At leading order in perturbation theory O($\alpha_s$), we can draw the
tree diagram corresponding to the quark-antiquark annihilation and the flavour creation diagrams shown in figure~\ref{fig:bb_prod}, that is the gluon fusion diagrams in the $t$-, $u$- and $s$-channel (from left to right).  The ${q \bar q}\to {b \bar b} g$ parton process and the processes described by the gluon splitting and  the flavour excitation diagrams depicted in figure~\ref{fig:bb_prod} are  order O($g_s\alpha_s$) in perturbation theory.

The different processes have different final state kinematics: the flavour creation yields ${b\bar b}$ pairs that are almost back to back and with symmetric transverse momentum $p_T$; the flavour excitation produces ${b\bar b}$ pairs with highly asymmetric $p_T$; the gluon splitting produces ${b\bar b}$ pairs with small opening angle and small $p_T$. In the forward (and backward) direction the gluon splitting is the dominant process. The LHCb detector is designed to take advantage of this feature \cite{Alves:2008zz}.

The particle acceptance region covered by the LHCb experiment is the very forward one with pseudorapidity $\eta$  in the range $2<\eta < 5$. The pseudorapidity $\eta$ of a particle is defined as $\eta=-\ln(\tan{\theta/2}))$, where $\theta$ is the angle of the particle three-momentum $\vec p$ relative to the positive direction of the beam axis. 
The acceptance region  at LHCb corresponds only to 4\% of the full solid angle, but the collected $pp\to b{\bar b}$ events represent about $25\%$ of the total cross section.
The visible $b$-hadron cross section in the pseudorapidity range $2<\eta<5$ has been measured to be $72~\upmu{\rm b}$ at $7~{\rm TeV}$ and almost double at $13~{\rm TeV}$, being about $144~\upmu{\rm b}$ \cite{Aaij:2016avz}. 

The general purpose experiments CMS and ATLAS have an acceptance limited to the more central region $|\eta|<2.2$, which corresponds to an efficiency of about 40\%  for the  $pp\to b{\bar b}$ processes. Semileptonic $B$ meson decays have not yet been studied at these experiments. 

Various important result on $B$ meson semileptonic decays have instead been provided by CDF and D0 experiments, that took data provided by $p\overline p$ collisions at $1.96$ TeV at Tevatron. In table \ref{tab:y4s_cs} we report a list of relevant cross sections, at different facilities.

The produced $b$-quark can hadronize, with different probabilities, called {\it production fractions}, into a full spectrum of $b$-hadrons, mainly $B^0$, $B^+$, $B_s$, $\Lambda_b$.
The measured fraction of $B^0$ and $B^+$, in the LHCb acceptance region, is about 36\% \cite{Amhis:2019ckw}, while the ratio
between the $B_s$ and the $B^0+B^+$ mesons production rate, $f_s/(f_d+f_u)$, is about $0.12$ and it has been observed to be slightly dependent on the $B_s$ transverse momentum itself \cite{Aaij:2019pqz}. The $\Lambda_b$ production fraction, compared to $B^0+B^+$ production, has been measured and it is $f_{\Lambda_b}/(f_u+f_d)\approx 0.26$. A strong dependence of the $f_{\Lambda_b}$ on the transverse momentum of $\Lambda_b$  has been observed \cite{Aaij:2019pqz}. In the LHCb acceptance range, the production rates  $B^0:B^+:B_s:\Lambda_b$ are approximately in the ratio $0.36:0.36:0.09:0.19$, with small fractions ($10^{-3}$) of $B_c$ and other $b$-baryons ($10^{-2}$). 

The production fractions  are crucial to determine the branching ratios of different hadron $B$ decays. For instance, the number of events $N(H_b)$ (produced in the LHCb acceptance) of a semileptonic process like $H_b\to H_c\ell\nu_\ell$, is given by
\beq
N(H_b)=2\, L\, \sigma(b {\overline b})\, \epsilon_{LHCb}\, f_{H_b}\, {\cal B}(H_b\to H_c \ell\nu_\ell)\,{\cal B}(H_c)\\   
\label{eq:lhcb}
\eeq
where $L$ is the integrated luminosity, $\sigma(b{\overline b})$ the total $b{\overline b}$ cross section, $\epsilon_{LHCb}$ the detector acceptance, $f_{H_b}$ the production fraction of the $H_b$ hadron species, ${\cal B}(H_b\to H_c \ell\nu_\ell)$ is the branching fraction of the process we are considering, and ${\cal B}(H_c)$ is the branching fraction of the $c$-hadron decay channel used to reconstruct $H_c$.

The precise absolute branching fraction measurements at hadron colliders using equation \eqref{eq:lhcb} would be affected by the large uncertainty in the $\sigma(b\bar b)$ and the knowledge of the detector acceptance for the decay analysed. In general, the branching fraction of a decay channel is measured relatively to a channel with a similar decay topology, which is often a decay of a neutral or charged $B$ meson, whose absolute branching ratio is well  known from $B$-Factories. Measuring ratios of branching fractions, most systematic uncertainties cancel, and the remaining uncertainties come from the knowledge of the ratio of production fraction $f(H_b)/(f_u+f_d)$ and the branching fraction of the normalization channel.

\subsection{Semileptonic measurements at $B$-Factories}
\label{sec:semilep_bfactories}

Generally speaking, the reconstruction of the  decays driven by the partonic decay $b \to c \ell \nu_\ell$ requires 
an efficient and reliable reconstruction of the lepton $\ell$, where the lepton can be an electron or a muon. In the case of the exclusive reconstruction of the final hadronic  state, an high efficiency reconstruction and identification of its decay products is also required. Some analyses also require the reconstruction of the other particles of the events, for example to infer the kinematics of the missing neutrino or reduce the combinatorial background in the signal reconstruction. 

For the study of semileptonic $B$ decays at $B$-Factories the  acceptance of the detectors is an important feature. 
The $B$-Factories detector geometry is solenoidal  around the interaction point, and asymmetric along the beam line. 
The geometric acceptance is slightly reduced compared to a symmetric detector like CLEO, which had a geometric acceptance close to $95\%$. To overcome this limitation, it is crucial to keep the detector performances very high and to exploit the high statistics as much as possible.
 Detailed descriptions of the $B$-Factories and of their detectors can be found in \cite{Aubert:2001tu, TheBaBar:2013jta} and \cite{Bondar:1998gu}. 
 Here we just briefly describe the most important subdetectors for the study of $B$ meson semileptonic decays:
\begin{enumerate}
\item a multilayer silicon detector allows the reconstruction of the tracks very close to the 
interaction point. This is crucial for the decay vertex reconstruction and for  
the tracking of very low momentum tracks; 
\item a low-mass drift chamber used for charged track reconstruction. The drift chamber 
allows a precise measurement of the momenta and the identification, through the measurement of 
the energy loss ($dE/dx$), of the charged particles; 
\item a specialized system to identify the nature of the charged particles based on the Cherenkov effect; 
\item a calorimeter for the measurement of the electromagnetic showers produced by photons and electrons;
\item an instrumented magnetic flux return used for the identification of muons and the detection of
$K_L$ mesons.
\end{enumerate}

Despite the relatively long lifetime of the $B$ mesons, their 
mean flight length transverse to the beam directions is only 30 $\upmu$m, and 
about 250 $\upmu$m in the beam direction. So the $B$ mesons decay in the beam pipe and only the decay products reach the various sub-detectors. 

The semileptonic decays are reconstructed starting from the identification of an high-momentum
lepton. Typically the minimum lepton momentum is required to be few hundreds of MeV.
For some analyses this momentum can be pushed down, but there is a minimum momentum under which 
the identification is not reliable. For example an electron has to reach the calorimeter to be clearly identified
from the measurement of $E/|\vec p|$, the ratio between the measured energy released 
in the calorimeter and the measured momentum of the associated track. 
A muon needs to reach the muon detector to be identified. Its identification relies on the calorimeter energy measurement, that needs to be compatible with the energy released by a minimum ionizing particle, and the number of the detecting plane traversed in the iron. A muon traverses more planes and releases less hits per plane than a pion. 
For both electrons and muons, the most relevant source of wrongly identified leptons are the pions. Pions that interact strongly in the calorimeter can mimic the energy released by an electron. The distribution of the energy released in the calorimeter is exploited to separate electrons from pions. 

Pions can mimic a muons because there is a finite probability that they go through the iron absorber without interacting (punch through). Moreover pions can decay in flight and generate
a muon which is identified in the main detector. Because of the small mass difference between pions
and muons, the kink resulting from the pion decay in flight is too small to be detected in most of the cases.
At small energy the background from pion decay in flight is dominant and prevents the reliability of the identified low momentum muons.

The performances of the lepton identification is done by using  control samples of electrons and muons. 
The cleanest sources of electrons and muons are
\begin{enumerate}
\item Bhabha and di-muon processes, $e^+e^-\to \e^+\e^-(\gamma)$ and $e^+e^-\to \mu^+\mu^-(\gamma)$;
\item decays of the $J/\psi$ into $e^+e^-$ and $\mu^+\mu^-$.
\end{enumerate}

\subsubsection{Soft pion from $D^{\ast}$}

The study of the exclusive $B\to D^*\ell \nu$ decays requires the reconstruction of the $D^{\ast+}$ or $D^{\ast0}$. These mesons are reconstructed usually through the decay chains 
$D^{\ast+}\to D^0\pi^+$ and $D^{\ast0}\to D^0\pi^0$.
Because of the little phase space available in $D^*\to D\pi$ decays, the emitted $\pi$ has a slow momentum in the $D^*$ rest frame. As a consequence the momentum of the $\pi$ is correlated with the variable $w$, and any inefficiency in reconstructing these pions in the low transverse momentum region affects the signal reconstruction in the zero-recoil phase space region. 

A low momentum $\pi^+$ does not cross the full tracking device, so its tracking efficiency is strongly dependent on the momentum. For transverse momenta of magnitude $p_T$ around 100\mev, the tracking relies entirely on the inner silicon trackers. Below 60\mev, the reconstruction is not possible because the track does not traverse enough layers. A good knowledge of the soft $\pi^+$ efficiency is required for precise measurements. 
At $B$-Factories the low $p_T$ track reconstruction efficiency is based on an approach used for the first time by the CLEO collaboration and described in detail in \cite{Allmendinger:2012ch}. This approach exploits the distribution of the $\pi^+$ helicity angle $\theta^*$ as a function of the $D^{*+}$ momentum. The helicity angle $\theta^*$ is defined as the angle between the slow $\pi^+$ momentum in the $D^*$ rest frame and the $D^*$ direction in the laboratory frame. The distribution of $\theta^*$ is expected to be symmetrical and can be described by $dN/d\cos{\theta^*}\propto (1+\alpha \cos^2\theta^*)$. The angle $\theta^*$ is connected with the slow $\pi^+$ momentum in the laboratory frame by $p_{\pi}=\gamma(p^*_{\pi}\cos \theta^* -\beta E_{\pi}^*)$ where $\beta$ and $\gamma$ are the $D^*$ boost parameters.
From the last relation, any asymmetry in the $\theta^*$ distribution can be related to the reconstruction efficiency in a region of the slow $\pi^+$ momenta.

Some analyses used also the reconstruction of $D^{\ast0}\to D^0\pi^0$ decays, where $\pi^0$ is reconstructed in $\gamma\gamma$ decays mode. At the B-Factories the photons can be reconstructed with high efficiency down to the energy of about 30-40~MeV, resulting in an efficiency almost uniform in the $\pi^0$ momentum and thus on $w$. 
One of the limitation on the usage of the soft $\pi^0$ is the difficulty to have a reliable control of the absolute efficiency to reconstruct the low momentum $\pi^0$. The approach used in Ref.\cite{Aubert:2007qs} exploits the $e^+e^-\to \tau^+\tau^- $ events. In the reconstruction of these events, one $\tau$ is  reconstructed either in one track and two clusters (sample dominated by $\tau\to \rho(\pi\pi^0)\nu$) or in one track and no cluster (sample dominated by $\tau\to \pi\nu, \mu\nu{\bar\nu}$). The other $\tau$ is reconstructed in the electron decay mode and used only to tag the $\tau$-pairs. From the comparison of these two samples, it is possible to measure the absolute efficiency to reconstruct a $\pi^0$ of momentum greater than 350~MeV. The efficiencies at lower momentum are obtained from the detailed simulation of the detector. The impact of the higher multiplicity of tracks and clusters present in $B{\bar B}$ events, compared to $\tau^+\tau^-$ events, is evaluated by comparing the rates of the reconstructed $D^0$ in $K^-\pi^+$ and $K^-\pi^+\pi^0$.
The systematic uncertainties on the soft $\pi^0$ reconstruction are typically larger than the corresponding uncertainty for charged pions.

\subsubsection{$B$ tagging}
\label{susubec:btagging}	

At the $B$-Factories the decay products of the two $B$ mesons originated from the decays of the $\Upsilon(4S)$ overlap and it can happen than one or more particles can be assigned to the wrong $B$ meson. This source of background can be relevant and the way to evaluate and eventually suppress its contributions depends strongly on the analysis. 

In the $\Upsilon(4S)\to \BBb$ decay, there are only two $B$ mesons in the final state. By reconstructing one of them in exclusive decay modes it is possible to reduce the combinatorial background and also the continuum. This technique, called {\it B tagging}, has been widely and successfully used at $B$-Factories. In addition to the background reduction, the information on the direction of the tagged $B$ ($B_{tag}$) can be used to constrain the kinematics of the full event and improve the resolutions in the study of the signal $B$ meson ($B_{sig}$) decay. The main disadvantage of the $B$ tagging approach is the small efficiency for the reconstruction of $B_{tag}$, usually well below 1\%. This is because there are many $B$ decay modes, all with small branching fractions and with high multiplicity in the final state, resulting in an overall small detection efficiency. 

The $B$ tagging approach can be classified according to two main categories: 
\begin{enumerate}
\item{\bf hadronic tagging}: the $B_{tag}$ is fully reconstructed in a mixture of many different hadronic decay modes. The reconstruction of the $B_{tag}$ starts reconstructing a set of charm mesons (called {\it seeds}), like $D^0$, $D^+$, $D^{*+}$, $D^{*0}$, $D_s$, $D_s^*$ or $J/\psi$ from their decay modes. Usually many decay modes of these seeds are added up together to increase efficiency. A seed is then combined with additional charmless mesons ($\pi^{\pm}$,  $K^{\pm}$, $\pi^0$ and $K_s$) to form a possible $B$ candidates. The two variables used to test the compatibility with a $B$ meson are
	\begin{enumerate}
	\item{$\Delta E=E_B^* - E_{beam}^*$}, the difference between the energy of the $B$ candidate in $\Upsilon(4S)$ and the expected $B$ candidate energy fixed by the energy of the beams;
	\item the energy substituted mass, $m_{ES{}}=\sqrt{ E_{beam}^{*2} - {|\vec{p}_B^{\,*}}|^2}$, where ${\vec{p}_B}^{\,\ast}$ is the momentum of the $B$ candidate.
	\end{enumerate}
\noindent A correctly identified $B$ meson gives $\Delta E\approx 0$ and $m_{ES}\approx m_B$. The quantity $m_{ES}$ exploits the feature that the energy of the $B$ mesons is precisely determined by the beam energy, which is known with a resolution better than 2~MeV. 
The tagging efficiency depends on the multiplicity and the kind of particles present in the analyzed final state. The purity, defined as the probability that a specific decay chain is correctly reconstructed,  varies considerably according to the decay mode considered. In case of more $B_{tag}$ candidates, the one with the highest purity is in general chosen. To gain in efficiency, usually more than a thousand possible decay modes are considered. The hadronic $B$-tagging approach has been improved over time by both BaBar and Belle. In BaBar, more decay modes and wider mass windows have been implemented according to the specific mode considered. 
Belle instead has made use of an algorithm described in \cite{Feindt:2011mr}. This algorithm uses a set of different neural-networks, properly trained, to estimate the probability that a seed has been correctly reconstructed. The output of the final neural network is used to rank the various $B_{tag}$ candidates. 
At the end, the average overall efficiency is about $0.3-0.5\%$ for the tagging $B^+$ and about $0.3\%$ for the $B^0$, with purity of about $10-30\%$. The reconstruction of the four-momentum of the $B_{tag}$ allows to determine clearly the four-momentum of the signal $B_{st}$, even in $B_{s g}$ with missing particles, using:
\begin{equation}
    p_{B_{sig}} = p_{\FourS} - p_{B_{tag}},\\
\end{equation}
\noindent where $p_\Upsilon(4S)= p_{e^+}+p_{e^-}$ is the four-momentum of the initial $\Upsilon(4S)$, determined by the energy of the initial electron and positron beams. 
The charge and the flavour of the reconstructed $B_{tag}$ are also exploited to clean the sample and reduce the backgrounds. 

\item{\bf semileptonic tagging}: the $B_{tag}$ is reconstructed using both \btodslnu and \btodlnu decays.  
The branching fractions of these decays are among the highest in $B$ decays. Moreover, the efficiency to reconstruct semileptonic decays is higher than the one to reconstruct fully hadronic $B$ decays. The final efficiency of the semileptonic $B$ tagging runs between $0.5-1.0\%$.
The final efficiency is higher than the hadronic $B$ tagging, but the background is also higher. Another disadvantage of the semileptonic tagging is that it does not allow tight kinematic constraints for the presence of neutrino in the tag side. 

\end{enumerate}

\subsection{Semileptonic measurements at LHCb}            

LHCb is a dedicated experiment that exploits the fact that the $b\overline b$ production rate is larger in the forward direction, as described in section \ref{subsub:had_prod}. 
Because the $c\overline c$ production  has similar production mechanism, and  has a cross section about twenty times higher than the $b\overline b$, also a huge amount of $c$-hadrons are produced in the forward direction.
The fact that all species of heavy hadrons are produced makes LHCb a unique facility for heavy flavour physics. 

The LHCb detector \cite{Alves:2008zz,Aaij:2014jba} is a single-arm forward spectrometer that covers the pseudorapidity range $2<\eta <5$.
It consists of the following subdetectors:
\begin{enumerate}
    \item a precise vertex detector for the identification of the  vertex (primary vertex, PV) where the inelastic $pp$ collision occurs, and the reconstruction of the decay vertex of the $B$ hadrons;
    \item two detectors specialized for the identification of protons, pions and kaons;
    \item an electromagnetic calorimeter for electrons and photon identification and energy measurement, and an hadronic calorimeter for the identification of high $p_T$ hadrons;
    \item a detector for muon identification.
\end{enumerate}
At LHCb, an  experimental challenge in the study of semileptonic $B$ decays is represented by the presence of the un-reconstructable  neutrino.
The momentum of the $B$ hadrons in production is not known. The identification of semileptonic events exploits the very good identification of the $B$ flight direction. The situation is complementary to that of the $B$-Factories, where, in untagged measurements, the magnitude of the $B$ momenta is known but not their direction. The reconstruction of the kinematics for semileptonic $B$ decays is described in  section \ref{subsub:lhcb_kine}.

In the forward direction, the $B$ hadrons are highly boosted so they have a mean flight length of about 1 cm. This property, associated with the great vertex resolution, is crucial for a clean reconstruction of the signal event. In particular, the large separation between the $B$ decay vertex and the PV reduces the combinatorial background. Moreover, the decay products of the second $B$ hadron, produced usually within the LHCb acceptance, are in general well separated in $\eta$, so the mis-assignment of tracks from $B$ hadrons is in general negligible. 

The majority, more than 99\%, of inelastic $pp$ collisions does not produce $b$-quarks and are a relevant source of backgrounds, so the triggering of the events is crucial: it has to be efficient for $B$-hadrons, and has to have a high reject rate for backgrounds. The trigger in LHCb exploits the fact that the $B$ hadrons are long lived, and that, having a relatively large mass, give decay products with an average $p_T$ larger than the typical particle produced in a $pp$ collisions. The trigger consists in a combination of an hardware trigger stage (L0) and a software one. The L0 trigger relies mainly on the muon detector and the calorimeters response. 

For the study of semileptonic decays in LHCb, the presence of a muon is very well suited because the L0 trigger line for muons is very efficient. The L0 muon trigger requires the presence of muons of $p_T$ greater than about $1.7$~GeV. This low threshold ensures a large efficiency for $B$ semileptonic decays. For comparison, at CMS and ATLAS this threshold is more than 5~GeV.
The trigger for electrons is not as efficient because its identification has to rely on the electromagnetic calorimeter where the trigger threshold has to be increased to avoid large backgrounds. Moreover, electrons are affected by bremsstrahlung that deteriorates their momentum reconstruction. For these reasons, usually only semileptonic decays into muons are exploited in LHCb.

At the luminosity of LHC, a large number of multiple $pp$ collision occurs in the same bunch crossing. On average about $40$ inelastic $pp$ collisions (called {\it pile-up}) are produced. The study of the $B$-hadron properties requires the detection of the decay vertex and the production vertex, and large pile-up can pollute the clear identification of the PV, and increase the occupancy in the various subdetectors, worsening the $B$ reconstruction. At LHCb, this problem  is overcome   decreasing locally the luminosity by about a factor $20$. That reduces the average number of visible collision per bunch crossing to about $1.8$. 


\subsubsection{Techniques for kinematic reconstruction}
\label{subsub:lhcb_kine}

As mentioned above, the precise determination of the flight direction, from the identification of the PV and the decay $B$ vertex, can be used to constrain the decay kinematics of semileptonic decays \cite{Dambach:2006ha}. In the hypothesis of a single missing particle with known mass, the unknowns are the components of the 3-momentum of the missing particle. Two constraints are obtained by the momentum conservation in the plane transverse to the decay flight, and the third is determined by the assumed mass of decaying $B$ hadron. Because this last constrain is quadratic, there are two possible solutions. 

\begin{figure}[tb!]
  \begin{center}
    \includegraphics[width=0.80\linewidth]{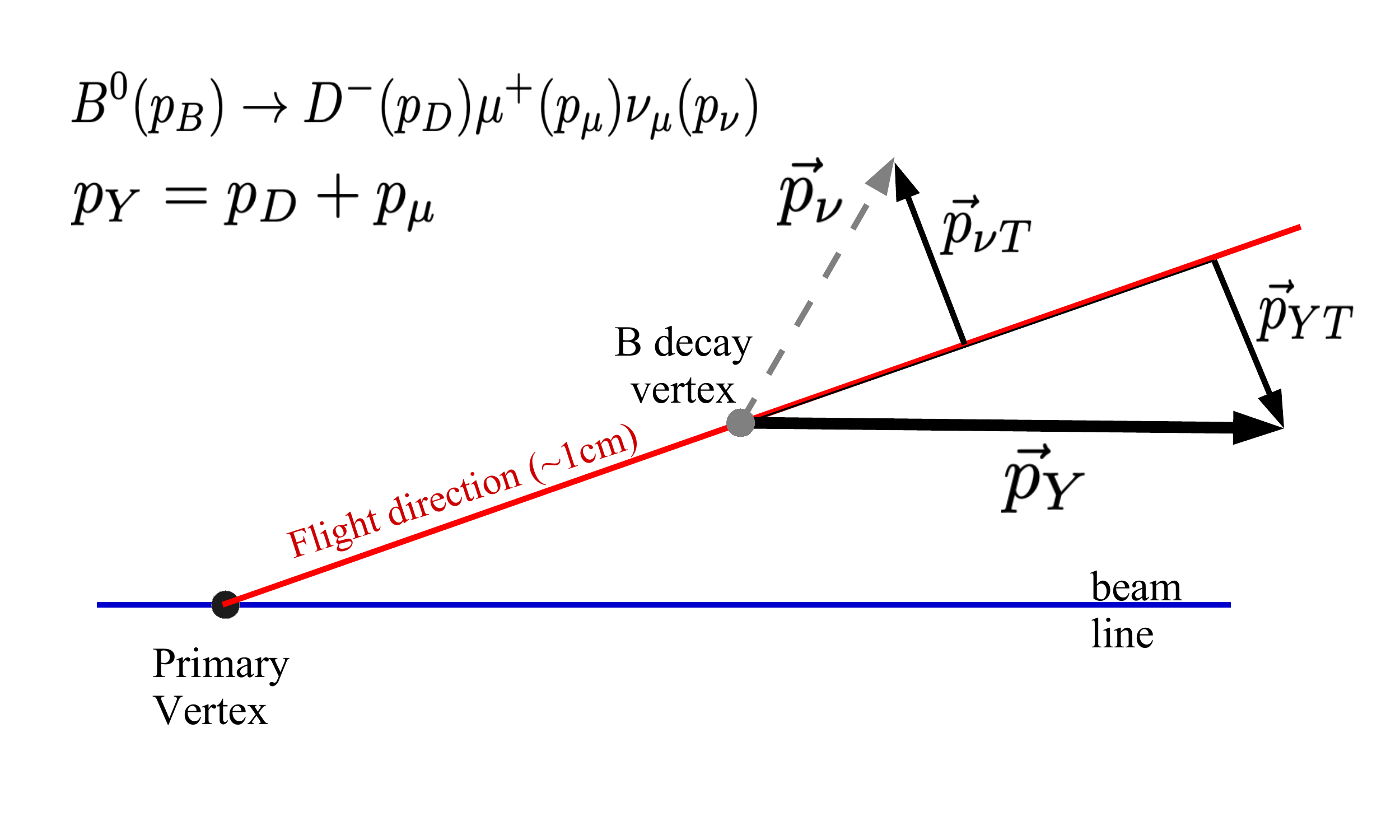} 

  \end{center}
   \vspace*{-10mm} 
  \caption{In LHCb the average flight length of the $B$ hadrons is about 1~cm. The good resolution in the vertex reconstruction allows to determine the flight direction.}
  \label{fig:lhcb_qsquare}
\end{figure}

In figure \ref{fig:lhcb_qsquare} we illustrate the ingredients exploited to constrain the kinematics. Lets consider the decay $B(p_B)\,\to D(p_D)\,\mu (p_\mu)\,\nu_\mu (p_\nu)$, where the four-momentum of the various particles are given in parenthesis. 
The visible system $Y\equiv D\mu$ has a four-momentum given by $p_Y=p_D+p_\mu$. It is useful to decompose $p_Y$ in the longitudinal ($p_{YL}$) and transverse ($p_{YT}$) components along the $B$ flight direction. If there is only a missing particle, like a neutrino, its transverse component is known just balancing the visible transverse component, ${\vec p}_{\nu T}=-{\vec p}_{Y T}$. The only unknown is the magnitude of the longitudinal component $|{\vec p}_{\nu L}|$, or, equivalently, the momentum $|{\vec p}_B|$ of the decaying $B$ meson. The $B$ four-momentum is given by  $p_B=p_Y+p_\nu$, so we can write 
\begin{equation}
p_\nu=p_B-p_Y \Longrightarrow m_\nu^2=m_B^2+m_Y^2-2\,(E_Y E_B - |\vec p_B||\vec p_Y|\cos\theta_{BY})\\
\label{eq:two_fold}
\end{equation}
\noindent where $\theta_{BY}$ is the angle of $Y$ respect to the flight direction. After setting the mass of the $B$ to its value, substituting $E_B=\sqrt{m_B^2+|\vec p_B|^2}$ in  \eqref{eq:two_fold}, squaring and solving for $|\vec p_B|$, we arrive to a simple second degree equation
\begin{equation}
(|\vec p_Y|^2\cos^2\theta_{BY}-E_Y^2) |\vec p_B|^2 + (2 M |\vec p_Y|\cos \theta_{BY})|\vec p_B|+(M^2-m_B^2 \, E_Y^2)=0
\label{eq:two_fold_final}
\end{equation}
\noindent where $M=[(m_B^2+m_Y^2)-m_\nu^2]/2$. The equation (\ref{eq:two_fold_final}) yields two solutions for the $B$ momentum, thus all the kinematic quantities we determine for the decay are affected by this ambiguity; for instance, in the example above, there are two possible $q^2=(p_B-p_D)^2$ values. Furthermore, the limited vertex resolution gives a fraction of decays with non-physical solution, which are usually excluded in the analyses. The fraction of these events depends by the decay considered but it is usually between $20$ and $30\%$. 

Without applying any requirement on the signal, the two solutions are equally probable and cannot be distinguished. However, after the signal selection requirements are applied, the one that gives the systematically smaller $|\vec p_B|$ usually has an higher chance to be the correct solution. Thus in practice, resolutions on $q^2$ of the order of 10-20\% can be achieved by selecting only this solution. 

Other approaches to improve the kinematic resolution have been used. In \cite{Ciezarek:2016lqu} it has been proposed a regression algorithm that uses the information of the flight decay length and the production angles to increase the chance to select the right solution. Another possibility is to consider $B$ decays that come from decays of narrow excited $B$ hadron states. The constraint that comes from the mass difference between the excited state and the $B$ meson removes the ambiguity. This approach has been described in \cite{Stone:2014mza} and it has been exploited for the first time 
in the analysis \cite{Aaij:2018unp}, where the $B^+\to D/D^*/D^{**}\mu\nu_\mu$ relative fraction have been measured by tagging the $B^+$ mesons from the $ B_{s2}^{*0}\to B^+ K^-$ decay. The price is a reduced signal efficiency due to the low rate of production of the excited $B_{s2}^{*0}$ state, and the low detection efficiency of the soft $K^-$ accompanying the $B^+$ meson. 

\section{Inclusive $|V_{cb}|$ determination} 
\label{InclusiveVcbdetermination}

In section~\ref{inclusivedecays} we have introduced the inclusive $B\to X_c\ell\nu$ decays.
The  total semileptonic rate for $B\to X_c\ell\nu$ decays is expressed in the framework of the HQE (see section~\ref{HeavyQuarkExpansion}), which allows to disentangle  coefficients and  corrections calculable in QCD perturbation theory from  genuinely non perturbative  parameters. The same holds for the moments of distributions of charged-lepton energy and hadronic invariant mass, defined in~(\ref{eq:lepton_moments}) and~(\ref{eq:hadron_moments}), respectively. As underlined in section~\ref{massschemes}, the
inclusive analysis requires a suitable definition of the quark mass in a  coherent framework, or scheme.  

The shapes of the kinematic distributions in the $B\to X_c\ell\nu$ decays are sensitive to 
the masses of the $b$ and $c$ quarks, and the non-perturbative HQE parameters, thus their knowledge is needed to 
 extract $|V_{cb}|$ from data. Non perturbative parameters can be extracted together with $|V_{cb}|$  in a simultaneous fit  (global fit)  based on experimentally measured distributions and momenta.
Global fit analyses  differ by the data sets they are based onto, the theoretical  scheme employed, the order of truncation of the HQE expansion.
Challenges are experimental selections applied to the data as well as to properly account for correlations.

In the following we describe the measurements of the moments of the charged lepton energy spectrum and the hadronic invariant mass distribution, which, together with the total rate, are the ingredients to extract $|V_{cb}|$ in global fits.

\subsection{Moment measurements}
The moments of the observables in $B\to X_c\ell\nu$ inclusive decays have been measured by various experiments. A list of the inputs included in the extraction of $|V_{cb}|$ performed by HFLAV \cite{Amhis:2019ckw} is reported in table~\ref{tab:inclusive_vcb_inputs}. 

\begin{table}[!htb]
\caption{Experimental measurements used in the global analysis of $\bar B\to
  X_c\ell^-\bar\nu_\ell$. $n$ is the order of the moment, $c$ is the
  threshold value of the lepton momentum in GeV.} \label{tab:inclusive_vcb_inputs}
\footnotesize
\begin{center}
\begin{tabular}{  >{\raggedright\arraybackslash}p{0.07\linewidth} 
>{\raggedright\arraybackslash}p{0.26\linewidth} >{\raggedright\arraybackslash}p{0.23\linewidth}  >{\raggedright\arraybackslash}p{0.36\linewidth}  }
  \hline \hline
  Exp.
  & Hadron moments $\langle m^{2n}_X\rangle$
  & Lepton moments $\langle E^n_\ell\rangle$
  & Remarks
  \\
  \hline 
  BaBar  ~\cite{Aubert:2009qda} \cite{Aubert:2004td}
              & $n=1$, $c=0.9,1.1,1.3,1.5$  $n=2$, $c=0.8,1.0,1.2,1.4$ $n=3$, $c=0.9,1.3$  
              & $n=0$, $c=0.6,1.2,1.5$  $n=1$, $c=0.6,0.8,1.0,1.2,1.5$ $n=2$, $c=0.6,1.0,1.5$ $n=3$, $c=0.8,1.2$ 
              & Lepton momentum spectrum is obtained with an inclusive measurements. The hadronic moments are determined in hadronic tagged B meson sample. \\
  Belle ~\cite{Schwanda:2006nf}  \cite{Urquijo:2006wd}
              & $n=1$, $c=0.7,1.1,1.3,1.5$  $n=2$, $c=0.7,0.9,1.3$  
              & $n=0$, $c=0.6,1.4$  \phantom{inv}   $n=1$, $c=1.0, 1.4$ \phantom{inv}   $n=2$, $c=0.6,1.4$, \phantom{inv}  $n=3$, $c=0.8,1.2$
              & Both lepton and hadronic moments measured using the hadronic $B$ tagged events.\\
  CDF  ~\cite{Acosta:2005qh} 
          & $n=1$, $c=0.7$ \phantom{invisible}  $n=2$, $c=0.7$ 
  	      &
  	      & Hadronic mass measurement obtained from the $D^{*}\pi$ mass distribution in $B\to D^{(\ast)}\pi\ell\nu$ decays, combined with the known
  	      $B\to D^{(\ast)}\ell\nu$ rates.\\
  CLEO  ~\cite{Csorna:2004kp}   
          & $n=1$, $c=1.0, 1.5$ \phantom{invis}  $n=2$, $c=1.0, 1.5$ 
          &
  	      & The kinematics of the hadronic part is inferred from the measurement of the neutrino momentum
  	       inclusively from the global event missing momentum.   \\	                 
DELPHI  ~\cite{Abdallah:2005cx} 	
          &  $n=1$, $c=0.0$ \phantom{invisible}  $n=2$, $c=0.0$  \phantom{invisible}  $n=3$, $c=0.0$  	      
  	      &  $n=1$, $c=0.0$ \phantom{invisible}  $n=2$, $c=0.0$  \phantom{invisible}  $n=3$, $c=0.0$ 
  	      & 
  	      Exploiting the large boost of the $B$ meason produced, the moments are measured without cuts on the lepton energy.\\
\hline\hline  	      
\end{tabular}
\end{center}
\end{table}

 We  have already introduced experimental techniques used for the study of semileptonic decays in section \ref{sec:semilep_bfactories}.  In the following we provide additional experimental details on some of the measurements performed at the $B$-Factories.

\subsubsection{Hadron moments}

The BaBar analysis \cite{Aubert:2009qda} uses the hadronic $B$ tagging technique. After the reconstruction of the $B_{tag}$, an identified lepton (electron or muon) is required in the event. The momentum of the lepton is required to be greater than $0.8\gev$ in the rest frame of the signal $B$ meson. All the tracks and clusters not associated with the $B_{tag}$ and the lepton are combined to reconstruct the four-momentum of the hadronic system $X_c$. The resolution on the resulting hadronic mass $m_X$ is improved using a kinematic fit of the full event, considering the conservation of the four-momentum and setting the missing mass to zero. The hadronic moments $\langle m^{2n}_X\rangle$ are reconstructed from the measured $m_X$ spectrum, for different cuts on the minimum lepton energy. The distribution of the mass spectrum for two minimum value of the lepton moments, are reported in figure \ref{fig:moments} (left). The different exclusive contributions to the $B\to X_c\ell\nu$ decay are not disentangled, because of the limited resolution, due mainly to lost or misidentified particles. The reconstructed $m_X$ distribution has to be corrected for the detector efficiency and resolution effects. The true values for the hadronic moments are obtained using a {\it per-event} corrections to the reconstructed moments, which are determined from simulations. The corrections depend on the lepton energy, the $X_c$ multiplicity and the missing mass in the event. 

The Belle analysis \cite{Schwanda:2006nf} also uses the hadronic $B$ tagging method. Belle sets the minimum lepton momentum at $0.7\gev$.
The true value of the hadronic moments, is extracted using an unfolding procedure based on the SVD algorithm \cite{Hocker:1995kb}. This approach requires the knowledge of the migration matrix that connects the reconstructed and the true values of $m_X$, which is obtained using  simulations.  

\subsubsection{Lepton moments}

In general the lepton energy momentum  $\langle E^n_\ell\rangle$ can be measured with higher precision than the hadronic mass moments. The BaBar analysis \cite{Aubert:2004td} uses an inclusive approach where the $B{\overline B}$ candidates are selected requiring two leptons in the event. In this analysis, to reduce the background due to the hadron misidentification for low energy leptons, only electrons are used. A tagging electron is required to have a momentum in the $1.4-2.3\gev$ range. The second electron in the event, the signal, is studied from momentum greater than $0.6\gev$. The background from the continuum is reduced with the global shape variables. The main source of background is due to lepton from secondary decay of charm mesons. This is reduced by requiring the charge correlation between signal and tagging lepton, and exploiting the kinematic properties of the two leptons. 
In general the moments have to be computed in the $B$ meson rest frame so in this inclusive analysis further corrections are needed to account for the small motion of the $B$ in the $\Upsilon(4S)$ rest frame.

\begin{figure}[t!]
  \begin{center}
    \includegraphics[width=0.47\linewidth]{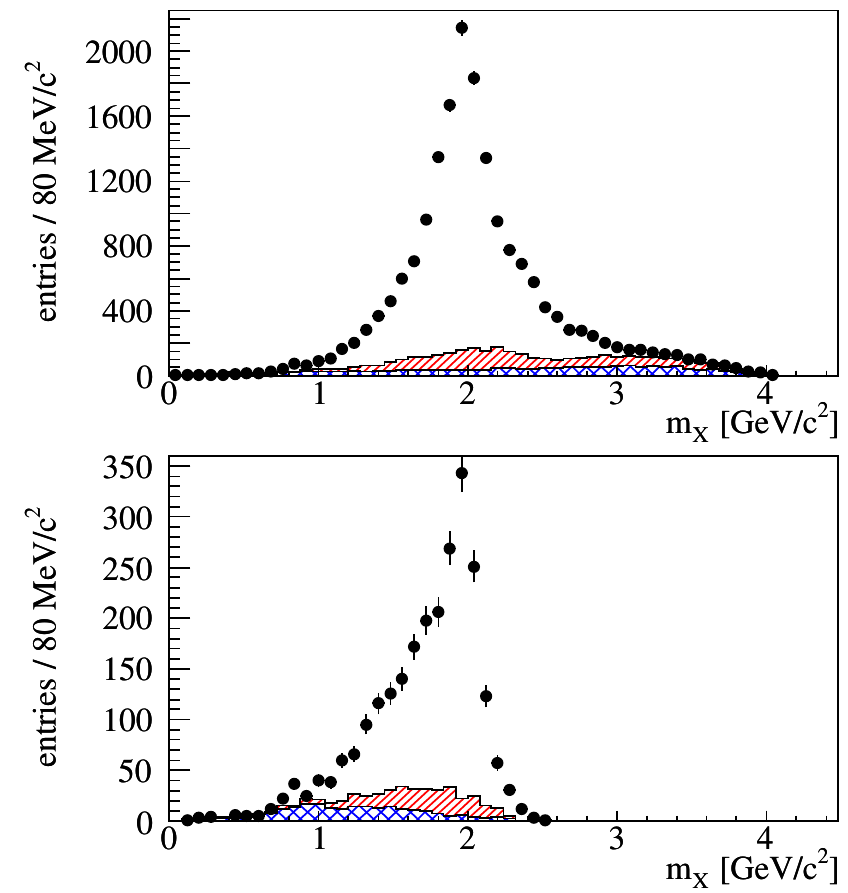}  \includegraphics[width=0.52\linewidth]{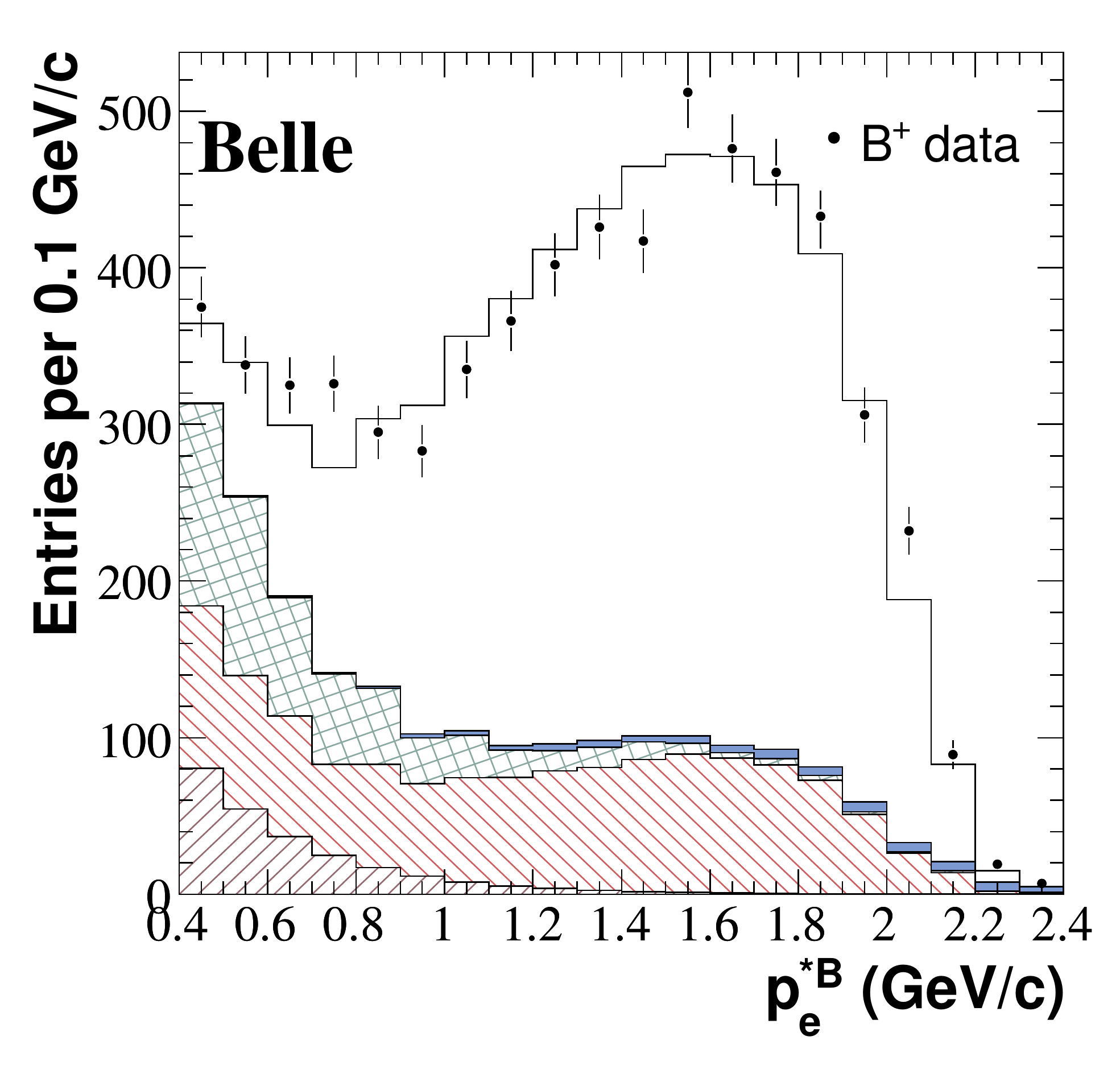}  \\
  \end{center}
  \vspace*{-5mm} 
  \caption{Left: Mass spectra for a lepton momentum cut of $p_\ell^*>0.6\gev$ (top) and $p_\ell^*>1.4\gev$ (bottom) obtained by BaBar prior the background subtraction. Plot from \cite{Aubert:2009qda}. Right: lepton momentum from Belle in the $B^+$ decays. Plot from \cite{Urquijo:2006wd}.}
  \label{fig:moments}
\end{figure}

The Belle analysis of the lepton moments \cite{Urquijo:2006wd} also is limited to electrons, and uses the hadronic $B$ tagging method. One advantage of the tagging approach, is that the four-momentum of the $B_{sig}$ is known from the fully reconstructed $B_{tag}$ so the moments are directly computed in the $B$ rest frame. The moments are extracted from the minimum lepton momentum cut of $p-\e^* >0.4 \gev$, computed in the $B$ rest frame. The distribution of the lepton momentum, for the $B^+$ sample, is reported in figure \ref{fig:moments} (right). 

\subsection{Results}
\label{sub:incl_results}

A recent global analysis of the inclusive $B\to X_c\ell\nu$ has been done by HFLAV
\cite{Amhis:2019ckw}. In this fit the hadronic mass moments $\langle m_X^{2n}\rangle$ of orders $n=1,2,3$ and the lepton energy moments $\langle E_{\ell}^n\rangle$ of order $n=0,1,2,3$ are used.  The lepton energy moments of order $n=0$ are just the partial branching fractions. The moments are determined with different lower values of the lepton energy ($E_{cut}$). Because the moments of the same order and with different $E_{cut}$ are strongly correlated, only a sub-sample of the measured moments are used in the global analysis. The list of the moments used is reported in table~\ref{tab:inclusive_vcb_inputs}. 

The moments of the $B\to X_c\ell\nu$ decay allow to determine a linear combination of the $b$ and $c$ quark masses.  Additional inputs can be used for a precise determination of $m_b$. The additional information can come from the moments of the photon energy in $B\to X_s\gamma$ decay, or from an external determination of the $c$ quark mass.   

\begin{figure}[ht!]
  \centering
  \includegraphics[width=0.32\textwidth]{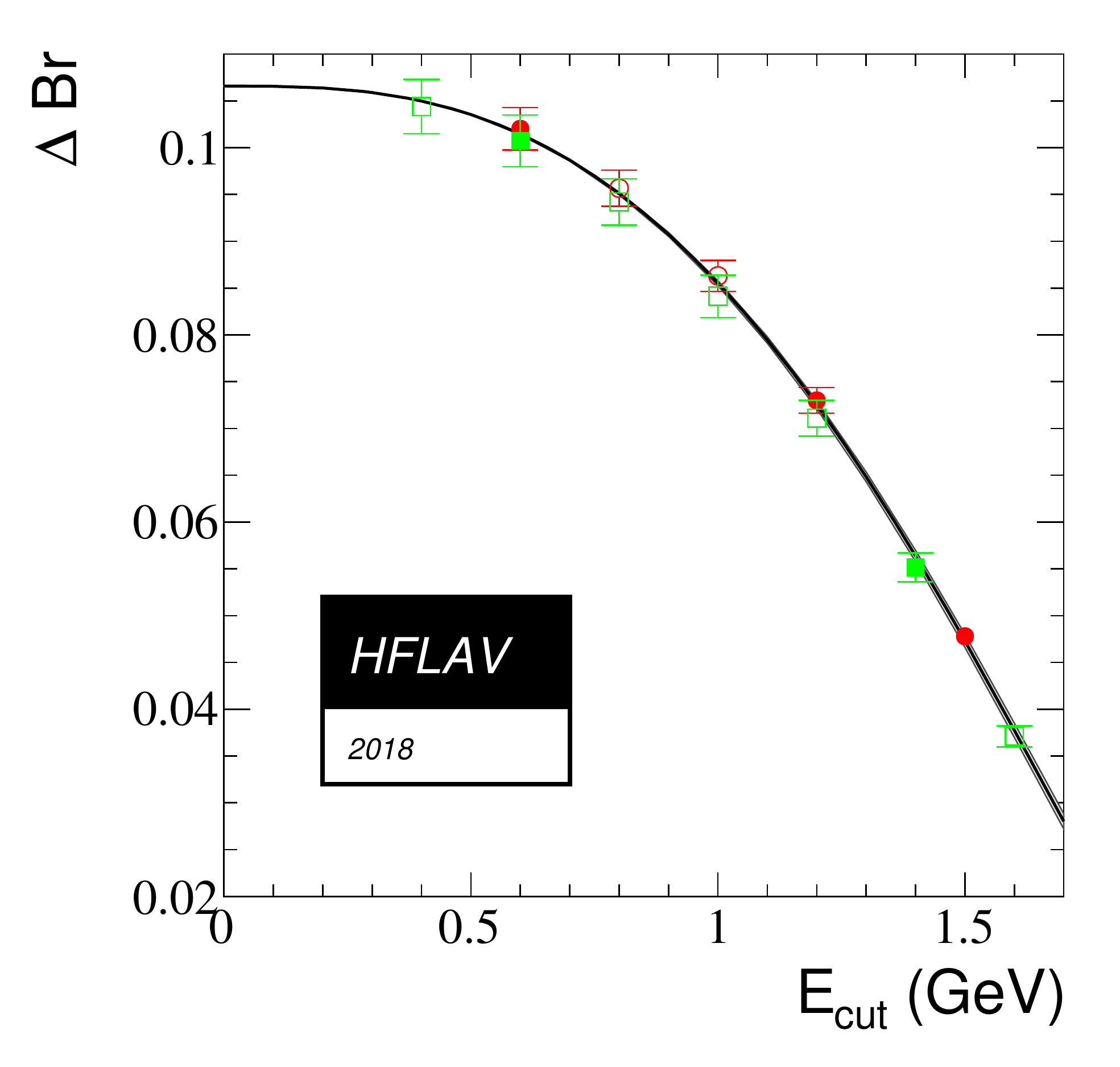}
  \includegraphics[width=0.32\textwidth]{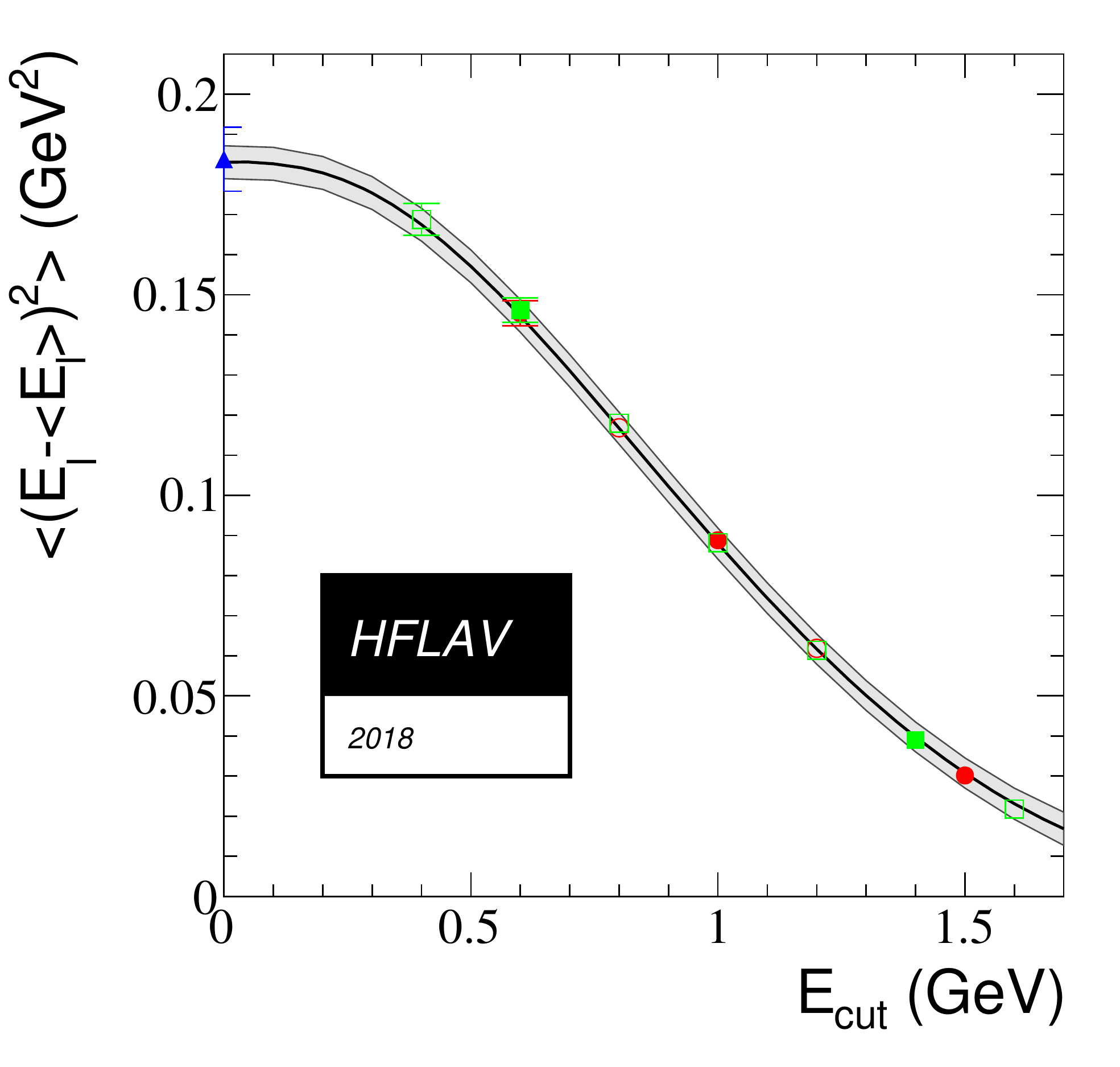}
  \includegraphics[width=0.32\textwidth]{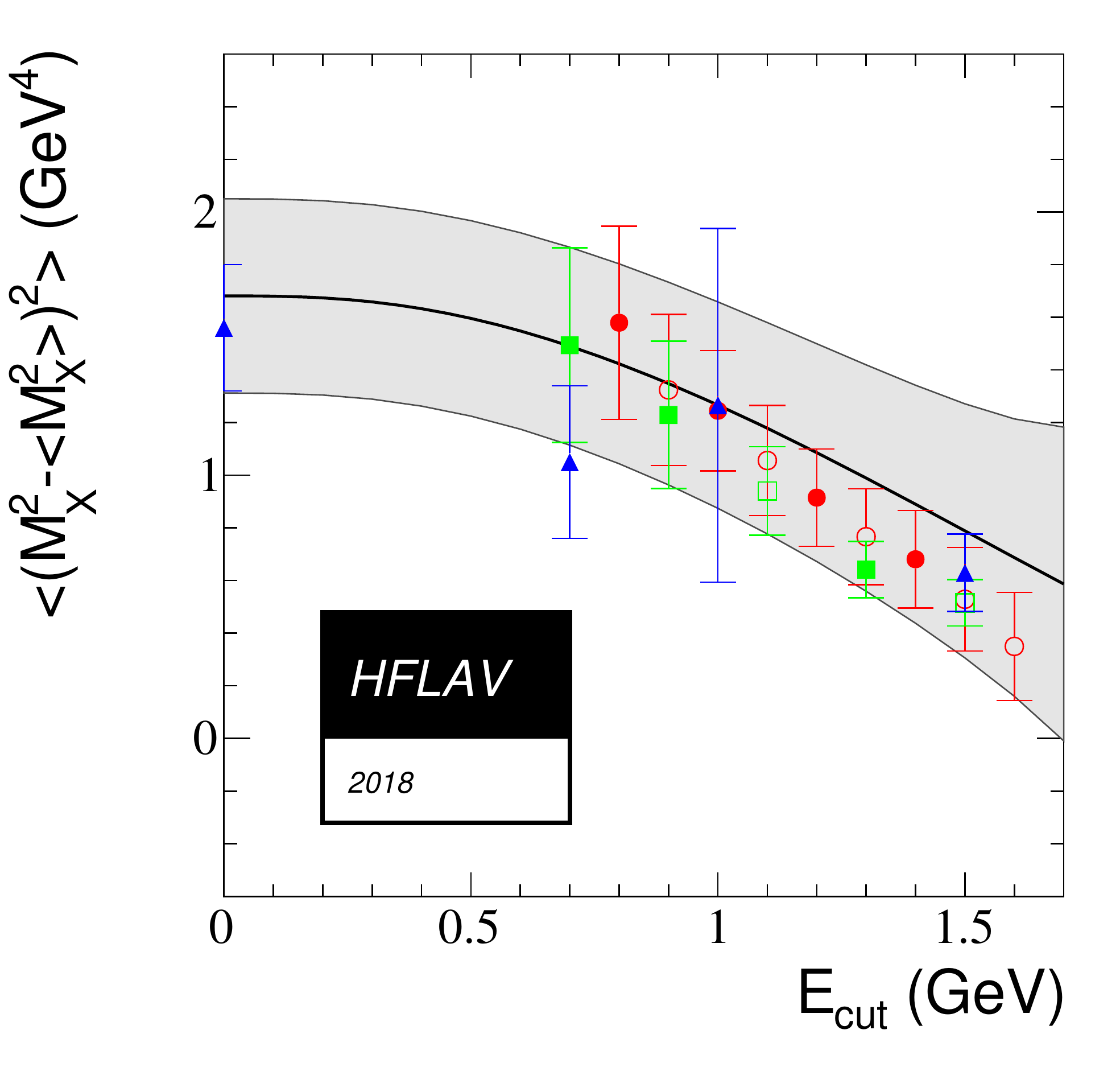}
  \caption{Distributions of the partial semileptonic branching fraction (left), one central lepton momentum (middle)
  and one hadronic mass central moment (right), with the result of the global fit in the kinetic mass scheme superimposed. The gray band is the theory prediction, fixing the HQE parameters at the fitted value, with the theory uncertainty. BaBar data are shown by circles (red), Belle by squares (green) and other experiments (DELPHI, CDF and CLEO) by triangles (blue). Open symbols (no internal color) are measurements not included in the fit. 
  }
  \label{fig:vcb_global}
\end{figure}

In the framework of kinetic scheme, $|V_{cb}|$ is extracted together with the $b$ and $c$ quark masses and 4 non-perturbative  parameters (namely $\mu^2_{\pi}$, $\mu^2_{G}$,  $\rho^3_{D}$ and $\rho^3_{LS}$). The subset of measurements used and the general approach follow the ones described in \cite{Gambino:2013rza}. The fit is based on theoretical calculations described in \cite{Gambino:2011cq,Alberti:2014yda}. In this analysis the $c$ quark mass is constrained to the value obtained in~\cite{Chetyrkin:2009fv}, which is $m_{c}^{\overline{\rm  MS}}(3\gev)=0.989\pm0.013\gev$. 
The result of the fit, projected on some of the lepton energy and hadronic mass moments, is shown in figure~\ref{fig:vcb_global}. Let us report also the resulting values for $|V_{cb}|$ and $m_{b}^{kin}$
\begin{eqnarray}
|V_{cb}|=(42.19\pm 0.78)\times 10^{-3}\nonumber\\
m_{b}^{kin}=4.554\pm 0.018\gev
\label{eq:inclvcb_result_ks}
\end{eqnarray}
\noindent where the quoted uncertainties include both the experimental and the theoretical uncertainties. It is worth to mention that the theoretical uncertainties are dominating. The excellent fit quality points toward the validity of the HQE fit, but the small $\chi^2$ per degree of freedoms of $\chi^2/ndf=0.32$, could be a signal of some overestimated theoretical uncertainties, or overestimated correlations between the various moments. These points have been discussed extensively for previous version of the global fit in \cite{Gambino:2013rza}.

An analysis performed in the framework of the 1S scheme, and based on the calculation of the lepton and hadron moments described in~\cite{Bauer:2004ve}, gives, for $|V_{cb}|$ and the $b$ quark 1S mass
\begin{eqnarray}
|V_{cb}|=(41.98\pm 0.45)\times 10^{-3} \nonumber\\
m_{b}^{1S}=4.691\pm 0.037\gev. \label{eq:inclvcb_result_1s}
\end{eqnarray}
\noindent This analysis uses the same list of lepton and hadron moments reported in table~\ref{tab:inclusive_vcb_inputs} and in addition the moments of the photon spectrum in $B\to X_s\gamma$ decays as further constraints. The central values of $|V_{cb}|$ in \eqref{eq:inclvcb_result_ks} and \eqref{eq:inclvcb_result_1s} are in good agreement, but the uncertainties are different. 
The uncertainty on $|V_{cb}|$ from the global fits is $1.8\%$ in the kinetic scheme and only $1.1\%$ in the 1S scheme. However, a direct comparison between these two results is not significant, since the   two schemes are not equivalent, as underlined in section~\ref{massschemes}. The 1S result~\cite{Bauer:2004ve} is at a disadvantage  compared to the one in the kinetic scheme, since it does not include all  contributions  of order $O(\alpha_s \Lambda_{QCD}^2/m^2_b)$.

All the analyses considered above include only the minimal set of four matrix elements which appear until order $O(1/m^{3}_b)$. 
At higher order, the large increase of HQE parameters complicates a great deal the extraction from data.  A model approach that estimates the effects of orders $O(1/m^{4}_b)$ and $O(1/m^{5}_b)$, in the so-called Lowest Lying State Approximation, was employed in a recent global fit \cite{Gambino:2016jkc}. Their results indicate  that  such  higher-order  terms induce a sub-percent reduction in $|V_{cb}|$, which is not appreciable at the current level of precision. Another recent suggestion is to use a symmetry within the HQE, the reparameterization invariance, to achieve a reduction  of  independent  parameters  in some  specific observables, that could be measured at $e^+ e^-$ colliders and used to extract $|V_{cb}|$ at order $O(1/m^4_b)$ \cite{Fael:2018vsp}.

\section{Exclusive $|V_{cb}|$ determination}
\label{ExclusiveVcbdetermination}
As discussed in section \ref{subsectionExclusive decays}, the $|V_{cb}|$ exclusive determination requires the theoretical knowledge of the decay form factors, together with the measurements of 
 the experimental decay rates. In the $B\to D^*\ell\nu_\ell$ channel, the form factors computed with the aid of heavy quark symmetries are currently available only at the zero recoil point $w=1$,  where the differential rates in  \eqref{diffrat0} vanish.
Therefore, a necessary step becomes to extrapolate the
 experimental measurements of the exclusive decays rates, yielding the  products $ |\eta_{EW}|^2 \: |{\cal F}(w)|^2\; |V_{cb}|^2$  or  $ |\eta_{EW}|^2 \: |{\cal G}(w)|^2\; |V_{cb}|^2$ at non-zero recoil points, to $w=1$, by using a parameterization of the dependence on $w$ of the form factors.
As outlined in sections \ref{parameterizations} and  \ref{parameterizations12}, the use of parameterizations introduces additional uncertainties, which could become significant at the current level of precision.
%

%
In  the \btodlnu channel, where 
 form factors calculated directly at  non-zero recoil points are already available, the role of parameterization becomes less relevant, because the extrapolation to $w=1$ 
reduces to an interpolation between experimental results and different theory points. 

In section \ref{BDstarchannel} we discuss  the exclusive determinations of $|V_{cb}|$ in the
$B\to D^*\ell\nu_\ell$ channel, presenting the results in section \ref{results633}. 
Two recent analyses by  Belle~\cite{Abdesselam:2018nnh} in 2018  and by  BaBar~\cite{Dey:2019bgc} in 2019, using both CLN and BGL parameterizations, are detailed
in section~\ref{Belleuntaggedmeasurement} and ~\ref{BaBartaggedmeasurement}, respectively.  
In section \ref{BDchannel} we discuss the $B\to D\ell\nu_\ell$ decays, detailing the most precise measurement (Belle \cite{Glattauer:2015teq})  in section \ref{Belletaggedanalysis1} and drawing the conclusions in section \ref{result1}.
Section \ref{sec:bs} is  devoted to a novel and promising method to determine $|V_{cb}|$, using the  $B_s \to D_s^{(*)}\mu\nu_\mu$  decays. 

The $B$ meson decays analyses can be complemented by analyses of bottom baryons. 
The measurement of the ratio of the branching fractions $\Lambda^0_b\to p\mu^-\bar\nu_\mu$ and $\Lambda^0_b\to \Lambda_c^+\mu\bar \nu_\mu$ by  the LHCb Collaboration \cite{Aaij:2015bfa} allows a direct measurement of the ratio $|V_{ub}|/|V_{cb}|$, that we discuss  in section
\ref{sec:lbtopmunu}.

	\subsection{The $B \to D^{\ast} \ell  \nu_\ell$ channel}
	\label{BDstarchannel}

\begin{figure}[tb!]
  \begin{center}
    \includegraphics[width=0.5\linewidth]{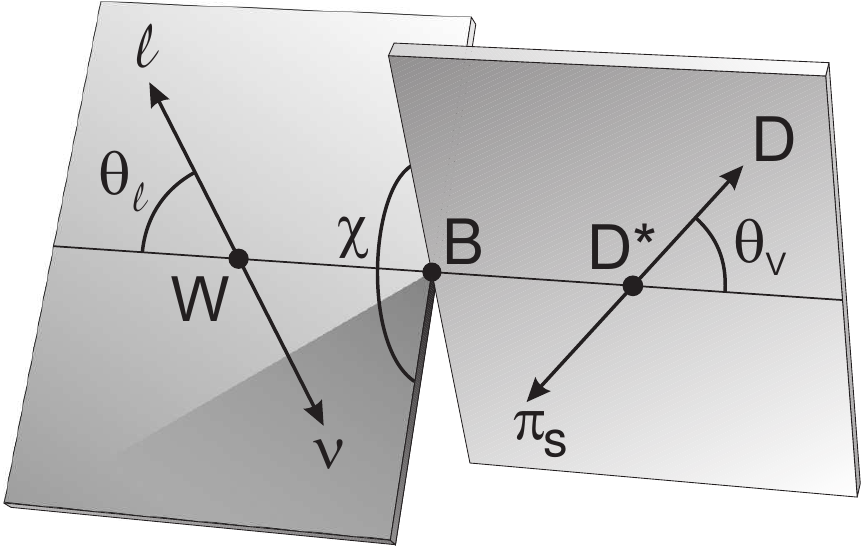} 
  \end{center}
  \caption{The helicity angles $\theta_V$, $\theta_\ell$ and $\chi$ in the  $B\to D^*\ell\nu_\ell$ with the subsequent $D^*\to D\pi$ decay.}
  \label{fig:helicity}
\end{figure}

The $B\to D^*\ell\nu_\ell$ channel, with a branching fraction of about $5\%$, is the most abundant semileptonic decays of the $B$ mesons. 
The $D^*$ is reconstructed in the $D^*\to D\pi$ or $D^*\to D\gamma$ decay modes, so the $B\to D^*\ell\nu_\ell$ decay can be seen as a four-body decay. A full description of this decay requires four independent kinematic variables. A customary choice of variables are $w$, the helicity angle of the $D$ meson ($\theta_V$), the helicity angle of  the charged lepton $\ell$ ($\theta_\ell$), and the angle $\chi$ between the hadronic and leptonic two-body decay planes. These angles are shown in figure \ref{fig:helicity}. Here the 
set $\cos\theta_V$, $\cos\theta_\ell$ and $\chi$
will be collectively called $\Omega$.

The determination of  $|V_{cb}|$ using the $B\to D^*\ell\nu_\ell$ decays has been performed by many experiments in various environments: CLEO, ALEPH, DELPHI, OPAL, and modern $B$-Factories, BaBar and Belle. The measurements performed by CLEO and LEP experiments, and also the first ones at the $B$-Factories, extracted $\eta_{EW}|V_{cb}|$ and some  parameters of  the form factor ${\cal F}(w)$  in (\ref{diffrat0}), 
by measuring only the differential decay rate $d\Gamma$ as a function of $w$ only. Since
the fully differential rate in  $w$ 
and $\Omega$ 
depends on the three helicity amplitudes,  
in these measurements  further assumptions are needed. For example, using the CLN parameterization, one relies on external determinations of the $R_1(1)$ and $R_2(1)$ ratios. 

The first measurement that extracted information on all the form factors was done by CLEO \cite{Duboscq:1995mv}. 
In this pioneering measurement, the joint distribution of $w$ and $\Omega$ 
was fit using an unbinned maximum likelihood method. 
By assuming a linear dependence on $w$ of 
$h_{A_1}(w)$, 
and $R_1(w)$ and $R_2(w)$ independents of $w$,
the following values were measured: $R_1=1.18\pm 0.30\pm 0.12$, $R_2=0.71\pm 0.22\pm 0.07$ and $\rho^2=0.81\pm 0.15\pm 0.06$, where the first uncertainty is statistical and the second is systematic. The parameters $R_1$ and $R_2$ 
were found consistent with the heavy quark symmetry limit of $R_1=R_2=1$. 
The measurement was limited by the statistics available, based only on  $2~{\rm fb}^{-1}$ only, but it was the first observation that the corrections to the heavy-quark symmetry limit are quite small.

BaBar and Belle have measured these form factors and $|V_{cb}|$ with significant improved precision thanks to the larger statistics, the improved analysis techniques and the better knowledge of the background from the decay into excited $D^{**}$ final states.
At $B$-Factories the $B\to D^*\ell\nu_\ell$ decay has been studied using both 
the untagged approach and the hadronic $B$-tagging technique.
In the following we describe in more details the two most recent measurements, one by Belle \cite{Abdesselam:2018nnh}, using the untagged approach, and one from BaBar \cite{Dey:2019bgc}, based on the hadronic $B$-tagging. 

\subsubsection{Belle untagged measurement}
\label{Belleuntaggedmeasurement}

The Belle experiment has measured the shape of the form factors and $|V_{cb}|$ 
from \bztodslnu using both the CLN and the BGL parameterizations~\cite{Abdesselam:2018nnh}.  This analysis, based on the full dataset of 711~${\rm fb}^{-1}$, extracts the parameters of interests from one-dimensional projections 
on $w$ and the angles $\Omega$. 

A positron or an anti-muon with momentum  in the range 0.3-2.4\gev or 0.6-2.4\gev in the laboratory frame, is combined with a $D^{*-}$ candidate. The $D^{*-}$ is reconstructed from a \dzb and slow pion $\pi^-$. The invariant mass difference between the $\dzb\pi^-$ combination and the \dzb candidates, $\Delta m= m(\dzb\pi^-)-m(\dzb)$, is required to be less than $165\mev$. To reduce the combinatorial background, the \dzb is reconstructed only in the $K^+\pi^-$ decay mode, which has a branching fraction of about $3.8\%$ and it is the experimentally cleanest mode. 

The most relevant backgrounds leftover, after the selection requirements, are
\begin{itemize}
\item Continuum background: $e^+e^-\to c{\bar c}$, where $\bar c$ gives a $D^{*-}$;
\item Combinatorial backgound: fake $D^{*-}$ candidates; 
\item $D^{**}$: resonant $B\to D^{**} \ell \nu_{\ell}$ decays, where $D^{**}$ decays to a $D^*$, and non-resonant $B\to D^{(*)}\pi \ell \nu_{\ell}$ decays;
\item Misidentified lepton: $D^{*-}$ candidate is combined with an hadron identified incorrectly as electron or muon;
\item Correlated background: when the $D^{*-}$ and the lepton come from the same $B$, like $B\to D^{*}\tau \nu_\tau$, $\tau\to \ell\nu\nu$; $B\to D^{*}X_c$ where $X_c\to \ell Y$;  
\item Uncorrelated background: when the $D^{*-}$ and the lepton come from different $B$'s.
\end{itemize}

\begin{figure}[tb!]
  \begin{center}
    \includegraphics[width=0.48\textwidth]{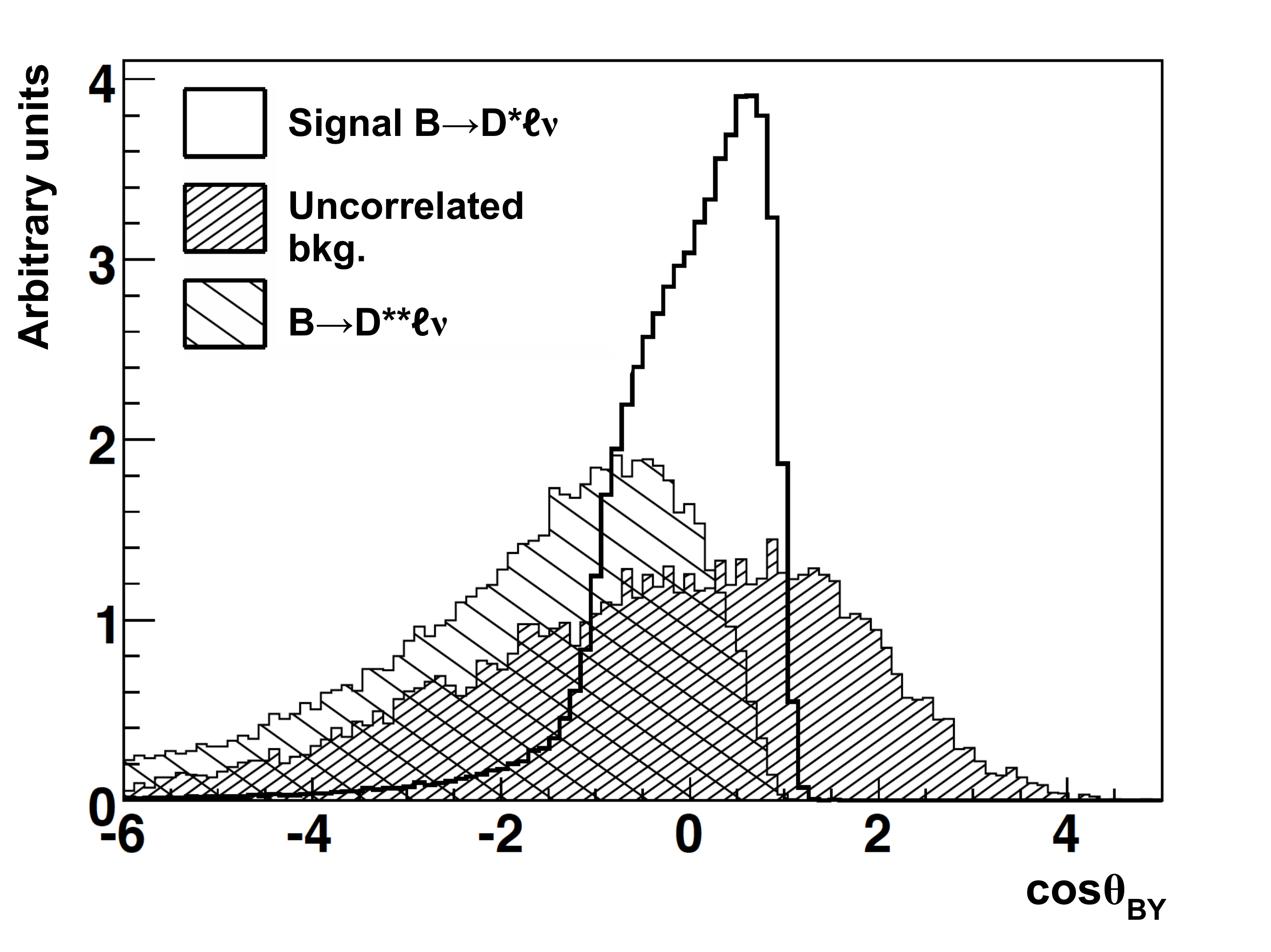}
    \includegraphics[width=0.50\textwidth]{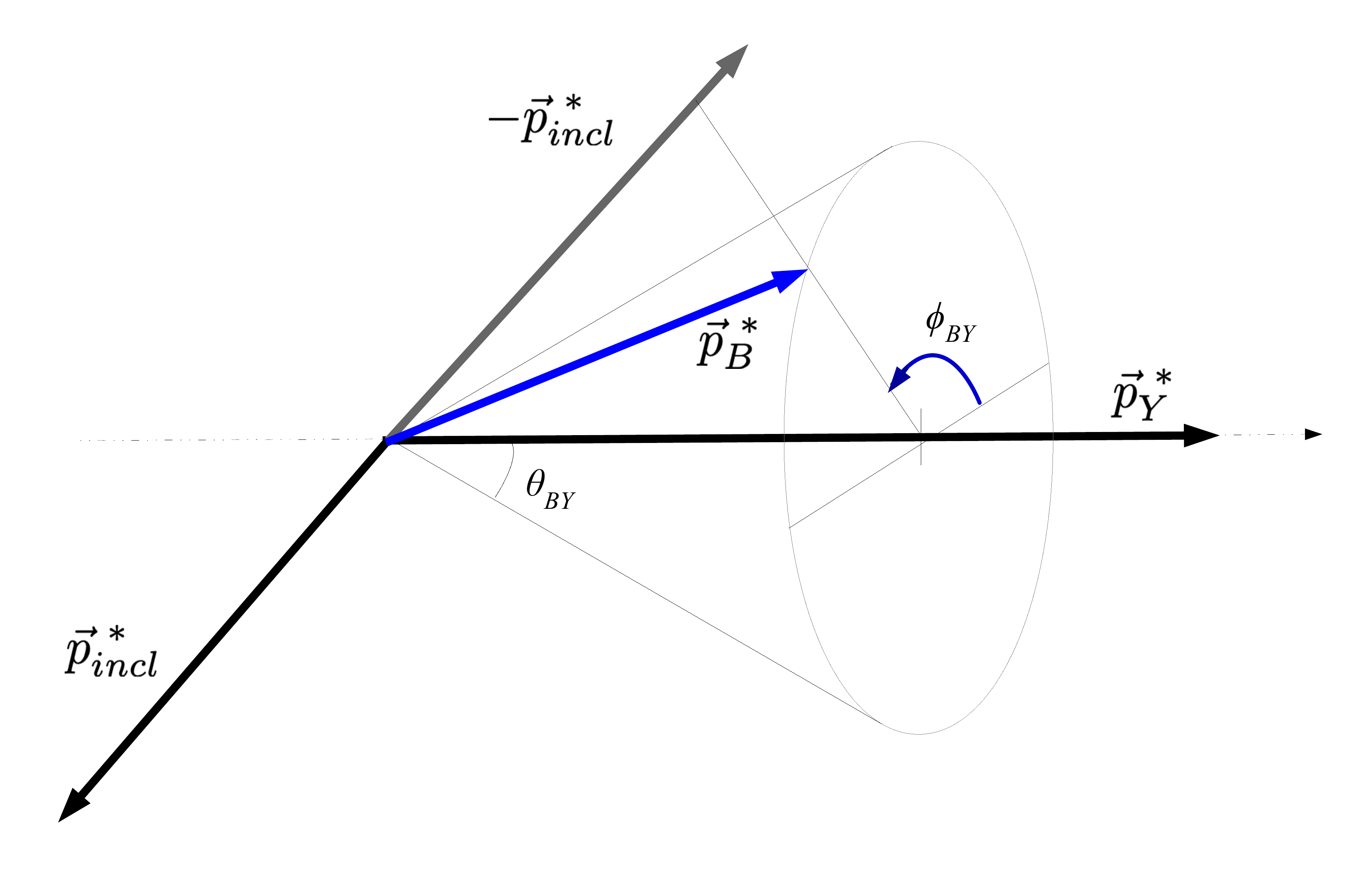}
  \end{center}
  \vspace*{-5mm} 
  \caption{Left: the $\cos\theta_{BY}$ distribution for the signal $B\to D^* \ell\nu_{\ell}$, the $D^{**}$ and the uncorrelated background. The differences in shapes show the discriminating power of this variable. Because of the bremsstrahlung the tail extend in the $\cos\theta_{BY}<-1$ region. Right: illustration on how the direction of the $B$ meson is determined. The quantity ${\vec p}^{~\ast}_{incl}$ is given by summing the three-momentum of the particles not associated to the signal. The direction of the signal $B$, ${\vec p}^{~\ast}_{B}$, is the one that minimizes the distance with the $-{\vec p}^{~\ast}_{incl}$ vector. }. 
  \label{fig:cosby_mc}
\end{figure}

The signal and the background yields for the various sources are extracted performing a binned maximum likelihood fit of the $D^*\ell$ candidates in the variables $\Delta m$, $\cos \theta_{BY}$ and $p_\ell$. The momentum of the lepton $p_\ell$ is sensitive to the form factors themselves, thus, to avoid biasing the measurement, it is divided only in two regions, below and above $0.6\gev$. This choice has been useful to constrain the residual lepton misidentification background that affects mainly the low lepton momentum region. The invariant mass difference $\Delta m$ is sensitive to the combinatorial background. The most powerful variable that allows to separate signal from the $D^{**}$ and the correlated background, is  $\cos\theta_{BY}$. In the assumption that the decay is  $B\to D^*\ell\nu$, in the $\Upsilon(4S)$ rest frame, the $B$ direction can be constrained in a cone around the axis given by the $Y\equiv D^*\ell$ direction

\begin{equation}
\cos\theta_{BY}=\frac{2 E_B^* E_Y^*-m_B^2-m_Y^2}{2|\vec{p}_B^{~*}||\vec{p}_Y^{~*}|}
\label{eq:cosby}
\end{equation}

\noindent where $E_B^*$ and $|\vec{p}_B^*|$ are given by the beam energy and all the other quantities are determined only by the visible system $Y$. In \eqref{eq:cosby} all the kinematic quantities are computed in the \FourS rest frame. For the signal, $\cos\theta_{BY}$ is constrained in the physical region $(-1.0\div 1.0)$, instead for the $B\to D^{**}\ell\nu$ and $B\to D^* \pi\ell\nu$, where one or more further particles are emitted, it is easy to show that $\cos\theta_{BY}$ is only constrained to be less than $+1.0$. 
Thus for $D^{**}$ background $\cos\theta_{BY}$ has a long tail below the $-1.0$ value. The uncorrelated background is also constrained from $\cos\theta_{BY}$ because its shape extends to the region with $\cos\theta_{BY}>1.0$. The distribution of $\cos\theta_{BY}$ for the most important physical backgrounds is shown in figure~\ref{fig:cosby_mc} (left). 

Because of the bremsstrahlung that affects the electrons, and the finite resolution in the momentum reconstruction of the visible energy,  $\cos\theta_{BY}$ for the signal also extends over the physical range. An important source of uncertainty in the computation of $\cos\theta_{BY}$ is due to the beam energy spread of few \mev that affects the computation of $p_B^*$ and $E_B^*$ hence smearing the $\cos\theta_{BY}$  variable. 

The signal yields extraction, from a simultaneous fit to $\Delta m$, $\cos \theta_{BY}$ and $p_\ell$, is performed for each bin of the kinematic variables considered ($w$ and the angles $\Omega$). After the background subtraction, a total number of $180 \times 10^3$ candidates is obtained.

\begin{figure}[ht]
  \centering
  \includegraphics[width=0.48\textwidth]{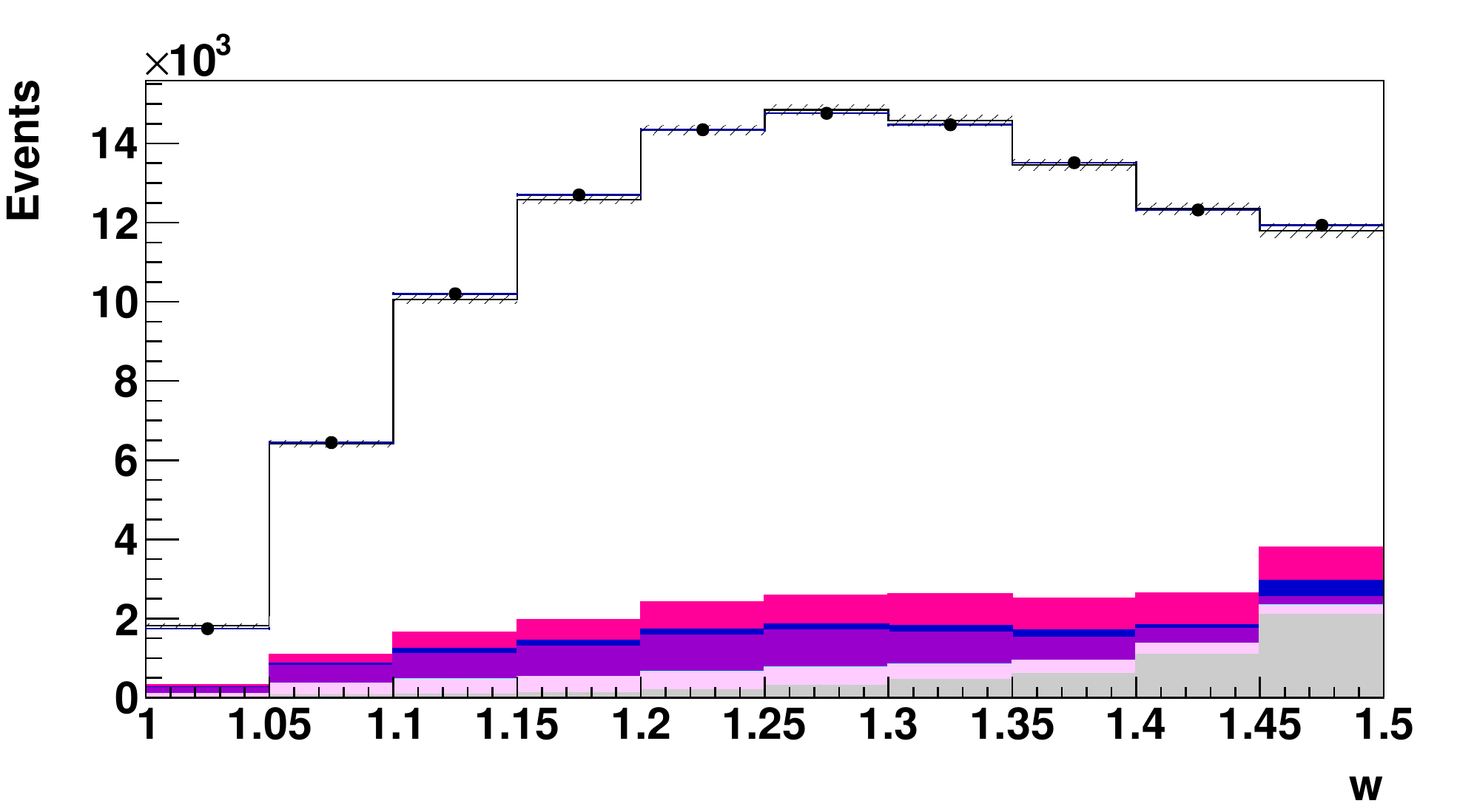}
  \includegraphics[width=0.48\textwidth]{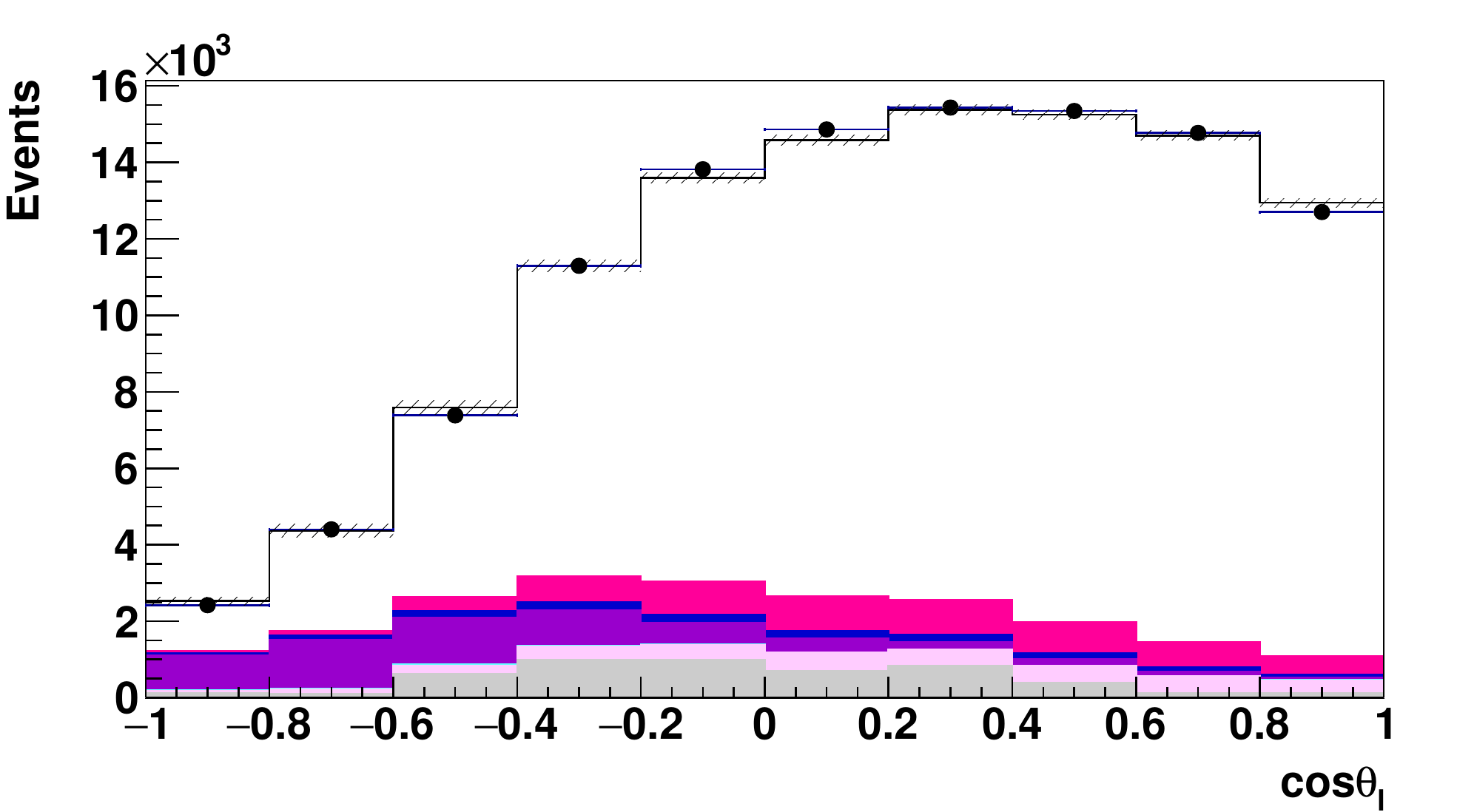}
  \includegraphics[width=0.48\textwidth]{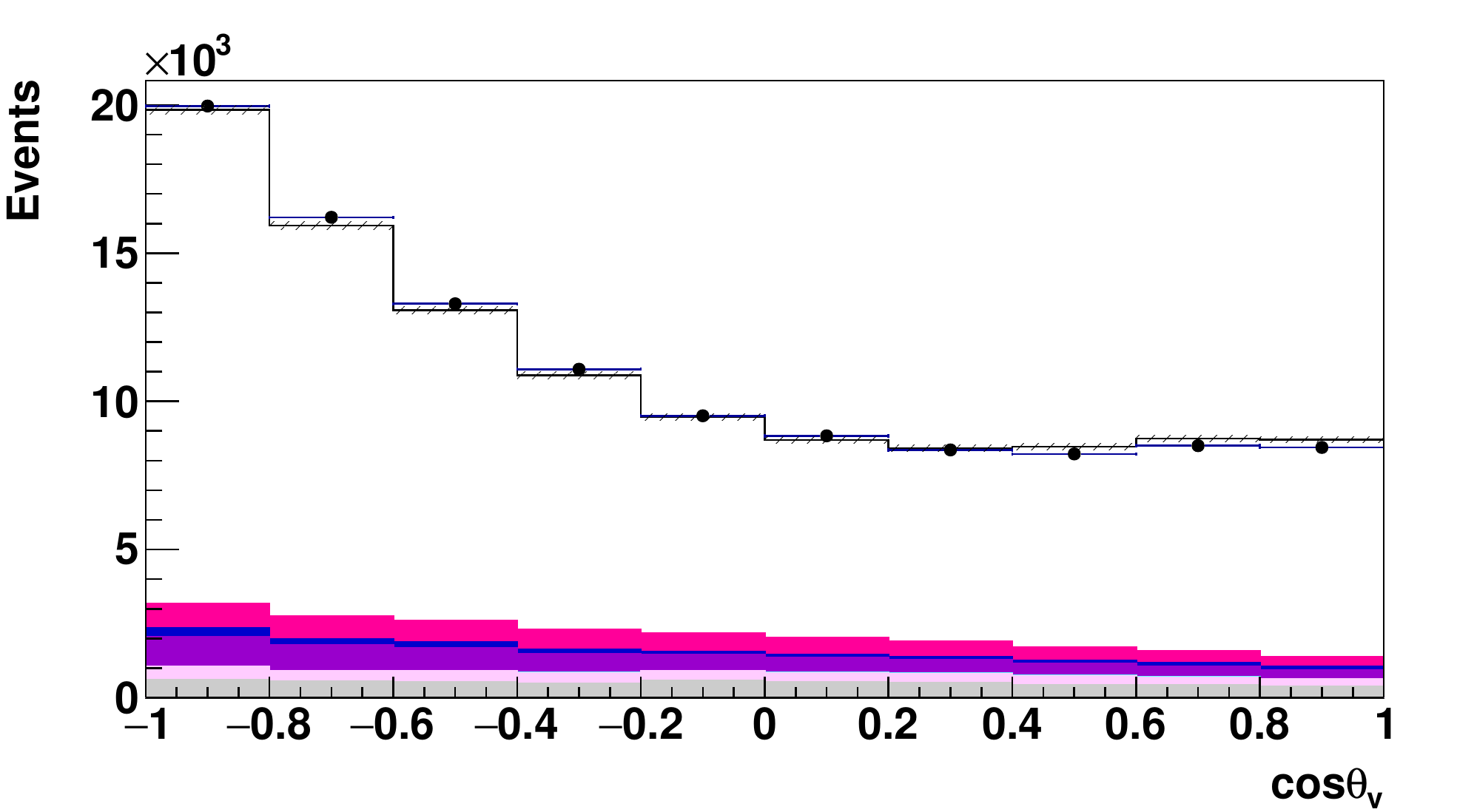}
  \includegraphics[width=0.48\textwidth]{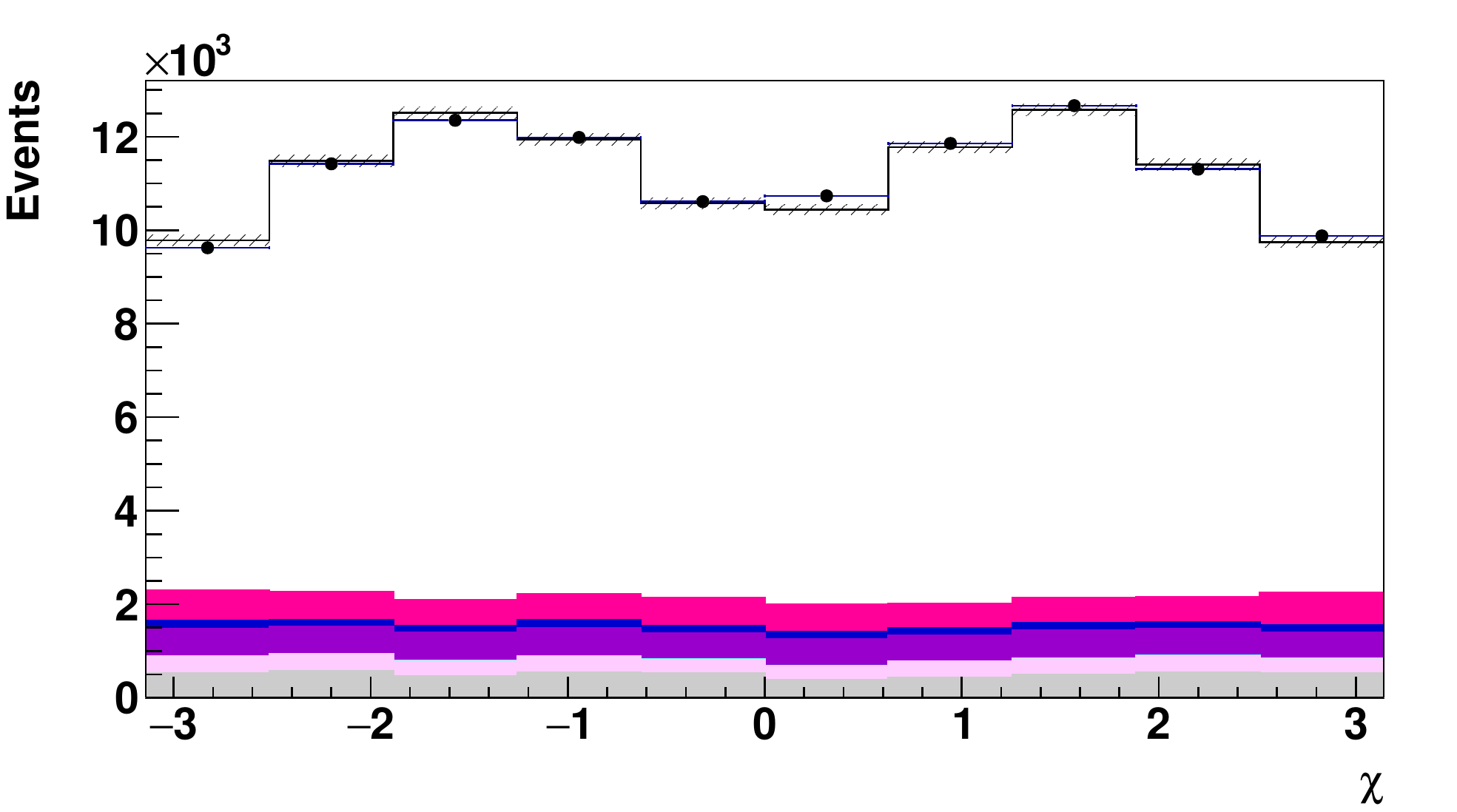}
  \caption{Distribution of the variables $w$, the cosine of the angles $\theta_V$, $\theta_\ell$, and the angle $\chi$ for the Belle analysis of  $B\to D^*\ell\nu_\ell$. The fit results using the CLN parameterization is superimposed. From \cite{Abdesselam:2018nnh}.}
  \label{fig:belle_kin}
\end{figure}

Because of the presence of the neutrino, the $B$-direction is not known, so these kinematic quantities cannot be calculated directly. From the value of $\cos\theta_{BY}$ per event, it is known only that the $B$ must lie on a cone around the direction of the $Y$ system. Various approaches have been used in different analyses to constrain the $B$ direction on the cone. In this analysis Belle exploits the rest of the events to 
built a rough estimation of the direction of the other $B$ inclusively. The $B$ direction is chosen as the one on the cone closest to opposite of the other $B$ meson direction. In figure~\ref{fig:cosby_mc}(right) is illustrated how the technique works. With this algorithm the resolutions of the kinematic variables are 0.020 for $w$, 0.038 for $\cos\theta_\ell$, 0.044 for $\cos\theta_V$ and 0.210 for $\chi$. The data are divided in 10 equidistant bins for each of four variables. The distributions of these variables, after the fit, are shown in figure \ref{fig:belle_kin}.
The signal yields in the four one-dimensional projections are simultaneously fitted to extract the shape of the form factors. Because the same events enter into the four projections, the correlation between all the various bins has to be carefully evaluated for both signal and backgrounds.

The measured yields are normalized to the total number of $B^0$ in the sample analyzed, which is given by $N_{B^0}=2\cdot f_{00} \cdot N_{BB}$, where $N_{BB}$ is the total number of $BB$ pairs collected by Belle, known with a precision of $1.4\%$, and $f_{00}$ is the branching ratio of $\Upsilon(4S)\to B^0 {\bar B}^0$ determined to be $f_{00}=0.486\pm0.006$ \cite{Tanabashi:2018oca}. 

The results of the fit on $\eta_{EW}{\cal F}(1)|V_{cb}|$ with the CLN parameterization are given in table~\ref{tab:dslnu_summary}, together with the results from other measurements obtained with the same parameterization.
This result, 
consistent with the other measurements, is the most precise and dominates the HFLAV average. 
It is dominated by the systematic uncertainties which give a contribution of $1.6\%$, while the statistic uncertainty is only $0.4\%$. The dominant source of systematics is the tracking efficiency, mainly the soft pion one, the lepton identification and the uncertainty on the total number of $\Upsilon(4S)$ candidates. Also the external parameters, ${\cal B}(D\to K\pi)$ and $f_{+-}/f_{00}$, give significant contributions.

The results of the fit with the BGL parameterization is reported in table \ref{tab:dslnu_bgl}. The series in the expansion of the form factors are truncated at $n=1$ for $f(z)$  and $g(z)$, instead ${\cal F}_1(z)$ is truncated at $n=2$. Following the notation used in \cite{Gambino:2019sif}, this BGL configuration is called BGL$^{(121)}$ and has five free parameters, one more than the CLN one. This parameterization describes the data very well and the data 
are not 
sensitive to higher orders coefficients. The unitarity constraints have not been applied. 
The result on $|V_{cb}|$ obtained with the BGL parameterization is compatible with the CLN one, but has a larger statistical uncertainty. The $\chi^{2}/\rm{ndf}$ of the fit to Belle data, in both CLN and BGL cases, are acceptable, so the available data  are not sensitive to the different parameterizations. 


\subsubsection{BaBar tagged measurement} 
\label{BaBartaggedmeasurement}

BaBar has measured the shape of the form factors of  \bztodslnu decays using both CLN and BGL parameterizations \cite{Dey:2019bgc}. This analysis is based on the full dataset of 450 fb$^{-1}$, and exploits a sample where one of the $B$ is fully reconstructed. The hadronic $B$-tagging is described in section \ref{sec:semilep_bfactories}. The knowledge of the kinematic of the $B_{tag}$ event by event, and the beam properties, allows to determine the four-momentum of the neutrino from the missing four-momentum $p_{miss}=p_{e^+}+p_{e^-}-p_{B_{tag}}-p_{D^*\ell}$. 

\begin{figure}[tb!]
  \begin{center}
    \includegraphics[width=0.45\textwidth]{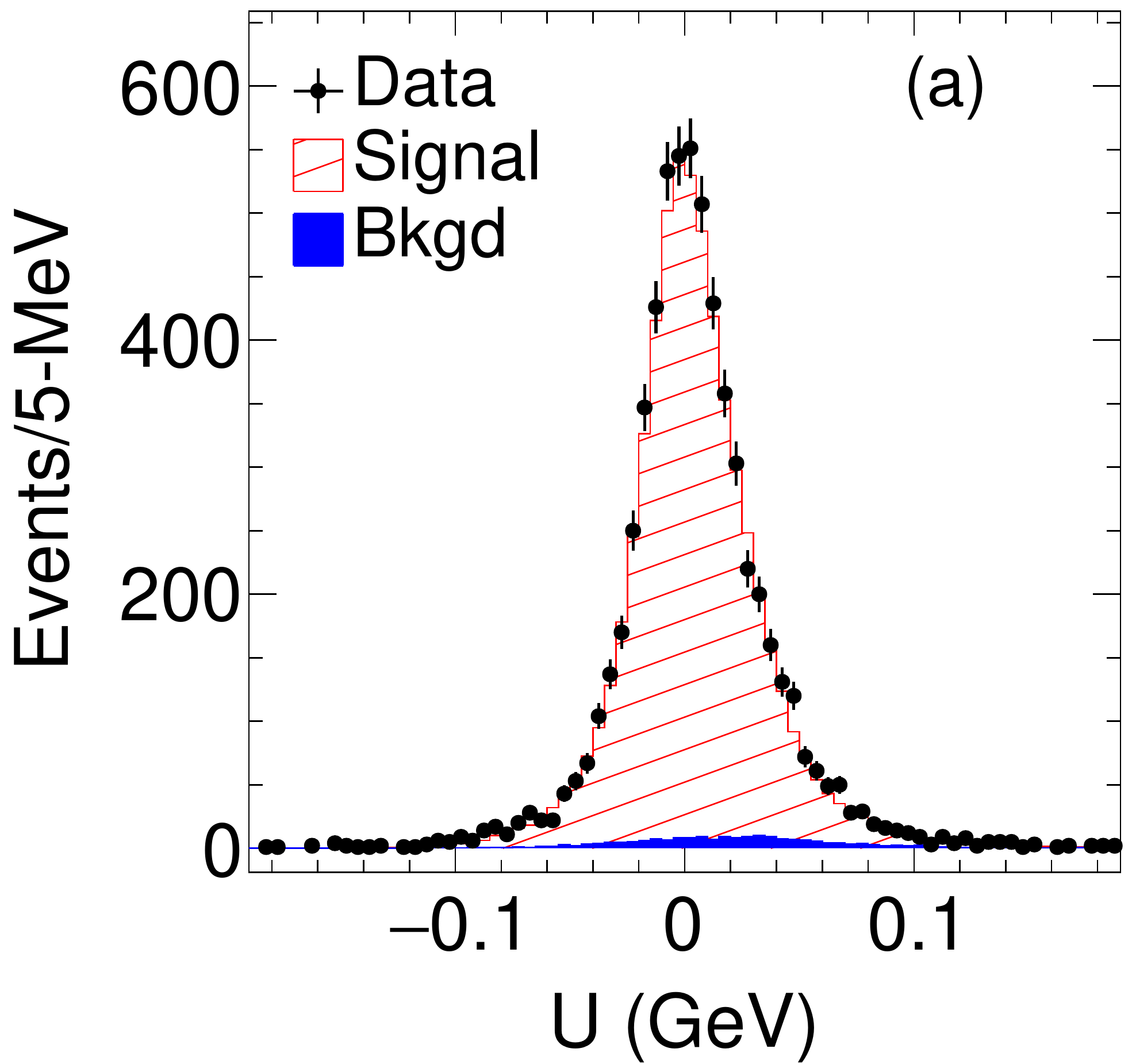}
    \includegraphics[width=0.45\textwidth]{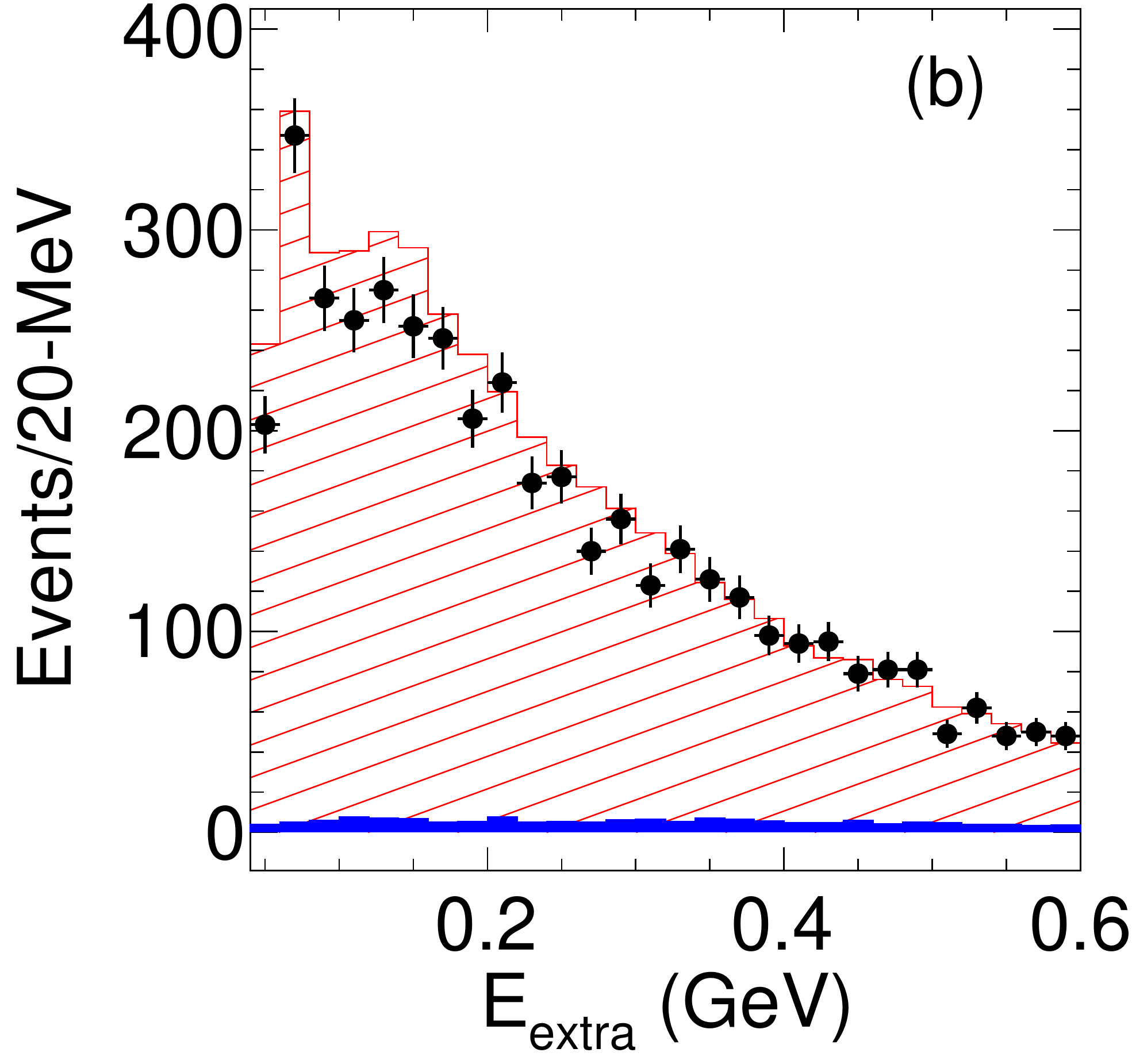}
  \end{center}
  \vspace*{-5mm} 
  \caption{Comparison between data and simulation in the variables $U$ (a) and $E_{extra}$ (b). From \cite{Dey:2019bgc}.}
  \label{fig:BaBar_tagged}
\end{figure}

In this analysis two decay chains are considered: \bztodslnu with $D^{*-}\to \dzb\pi^-$, and \bptodslnu with $D^{*0}\to \dzb\pi^0$.
The \dzb is reconstructed in the three cleanest modes $K^+\pi^-$,  $K^+\pi^-\pi^0$, $K^+\pi^-\pi^+\pi^-$. As usual the $D^*$ is selected requiring $\Delta m$ to be consistent with the expectations. The lepton is required to have momentum in the laboratory frame greater than 0.2\gev or 0.3\gev, if it is an electron or a muon, respectively. Besides the $B_{tag}$, the $D^*$ and the lepton, no additional tracks are allowed in the event. The full decay chain $e^+e^-\to \Upsilon(4S)\to B_{tag} B_{sig}(\to D^*\ell\nu_\ell )$ is considered in a kinematic fit that includes constraints on the beam spot, the secondary vertices, the masses of $B_{tag}$, $B_{sig}$, $D^*$ and the missing neutrino. The probability of the $\chi^2$ of this constrained fit is the main discriminating variable against the backgrounds. 
The sample is further cleaned rejecting candidates with large values for $E_{extra}$, which is defined as the sum of the energy of the photons not associated with the signal. The overall background level is only $2\%$ and it is due to $B\bar B$ events decaying generically. The agreement between the signal and simulations for the $E_{extra}$ and the variable $U=E_{miss}-|\vec{p}_{miss}|$ is very good, as can be seen in figure \ref{fig:BaBar_tagged}. After all the selection requirements, a total of about $5900$ signal candidates is obtained.  

The shape of the form factors is extracted using an unbinned maximum likelihood fit where the signal events are described by the four dimensional decay rate $d\Gamma/dw d\Omega$. All events in the signal region, defined by $|U|<90\mev$, are considered in the likelihood as signal, and the small residual background is subtracted using information from large sample of $B{\bar B}$ simulated events. 

The extraction of $|V_{cb}|$ is performed indirectly by adding to the likelihood the constraint that the semileptonic decay width $\Gamma$ is given by  $\Gamma={\cal B}/\tau_B$, where ${\cal B}$ is the $B\to D^*\ell\nu$ branching fraction and $\tau_B$ is the $B$ meson lifetime. The values of these external inputs are taken from HFLAV \cite{Amhis:2019ckw}. 

The result with the BGL parameterization is reported in table~\ref{tab:dslnu_bgl}  and with the CLN one  in table~\ref{tab:dslnu_summary}. They are perfectly compatible and also compatible with the HFLAV average. 
Because of the limited signal statistics, 
The form factors are truncated at $n=1$, BGL$^{(111)}$, to avoid the violation of the unitarity constraints due to poorly determined parameters. Higher order terms in BGL are checked and found to have a negligible effect on the shape of the form factors. 


The dominant source of systematic uncertainty on the measurement of $|V_{cb}|$ is due to the remnant background that contaminates the angular distributions. The resolution on the kinematic variables is about a factor five better than the one possible with the untagged measurement. The impact of the finite resolution is evaluated using the simulation and turns out to be negligible. 


\begin{table}[hb]
\centering
\caption{Fit results with the BGL parameterization of some recent analyses. The notation is BGL$^{n_f,n_{F1},n_g}$ where $n_f$ $n_{F1}$ and $n_g$ are the order of the $z$-expansions for the $f(z)$, $F_{1}(z)$ and $g(z)$ respectively. The parameters cannot be compared directly because different constants are used, in particular the $B_c^*$ masses value in the Blasckhe factor. Furthermore BaBar assumes $t_0=t_+-\sqrt{t_+(t_+-t_-)}$, while the other fits assume $t_0=t_-$.}
  \vspace*{3mm} 
  \label{tab:dslnu_bgl}
  \begin{footnotesize}
  \begin{tabular}{l c c c c}
   \hline
    BGL                  & Bigi et al.\cite{Bigi:2017njr}& Belle \cite{Abdesselam:2018nnh} & BaBar \cite{Dey:2019bgc} & Gambino et al.\cite{Gambino:2019sif}\\
    			    &  BGL$^{222}$       & BGL$^{120}$  & BGL$^{111}$  & BGL$^{222}$  \\
    \hline                                                           
    $|V_{cb}|\times 10^{-3}$& $41.7^{+2.0}_{-2.1}$ 	       	&	$38.3\pm0.97$ 		&	$38.36\pm0.90$		& 	$39.6^{+1.1}_{-1.0}$	 \\
     $a_0^{f}$&			$0.01223\pm0.00018$         		&	$0.0131\pm0.0002$ 	&	$0.0129\pm0.0003$	& 	$0.01221\pm0.00016$	 \\
     $a_1^{f}$&			$-0.054^{+0.058}_{-0.043}$          &	$0.0169\pm0.0050$	&	$0.0163\pm0.0010$	& 	$0.006^{-0.032}_{-0.045}$	 \\
     $a_2^{f}$&			$0.20^{+0.7}_{-1.2}$           		&	                          - 		&	                            		& 	$-0.2^{+1.2}_{-0.8}$	 \\
     $a_1^{F_1}$& 		$-0.0100^{+0.0061}_{-0.0056}$	&	$0.0070\pm0.0018$ 	&     $0.0003\pm0.0011$       & 	$0.0042\pm0.0022$	 \\
     $a_2^{F_1}$& 		$0.012\pm 0.010$	              		&	$0.085\pm0.034$ 		&	-	                       		& 	$-0.069^{+0.041}_{-0.037}$	 \\
     $a_0^{g}$&			$0.012^{+0.011}_{-0.008}$        	&	$-0.0241\pm0.0058$ 	&	$0.0274\pm0.0011$	&  	$0.024^{+0.021}_{-0.009}$	 \\
     $a_1^{g}$&			$0.7^{+0.3}_{-0.4}$         		&	- 	                        	&	$0.0833\pm0.0667$	&  	$0.05^{+0.39}_{-0.72}$	 \\   
     $a_2^{g}$&			$0.8^{+0.2}_{-1.7}$       			&	- 	                        	&	-	                        	& 	$1.0^{+0.0}_{-2.0}$	 \\   
  \hline
  \end{tabular}
  \end{footnotesize}
\end{table}

\subsubsection{Results}
\label{results633}

\begin{table}[hb]
\centering
\caption{Results of \btodslnu measurements with the CLN parameterization and the current HFLAV average \cite{Amhis:2019ckw}. Only $\eta_{EW} {\cal F}(1) |V_{cb}| (\times 10^3)$ and $\rho^2$ are reported.} 
\label{tab:dslnu_summary}
\begin{footnotesize}
\begin{tabular}[t]{l >{\raggedright\arraybackslash}p{0.18\linewidth}  >{\raggedright\arraybackslash}p{0.65\linewidth} }
\hline\hline
  & $\eta_{EW} {\cal F}(1) |V_{cb}| (\times 10^3)$\newline $\rho^2$  & Remarks \\
 \hline
 
BaBar\cite{Aubert:2007rs} & 33.77$\pm$0.29$\pm$0.98  1.184$\pm$0.048$\pm$0.029 & Untagged measurement of the \bztodslnu decay. Fit to the four projections: $w$ and the three helicity angles. Data fitted with the CLN. Extracted also $R_1(1)$ and $R_2(1)$, together with $\rho^2$. The form factors are further constrained to a dedicated measurement performed by BaBar which uses only clean $B^0\to D^{*-} e^+ \nu_{e}$  data samples \cite{Aubert:2006cx}. \\

BaBar\cite{Aubert:2007qs}  &34.81$\pm$0.58$\pm$1.06  1.125$\pm$0.058$\pm$0.053 & Untagged measurement of \bptodslnu with $D^{*0}$ reconstructed in $\dzb\pi^0$ decay mode. One-dimensional fit 
of $w$ using only CLN. Parameters $R_1(1)$ and $R_2(1)$ taken from external inputs.\\

BaBar\cite{Aubert:2008yv}  &35.75$\pm$0.20$\pm$1.09  1.180$\pm$0.020$\pm$0.061 & Global analysis of \btodslnu and \btodlnu using inclusive samples of $B\to D^-\ell\nu_\ell X$ and 
$B\to \dzb\ell\nu_\ell X$ decays. The fit is performed multidimentional on $p_\ell^*$, $p_D^*$ and $\cos\theta_{BY}$ variables. Only the CLN parameterization was used.\\
 
Belle\cite{Abdesselam:2018nnh}  & 35.07$\pm$0.15$\pm$0.56  1.106$\pm$0.031$\pm$0.008 & Untagged measurement of \bztodslnu. Fit to the four projections. Data fitted with the CLN. Extracted also $R_1(1)$ and $R_2(1)$, together with $\rho^2$. 
Results also using the BGL. Published also the background subtracted spectra of the four projections with all the information, like efficiencies and migration matrix, needed for subsequent refitting.\\

\hline
HFLAV\cite{Amhis:2019ckw}  & 35.27$\pm$ 0.11$\pm$0.36  1.122$\pm$0.015$\pm$0.019 & The average includes also older measuremenets from CLEO and LEP experiments: DELPHI, ATLAS and OPAL. The average Confidence Level is only 0.8\%. \\

\hline\hline
Belle\cite{Abdesselam:2017kjf} & 34.93$\pm$0.23$\pm$0.59  & Tagged measurement of \bztodslnu, not published. For the first time the spectrum of the projections on $q^2$ and the angular variables was released. The spectrum unfolded and corrected for the efficiency was also released.\\

BaBar \cite{Dey:2019bgc}& $34.94\pm0.50$ $0.96\pm0.08$ & Tagged measurement of \bptodslnu and \bztodslnu decays. Not included yet in the HFLAV average. The uncertainties include bot the statistical and systematics. Fit is unbinned to $q^2$ and the angular variables. Data fitted using both CLN and BGL. This measurement is not normalized, so $|V_{cb}|$ is extracted from the measured $B^0\to D^{*-}\ell^+\nu_\ell$ branching fractions \cite{Amhis:2019ckw}.\\
\hline\hline
\end{tabular}
\end{footnotesize}
\end{table}

Table~\ref{tab:dslnu_summary} reports a summary of the measurements of $\eta_{EW}{\cal F}(1)|V_{cb}|$ obtained with the CLN parameterization, together with the HFLAV average. Using the FLAG 2019 value for the normalization of the form factor $\eta_{EW}{\cal F}(1)=0.910\pm 0.013$, the HFLAV average for $|V_{cb}|$ is 
\beq
|V_{cb}|=(38.76\pm 0.42 \pm 0.55)\times 10^{-3}
\label{eq:vcb_dslnu}
\eeq
\noindent where the first error is experimental and the second is due to the form factor normalization. The results obtained with CLN and BGL parameterizations are consistent.

In the 2017 the Belle collaboration released an analysis, not published, of $B\to D^*\ell\nu$ using the hadronic $B$-tagging \cite{Abdesselam:2017kjf}. The form factors and $|V_{cb}|$ were extracted from the projections on $w$ and the angles $\Omega$ with a fit similar to the one described before for the untagged analysis. The result of the fit, 
performed using the CLN parameterization, 
was consistent with 
previous measurements. The Belle collaboration released also the spectra of the projections on the four kinematic variable, unfolded for the resolution and corrected for the efficiencies. 
Some groups took this opportunity to fit the Belle data using not only the CLN parameterization but also the BGL one~\cite{Grinstein:2017nlq, Bigi:2017njr, Bernlochner:2017xyx, Jaiswal:2017rve}. They observed    that the 
central 
value of $|V_{cb}|$  using the BGL parameterization was systematically higher than the 
value obtained with the CLN one, 
and that, depending on the choice of constraints and inputs of the analysis, could be lifted  up to $6-7\%$.
For illustration, 
we report in table \ref{tab:dslnu_bgl} the BGL results of \cite{Bigi:2017njr}.
The fact that the BGL result 
became 
compatible with the inclusive determination of $|V_{cb}|$ 
disclosed the possibility that 
a suitable choice of the parameterization 
could be enough to solve the $|V_{cb}|$ puzzle. 
%
However, some inconsistencies 
were observed  in the fits exploiting the BGL approach; for example  it was shown
in \cite{Bernlochner:2017xyx} that the form factor ratio $R_{1}(w)$   determined from the results of the  fits
 strongly contradicts the HQS predictions.

More data were eagerly needed. They have been provided in 2018 by  Belle~\cite{Abdesselam:2018nnh}  and in 2019 by  BaBar~\cite{Dey:2019bgc}.
These analyses, detailed
in section~\ref{Belleuntaggedmeasurement} and ~\ref{BaBartaggedmeasurement}, respectively, show no sign of discrepancy on $|V_{cb}|$ between the BGL and CLN parameterizations, within the uncertainties. Belle also in this case released the data in a format that allows them to be fitted by 
outside groups, prompting a new analysis by some among the authors of  the 2017 fits~\cite{Gambino:2019sif}. The  new fits, which include also the  previous Belle analysis, have been performed
with both CLN and BGL parameterizations,
in different configurations,  and the results found to be consistent. 
Also the BGL  discrepancy with HQS mentioned  before seems to be overcome.
The BGL value  is reported in table \ref{tab:dslnu_bgl}, for comparison with the results of Belle and BaBar.

Nevertheless the initial discrepancies have been useful to revisit the assumptions under the widely used CLN parameterization. The possible systematics due to the parameterization itself 
had never been considered in the $|V_{cb}|$ extraction. Theoretical analyses have investigated constraints and subtleties of the different approaches, including the studies on
 the optimal number of parameters of the BGL fit, and the risk of overfitting~\cite{Bernlochner:2019ldg, Gambino:2019sif}.
 Moreover, with only few exceptions, most of the 
experimental
analyses 
were using 
only the CLN parameterization. With the increasing precision, it is crucial 
to describe the shape of the form factors in a model independent way.
It is worth to mention now that, when calculation of the form factor at $w>1$ will be available, the role of
parameterizations 
will become less relevant, because the extrapolation to $w=1$ will 
reduce to an interpolation between experimental results and different theory points. 

	\subsection{The \btodlnu channel}
	\label{BDchannel}

The analysis of \btodlnu decays is difficult because of the large background from \btodslnu where the $D^{\ast}$ decays in $D\pi$ or $D\gamma$, with the soft pion or gamma lost or not detected. This kind of background is usually called {\it feed-down} in the literature.
In the past, untagged approaches have been used, similar to the one described above for the \btodslnu, by the CLEO \cite{Bartelt:1998dq} and Belle \cite{Abe:2001yf} collaborations. CLEO has used both \bptodlnu and \bztodlnu  decays. In general, the signal selection relies mostly on the selection of a good $D$ meson candidate, so the combinatorial background is large. The consequence is that only few, low multiplicity $D$ decay modes can be exploited. CLEO in fact uses the $\dzb\to K^+\pi^-$ and $D^-\to K^+\pi^-\pi^-$ decay modes, which are the cleanest.
The \bptodlnu decays with the untagged approach is the most difficult because of the large feed-down: the \dzb can come from both \bptodslnu, with $D^{*0}$ decaying in \dzb almost all the time, and \bztodslnu with with $D^{*-}\to \dzb\pi^-$, which has a branching fraction of $68\%$. The \bztodlnu instead is easier because the $D^-$ comes only from \bztodslnu with $D^{*-}\to D^-\pi^0$, which has a branching fraction of only $31\%$. For this reason Belle analyzed only the \bztodlnu. The larger phase space suppression in the region close to $w\to 1$ for the $B\to D$ decays, compared with the $B\to D^*$, 
implies a large background in the region crucial for the $|V_{cb}|$ extraction. 

The hadronic $B$-tagging is particularly suitable for the \btodlnu, as 
shown in the BaBar analysis \cite{Aubert:2009ac}, where both \bztodlnu and \bptodlnu are studied. The hadronic tagging allows to reduce the combinatorial background in the \dzb and $D^-$ reconstruction, and also the feed-down from $D^{*}$ and $D^{**}$ decays, because the tagging allows to separate clearly $\Upsilon(4S)\to B^+B^-$ from $\Upsilon(4S) \to B^0{\overline B}^0$ decay modes.

\subsubsection{Belle tagged analysis}
\label{Belletaggedanalysis1}
The most precise measurement has been done by Belle \cite{Glattauer:2015teq} and uses an improved hadronic $b$-tagging approach. The tracks and the clusters of the event, remaining after the identification of the $B_{tag}$, are used to identify the
\btodlnu
signal decay. The lepton is required to have a momentum greater than 0.3\gev for the electron case, and 0.6\gev for the muon case.
The low signal efficiency, due to the reconstruction of the hadronic tag, is partially compensated by the possibility to reconstruct $D$ mesons in many different decays modes, also including $\pi^0$ and $K_s$ particles. In particular, the $D^-$ meson is reconstructed in 10 possible final states, covering about $29\%$ of the total rate, and the \dzb is reconstructed in $13$ final states, corresponding to more than $40\%$ of the total rate. 

\begin{figure}[ht!]
  \centering
  \includegraphics[width=0.48\textwidth]{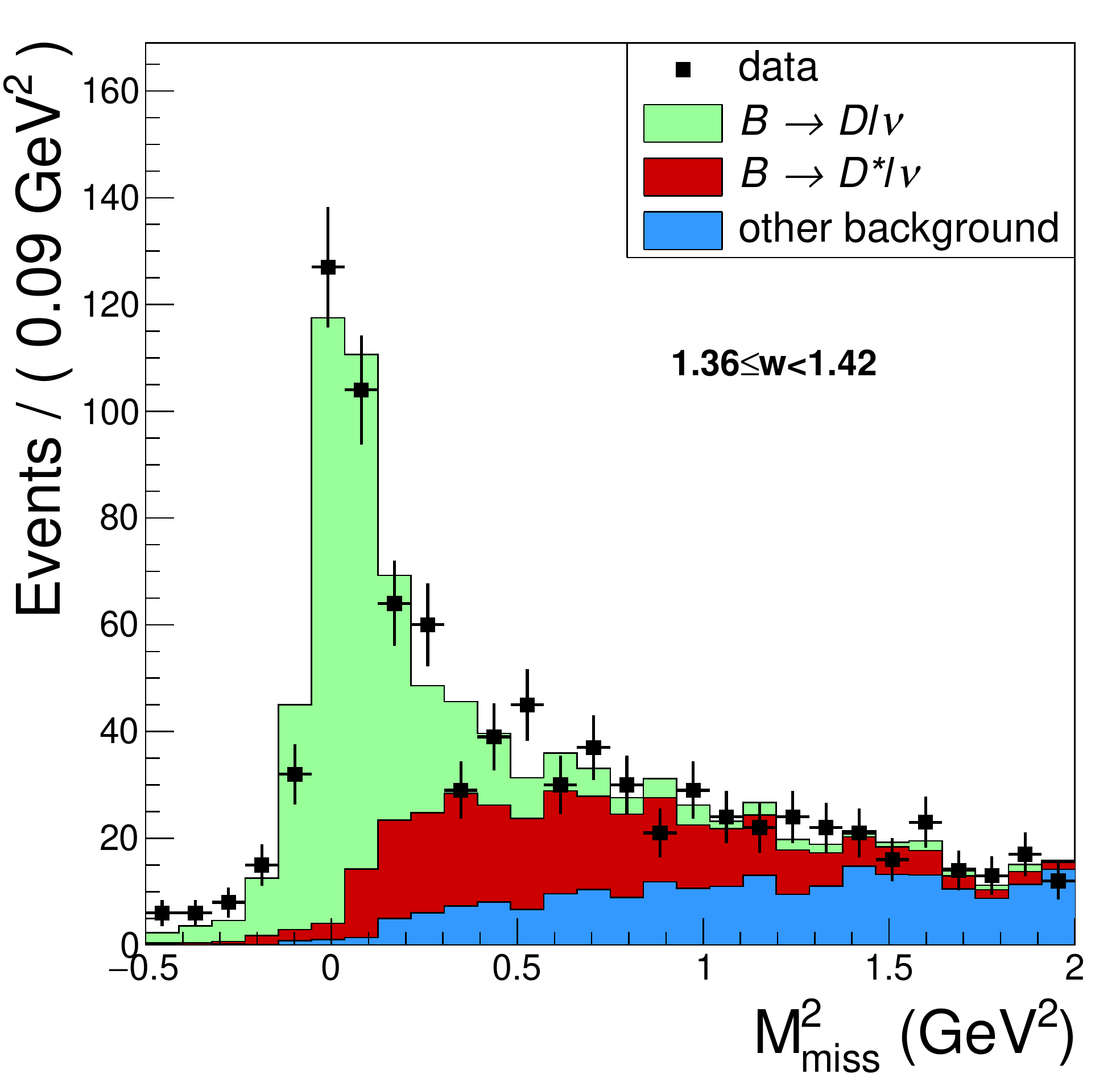}
  \includegraphics[width=0.48\textwidth]{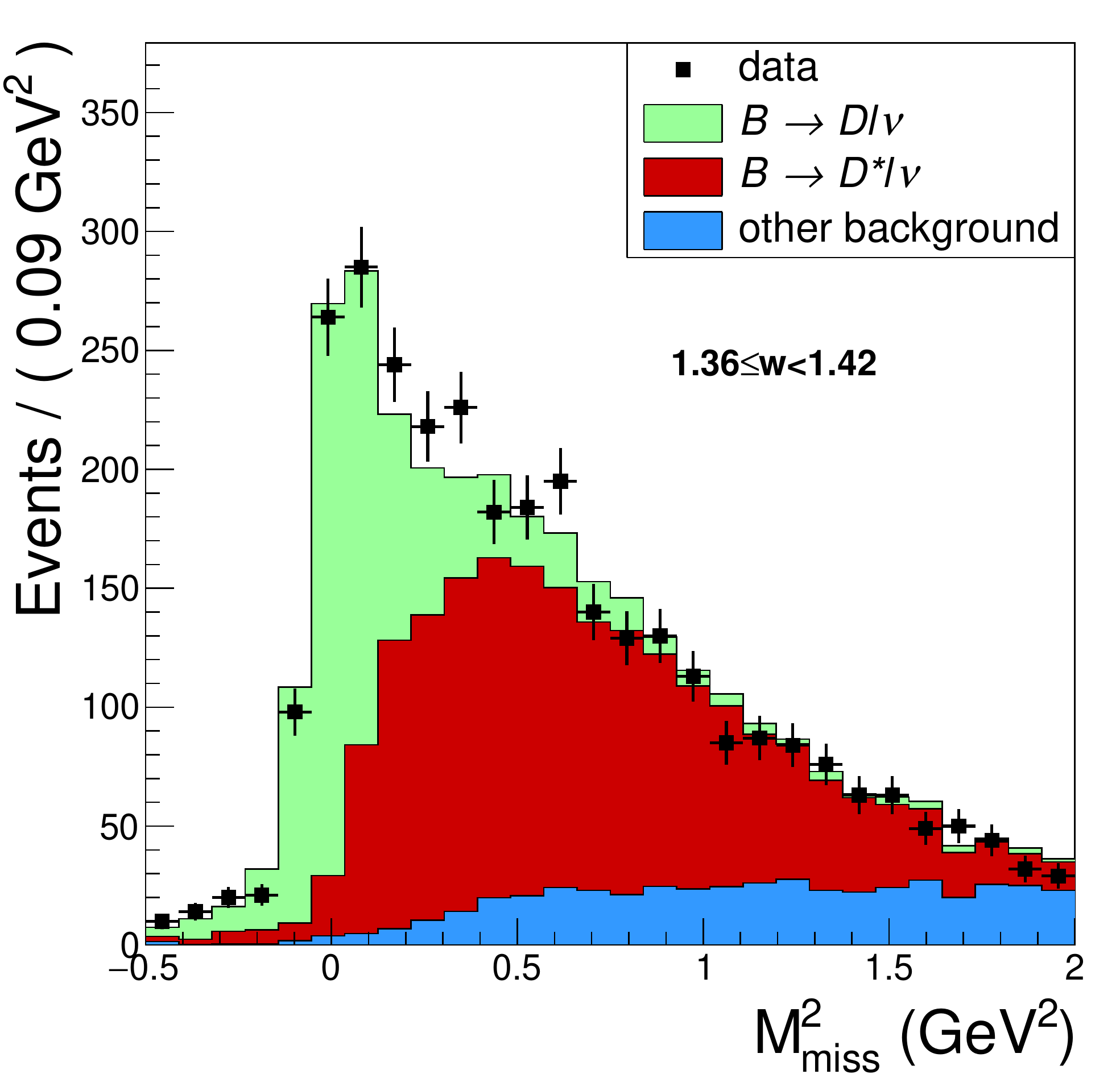}
  \caption{Distribution of \mmsq from the Belle tagged analysis \cite{Glattauer:2015teq}. The distribution corresponds to the bin with $1.36<w<1.42$ separately for $B^0\to D^- e^+\nu_e$ decays (left) and $B^+\to \dzb e^+\nu_e$ decays (right). The larger feed-down due to $D^*$ background present in the $B^+$ sample is clearly visible.} 
  \label{fig:belle_dlmnu_mm2}
\end{figure}

The discriminating variable used to separate the signal \btodlnu from background is the missing mass squared \mmsq. The distribution of \mmsq for a bin in $w$ is reported in figure \ref{fig:belle_dlmnu_mm2} separately for $B^0$ and $B^+$ decays. The signal extraction is performed separately in ten bins in $w$, in the range from 1 to 1.6, with the Barlow and Beeston algorithm \cite{Barlow:1993dm}, that accounts for statistical uncertainties in both data and simulation.
The shapes of the backgrounds and the signal are determined from simulations and fixed in the fit. The fit to extract the signal yields is simultaneous in the four samples: $B^0\to D^- e^+\nu_\e$, $B^+\to \dzb e^+\nu_\e$, $B^0\to D^- \mu^+\nu_\mu$, $B^+\to \dzb \mu^+\nu_\mu$.
The largest source of systematic uncertainty is due to the calibration of the hadronic $B$-tagging sample. This calibration is required because the composition and the efficiency of the various hadronic $B$ decay modes used in the $B$-tagging definition have to be adapted to the data. The other relevant sources of uncertainties
are  the knowledge of the branching ratios of $D^-$ and ${\overline D}^0$ mesons, and   of the tracking efficiency.

\begin{figure}[ht]
  \centering
  \includegraphics[width=0.70\textwidth]{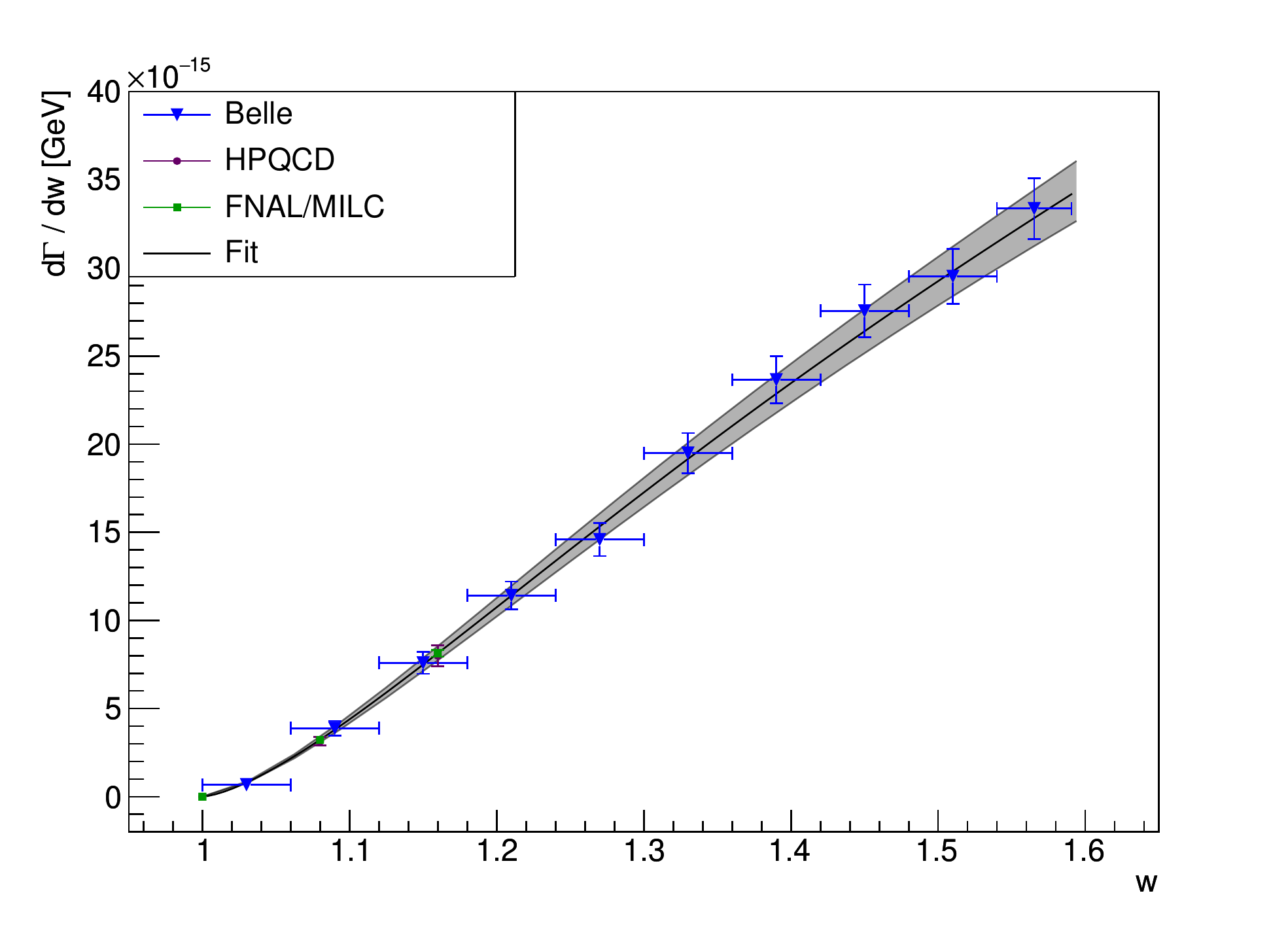}
  \vspace*{-5mm} 
  \caption{Differential decay width of \btodlnu  decay obtained by Belle \cite{Glattauer:2015teq}, and  results of the combined fit between  data and  lattice calculations from FNAL/MILC and HPQCD.} 
  \label{fig:dgdw_fit}
\end{figure}

The distribution of the measured differential decay width $d \Gamma/dw$, is shown in figure~\ref{fig:dgdw_fit} (left) with the result of the fit using the BGL parameterization, superimposed. The fit also exploits the available lattice calculations from FNAL/MILC \cite{Lattice:2015rga} and HPQCD \cite{Na:2015kha} for the values of $w\in (1, 1.08, 1.16)$. The lattice calculations are obtained for both $f_+(z)$ and $f_0(z)$, while the experimental $d \Gamma/dw$ depends only on $f_+(z)$. Nevertheless, exploiting the kinematic constraint between $f_+$ and $f_0$ at maximum recoil, $f_0(\qsq_{min})=f_+(\qsq_{min})$, the lattice data on $f_0(z)$ help to reduce the uncertainties on $|V_{cb}|$. The fit result depends on the truncation order $n$ of the $f_{+,0}(z)$ series. The default result is obtained with $n=3$ because the fit stabilizes for $n\ge 3$. The result is $|V_{cb}|=(40.83\pm 1.13)\times 10^{-3}$. 
By fitting $d\Gamma/dw$ with the CLN parameterization, and taking ${\cal G}(1)=1.0541\pm 0.0083$~\cite{Bailey:2014tva}, the results is $|V_{cb}|=(39.86\pm 1.33)\times 10^{-3}$. 
The result with the CLN parameterization is less precise than BGL one because  in the latter additional lattice point are used.

\begin{table}[h!]
\centering
\caption{Results of \btodlnu measurements with the CLN parameterization and the current HFLAV average \cite{Amhis:2019ckw}. Both  $\eta_{EW} {\cal{G}}(1) |V_{cb}|$  and $\rho^2$ are reported.}
\label{tab:dlnu_summary}
\begin{footnotesize}
\begin{tabular}[t]{l  >{\raggedright\arraybackslash}p{0.20\linewidth} >{\raggedright\arraybackslash}p{0.63\linewidth} }
\hline\hline
  & $\eta_{EW} {\cal{G}}(1) |V_{cb}|$ \newline $\rho^2$  & Remarks \\
\hline

BaBar\cite{Aubert:2008yv} & 42.76$\pm$1.71$\pm$1.26 $1.200\pm 0.088\pm 0.043$ & Global analysis of \btodslnu and \btodlnu using inclusive samples of $B\to D^-\ell\nu_\ell X$ and $B\to \dzb\ell\nu_\ell X$ decays. The fit is multidimentional on $p_\ell^*$, $p_D^*$ and $\cos\theta_{BY}$ variables. Only the CLN parameterization was used.\\

BaBar \cite{Aubert:2009ac}& 43.84$\pm$0.76$\pm$2.19 $1.215\pm 0.035\pm 0.062$ & Tagged measurement using both $B^0$ and $B^+$. The sample is normalizaed to the inclusive $B\to X\ell\nu$ which is known with an uncertainty of only 1\%. The fit is based  on CLN. \\

Belle \cite{Glattauer:2015teq}& 42.22$\pm$0.60$\pm$1.21 $1.090\pm 0.036\pm 0.019$ & Tagged measurement using both $B^0$ and $B^+$. Both fit use CLN and  BGL, as well as the lattice data points at non-zero recoil from FNAL/MILC \cite{Lattice:2015rga} and HPQCD \cite{Na:2015kha}. This analysis published also the unfolded $w$ spectrum corrected for the efficiency.  \\

\hline
HFLAV \cite{Amhis:2019ckw}  & 42.00$\pm$ 0.45$\pm$0.89  1.131$\pm$0.024$\pm$0.023 & The average includes also an older measurement from CLEO.  \\
\hline\hline
\end{tabular}
\end{footnotesize}
\end{table}

\subsubsection{Results}
\label{result1}
A summary of the measurements of $\eta_{EW}{\cal{G}}(1)|V_{cb}|$ obtained with the CLN parameterization is reported in table~\ref{tab:dlnu_summary}, together with the HFLAV average. Using the ${\cal{G}}(1)$ from \cite{Bailey:2014tva}, the HFLAV average is 
\beq
|V_{cb}|=(39.58\pm 0.94 \pm 0.37)\times 10^{-3} 
\label{eq:vcb_dlnu}
\eeq
\noindent where the first error is experimental and the second is due to the form factor normalization. This result is compatible with the result from \btodslnu given in  section \ref{eq:vcb_dslnu}. 
A fit of both BaBar and Belle data, combined with lattice calculation and performed  using both BGL and CLN parameterizations, gives consistent results, even if  the BGL value, $|V_{cb}|=(40.49\pm 0.97)\times 10^{-3}$, is slightly higher than the one obtained with  CLN~\cite{Bigi:2016mdz}. 

\subsection{The $B_s \to D_s^{(*)}\mu\nu_\mu$ channel}
\label{sec:bs}

LHCb has recently extracted $V_{cb}$ from semileptonic $B_s^0$ decays for the first time~\cite{Aaij:2020hsi}. The measurement uses both $B_s^0\rightarrow D_s^{-}\mu^+\nu_{\mu}$ and $B_s^0\rightarrow D_s^{*-}\mu^+\nu_{\mu}$ decays using $3$~fb$^{-1}$ collected in 2011 and 2012. The value of $|V_{cb}|$ is determined from the observed yields of $B_s^0$ decays normalized to those of $B^0$ decays after correcting for the relative reconstruction and selection efficiencies. The normalization channels are $B^0\to D^-\mu^+\nu_{\mu}$ and $B^0\to D^{*-}\mu^+\nu_{\mu}$ with the $D^-$ reconstructed with the same decay mode of the $D_s$ ($D_{(s)}^-\to [K^+K^-]_{\phi}\pi^-$). With this choice the signal and the reference channels have the same particles in the final state and similar kinematics, minimizing in this way the systematic uncertainties.  

The shape of the form factors are extracted as well, exploiting the kinematic variable $p_{\perp}(D_s)$, which is the component of the $D_s^-$ momentum perpendicular to the $B_s^0$ flight direction. This variable is highly correlated with  $q^2$ and also slightly correlated with the helicity angles in the $B_s^0\rightarrow D_s^{*-}\mu^+\nu_{\mu}$ decay. 

The $D_s^*$  is not explicitly reconstructed, but its contribution is disentangled from the $D_s$ using the corrected mass $m_{corr}$, which is defined as \mbox{$m_{corr}=\sqrt{m_Y^2+|{p}_{\perp}(Y)|^2}+|{p}_{\perp}(Y)|$}, where ${p}_\perp(Y)$ is the transverse momentum (to the flight direction) of the visible system $Y\equiv D_s^-\mu^+$ and $m_Y$ is its invariant mass. The variable $m_{corr}$ is useful to discriminate $D_s$, $D_s^*$ and the feed-down background categories: it peaks at the mass of the $B_s$ when there is a single massless particle missing, and peaks at lower values when there are other missing particles associated with the signal candidate, like in $B_s\to D_s\mu\nu_{\mu} X$ decays. 
\footnote{It is interesting to mention that $m_{corr}$ can be generalized also to the case of a massive missing particle, assuming $B\to YX$ where $Y$ is the visible system, and $X$ a single particle with mass $m_X$. In this case,  under the same assumptions behind the standard formula, one has \mbox{$m_{corr}=\sqrt{m_Y^2+|{p}_{\perp}(Y)|^2}+\sqrt{m_X^2+|{p}_{\perp}(Y)|^2}$}.}. 

Analogously to the $B\to D\ell\nu_\ell$ and $B\to D^*\ell\nu_\ell$ decays described before, one of the most relevant backgrounds to $B_s^0\to D_s^-\mu\nu_{\mu}$ signal decays is due to the semileptonic $B_s^0$ decays into excited 
strange charmed states with $L=1$, which in turn decay into $D_s$ and $D_s^*$ with the emission of pions and photons. There are no experimental measurements of the semileptonic $B_s$ decays into these excited states. Of the four $L=1$ excited states $D_s^{**}$, only the states with $j_l=1/2$, namely $D_{s0}(2317)$ and $D_{s1}(2460)$, are known to decay into $D_s$ in the final state because they have a mass below the kinematic threshold needed to decay strongly in $DK$ and $D^*K$. The two states with $j_l=3/2$, $D_{s1}(2536)$ and $D_{s2}(2575)$, instead, do not contribute significantly to the signal because they have a mass high enough that their dominant decay modes are the strong decays into $D^*K$ and $DK$, respectively. Only the $D_{s1}(2536)$ has been observed to decay into a $D_s$ meson. For the same reasons, also higher orbitally or radially excited states do not give $D_s$ in the final state. After the full selection, the background due to the decay of excited state is only few percent of the full $D_s\mu$ sample. Also the normalization channels $B^0\to D^-\mu\nu_\mu$ and $B^0\to D^{*-}\mu\nu_\mu$ suffer of similar kind of background. This background results to be about $9\%$ of the $D^-\mu$ sample. 

\begin{figure}[t!]
  \centering
  \includegraphics[width=0.48\textwidth]{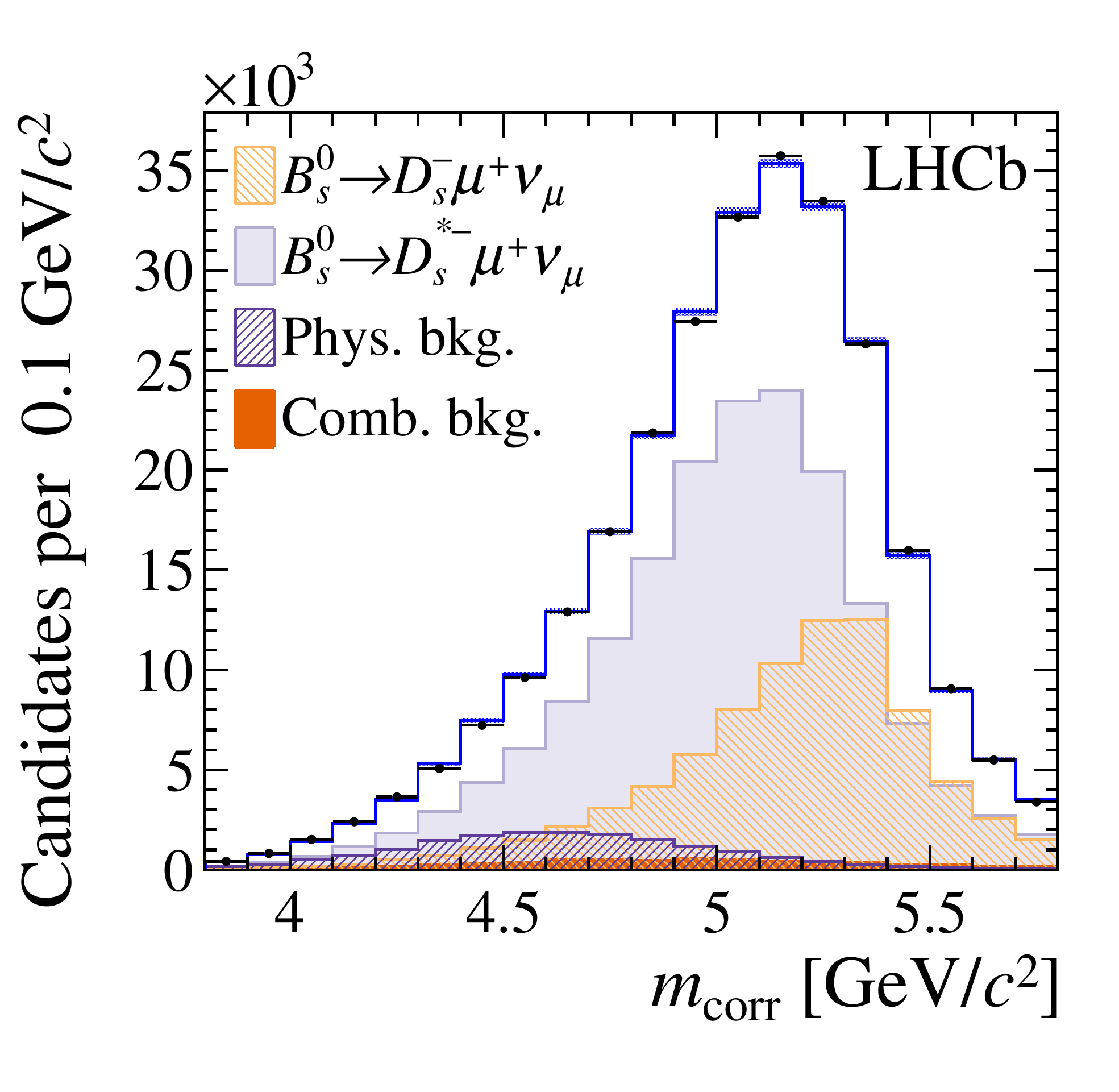}
    \includegraphics[width=0.48\textwidth]{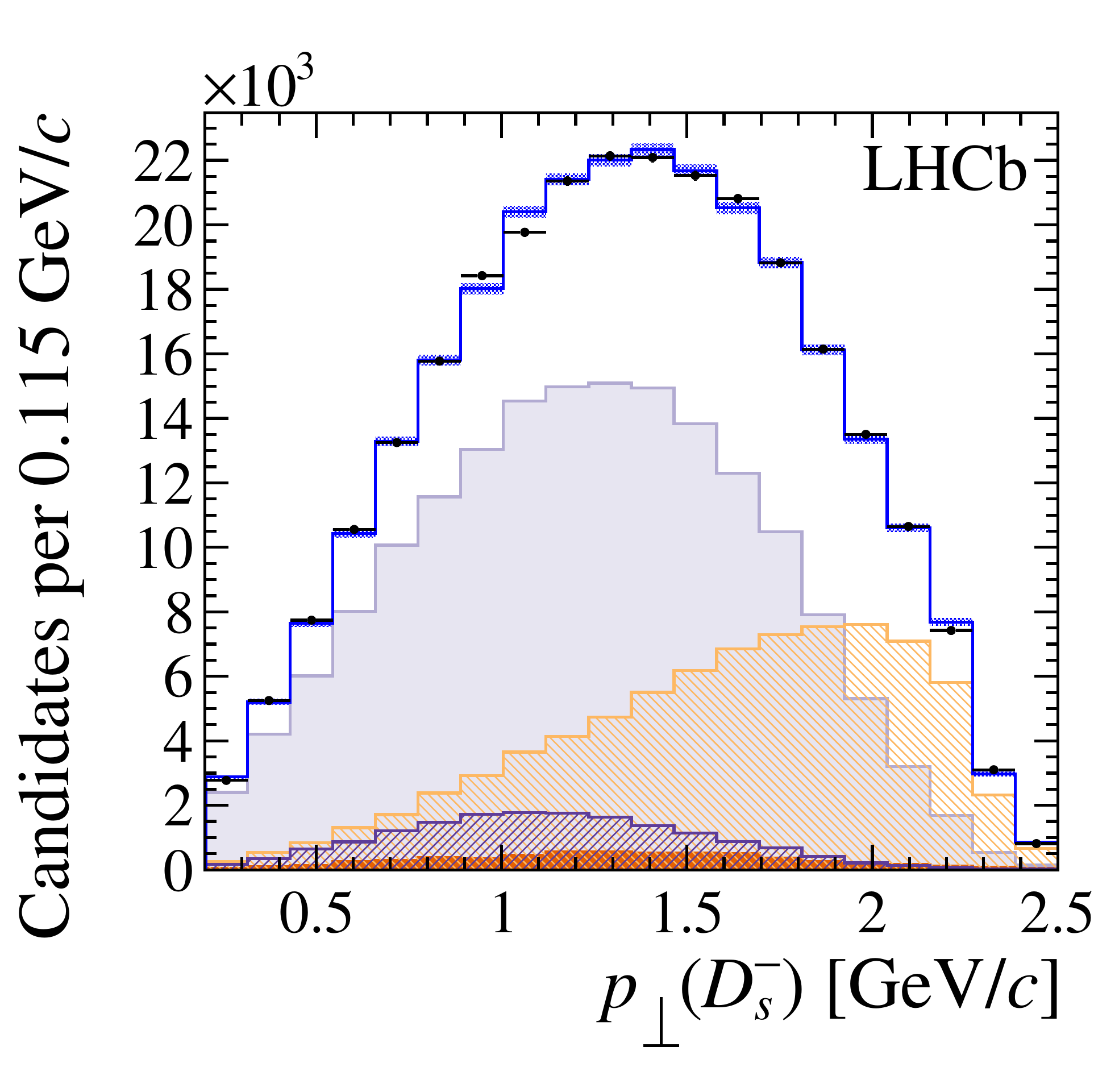}
  \caption{ Distribution of $m_{corr}$ (left) and $p_{\perp}(D_s^-)$ (right) for the inclusive sample of $D_s^-\mu^+$ signal candidates, with fit projections based on the CLN parameterization superimposed.} 
  \label{fig:bsvcb_fit}
\end{figure}

A fit to the $D_s^-\mu$ sample of the two dimensional distribution $m_{corr}$ and $p_\perp(D_s)$ allows to identify $B_s^0\rightarrow D_s^{-}\mu^+\nu_{\mu}$ and $B_s^0\rightarrow D_s^{*-}\mu^+\nu_{\mu}$ signal yields, providing at the same time a measurements of the form factors parameters. The projections of the fit to the signal sample is reported in Figure \ref{fig:bsvcb_fit}. Analogously the yields of the normalisation channels are extracted from a fit to the $D^-\mu$ sample.  
For the $B_s\to D_s\mu\nu_\mu$ decay, $|V_{cb}|$ is connected with the measured ratio of signal yields, $N_{sig}$, and the normalization channel yields, $N_{ref}$, through the relation 
\begin{equation}
\frac{N_{sig}}{N_{ref}}={\cal K}\tau_s \int{\frac{d{\Gamma}(B_s\to D_s\mu\nu_\mu)}{d w}}dw\nonumber \\
\end{equation}
\noindent where $\tau_s$ is the $B_s$ lifetime, and the constant $\cal{K}$ depends on the external inputs as 
\begin{equation}
{\cal K}=\xi\frac{f_s}{f_d}\frac{{\cal B}(D_s^-\to K^+K^-\pi^-)}{{\cal B}(D^-\to K^+ K^-\pi^-)}  \frac{1}{{\cal B}(B^0\to D^-\mu\nu_{\mu})} \nonumber \\
\end{equation}
\noindent where $\xi$ is the efficiency ratio between the signal and the normalization. 
In the analogous expression for the $B_s\to D_s^*\mu\nu_\mu$ decay, the integral of the decay width is done on the variables $(w, \cos\theta_\ell,\cos\theta_V,\chi)$, and there is an explicit dependence on the branching fraction of the $D^{*-}\to D^-\pi$ decay.

This analysis takes advantage of the recent results from lattice on the $B_s\to D_s^-$ and $B_s\to D_s^{*-}$ form factor calculations, summarized briefly in section \ref{Formfactors2}. In particular for the $B_s\to D_s^{*-}$ only the calculations at zero recoil, $h_{A1}^{B_s}(1)$ is available, and the most recent result from Ref.\cite{McLean:2019sds} is used.  
For the $B_s\to D_s\mu\nu_{\mu}$ it has been exploited the very recent calculation of the $B_s\to D_s$ form factors performed in the full $w$-range \cite{McLean:2019qcx}. 

In this analysis both the CLN parameterization and a 5-parameter version of BGL  have been used. 
In the analysis with the CLN parameterization, the form factor parameters $\rho^2(D_s^*)$, $R_1(1)$ and $R_2(1)$ are free to float in fit, while $\rho^2(D_s)$ and the normalizations  $h_{A1}^{B_s}(1)$ and ${\cal G}^{B_s}(1)$ are constrained from the theory calculations. 
The results of the form factors are affected by large statistical uncertainty, but are consistent with the results from the $B$ decays. The result for $|V_{cb}|$ is
\bea
|V_{cb}|_{CLN}&=(41.4\pm0.6\pm 0.9\pm 1.2)\times 10^{-3},\nonumber
\eea
\noindent where the first uncertainty is statistical, the second systematic and the third due to the limited knowledge of the external inputs, in particular the constant $f_s/f_d\times {\cal B}(D_s^-\to K^+K^-\pi^+)$, which is known with an uncertainty of about $3\%$. It is worth to mention that the formulation of the CLN parameterization used is the same obtained for the $B$ meson case. The constants that appear in the equations \eqref{eq:CLN:param} could be slightly different for the $B_s$ case, because the coefficients include the Blaschke factor, which depends on the masses of the initial and final mesons.

In the analysis with the BGL parameterization the fitted parameters are the coefficients of the series of the $z$ expansions. For the $B_s\to D_s^*\mu\nu_{\mu}$ decays the expansion of the form factors $f(z)$, ${\cal F}_1(z)$ and $g(z)$ are truncated at the first order in $z$. For the $B_s\to D_s\mu\nu_{\mu}$ decays, the expansion of $f_+(z)$ is truncated at the second order in $z$, and the three coefficients constrained to the values obtained from Ref.\cite{McLean:2019sds}. The results for $|V_{cb}|$ is
\bea
|V_{cb}|_{BGL}&=(42.3\pm0.8\pm 0.9\pm 1.29)\times 10^{-3},\nonumber
\eea
which is consistent with the result based on CLN parameterization.

The results obtained are in agreement with the exclusive determinations of $|V_{cb}|$ with $B^0$ and $B^+$, and also consistent with the inclusive determination. Although not competitive with the results obtained at the $B$-Factories, the novel approach used can be extended to the semileptonic $B^0$ decays in LHCb. 

\subsection{Direct measurement of $|V_{ub}|/|V_{cb}|$}
\label{sec:lbtopmunu}

The LHCb collaboration has measured the ratio of the branching fractions $\Lambda^0_b\to p\mu^-\bar\nu_\mu$ and $\Lambda^0_b\to \Lambda_c^+\mu^-{\bar \nu}_\mu$ \cite{Aaij:2015bfa}, from which they have determined the first direct measurement of the ratio $|V_{ub}|/|V_{cb}|$. The measured ratio of branching fractions is related to $|V_{ub}|/|V_{cb}|$ through the relation
\begin{equation}
\frac{|V_{ub}|}{|V_{cb}|}=\sqrt{R_{FF} \frac{{\cal B}(\Lambda^0_b\to p\mu{\bar\nu}_\mu)}{{\cal B}(\Lambda^0_b\to \Lambda_c^+\mu{\bar\nu}_\mu)}}\\
\label{eq:vubvcb}
\end{equation}
\noindent where $R_{FF}$ is the ratio of the relevant form factors, which have to be calculated using non perturbative approaches.
In lattice QCD, unquenched results for the form factors away from the static limit have been performed in 2015 \cite{Detmold:2015aaa}. We  discuss $B$-baryon form factors in section \ref{Comparisonwithbaryondecays}.

In the normalization channel $\Lambda^0_b\to \Lambda_c^+\mu^-{\bar \nu}_\mu$, the baryon  $\Lambda_c^+$ is reconstructed in the $\Lambda_c^+\to {p} K^- \pi^+$ decay mode.
With the choice of this normalization channel,
many experimental uncertainties cancel, in particular the large uncertainty on the $\Lambda_b^0$ production rate and on the muon and proton identification efficiency. 
The remaining source of systematic uncertainty that has to be properly accounted for is mainly due to the reconstruction efficiency of the further $K$ and $\pi$ particles required to build the $\Lambda_c^+$ candidates. 

The signal selection exploits the long lifetime of the $\Lambda_b^0$ baryon. The $p\mu^-$ and $\Lambda_c^+\mu^-$ vertexes are required to be displaced from the primary vertex and they are further required to be {\it isolated}, which means that there are no additional tracks that make a good vertex with the signal and normalization candidates.
The isolation reduces most of the combinatorial background and feed-down from $B$-hadron decays with additional charged tracks. The remaining background comes from feed-down events with neutral  or unreconstructed charged particles.  For example, for the signal $\Lambda^0_b\to p\mu^-{\bar\nu}_\mu$, it comes  mainly from $\Lambda_b^0\to \Lambda_c^+\mu^-{\bar\nu}_\mu$, where $\Lambda_c\to pX$ and $\Lambda_b^0\to N^{*+} \mu^-{\bar\nu}_\mu$, with an excited baryon $N^*$ decaying in proton and missing particles. 

The available lattice QCD calculation \cite{Detmold:2015aaa} is more accurate in the high $q^2$ region, in particular the predicted ratio 
$1/R_{FF}=1.471\pm 0.095\pm 0.109$, where the first uncertainty is statistical and the second systematic, is given in the regions $q^2>15\,  {\rm GeV}^2$ for $\Lambda^0_b\to p\mu^-{\bar\nu}_\mu$  and $q^2>7\, {\rm GeV}^2$  for $\Lambda^0_b\to \Lambda_c^+\mu^-{\bar\nu}_\mu$.
The measurement is performed in both regions, where also the signal extraction is cleaner. 
The $q^2$ reconstruction is  described in section \ref{subsub:lhcb_kine}. The problem of having two equally probable solutions (see section \ref{subsub:lhcb_kine}) for each reconstructed $\Lambda_b^0$ momentum affects the resolution of $q^2$.
As a consequence, the measured partially fractions computed in the high $q^2$ range have to be corrected for the effect of the limited $q^2$ resolution. 
To avoid biases in the measurement, the ratio of branching fractions is extracted only for events where both the solutions are within the $q^2$ ranges considered. Even if this choice results in a loss of efficiency, it is beneficial for the control of the systematic uncertainties. 

The measurement of the branching fraction of the normalization channel relies on the known absolute branching fraction ${\cal B}(\Lambda_c^+ \to pK^-\pi^+)=(6.28\pm 0.32)\%$ \cite{Tanabashi:2018oca}, whose value is based on the average of the two most precise available measurements, performed   by  Belle \cite{Zupanc:2013iki} and BESIII
\cite{Ablikim:2015flg}.
It is worth to remark that these two measurements are only marginally consistent, and more effort should be pursued, using also BaBar and LHCb data.

Updating the measured ratio in \cite{Aaij:2015bfa} with the most recent value of ${\cal B}(\Lambda_c^+\to pK^-\pi^+)$\cite{Tanabashi:2018oca}, that we just mentioned, one obtains
\begin{equation}
\frac{|V_{ub}|}{|V_{cb}|}=0.079\pm 0.004\pm 0.004\\
\label{eq:lhcb_vubvcb}
\end{equation}
\noindent where the first uncertainty is experimental and the second one is from the lattice QCD calculation. Even if this is not a direct measurement of $|V_{cb}|$, by taking $|V_{ub}|$  from external inputs  it is possible to determine $|V_{cb}|$. 
For instance, using the  exclusive determination of the $B\to \pi\ell\nu_\ell$ decay rate from HFLAV \cite{Amhis:2019ckw}, one obtains $|V_{cb}|=(46.4\pm 3.8)\times 10^{-3}$, which is compatible with the inclusive measurement, as reported in section~\ref{sub:incl_results}. 
This  measurement  of $|V_{ub}|/|V_{cb}|$ relies only on a single lattice QCD calculation, but the predicted $q^2$ shape for the normalization channel has been validated by a  LHCb measurement of the $q^2$ spectrum of $\Lambda_b^0\to \Lambda_c^+\mu^-{\bar\nu}_\mu$ decays \cite{Aaij:2017svr}. 

The $\Lambda_b^0\to \Lambda_c^+\mu^-{\bar\nu}_\mu$ decays, with a proper normalization channel, would allow a theoretically clean extraction of $|V_{cb}|$. Using semileptonic $B$ meson decays as normalization, like $B^0\to D^+\mu^-{\bar\nu\nu}$ decays, will have as limiting factor the uncertainty on the external parameters, analogously to the $B_s^0$ case described in section \ref{sec:bs}. In particular, it will be limited by the uncertainties on the production fraction ratio $f_{\Lambda_b}/f_d$ and on the $B^0\to D^+\mu^-{\bar\nu\nu}$ branching fraction.

The LHCb analysis, besides being the first one made at hadronic colliders, and the first one to use $B$-baryon decays, opened the possibility to extract $|V_{ub}|/|V_{cb}|$ from the ratio ${\cal{B}}(B_s^0 \to K^-\mu^+\nu_\mu)/{\cal{B}}(B_s^0\to  D_s^-\mu^+\nu_\mu)$, a measurement which is ongoing at LHCb.

\subsection{The $|V_{cb}|$ puzzle}
\label{sub:vcb_puzzle}

As we have seen, the   inclusive and exclusive semileptonic  searches rely on different theoretical tools  and  experimental techniques. The agreement among  $|V_{cb}|$ values from inclusive and exclusive decays can be regarded as an interesting test of our capability to investigate weak interactions and  QCD  dynamics. From this prospective, a lot of attention has been devoted to  a discrepancy which, since more than three decades, is observed between the values extracted from exclusive and inclusive decays. It is referred as the  $|V_{cb}|$ puzzle.
\begin{figure}[t!]
  \centering
  \includegraphics[width=1.00\textwidth]{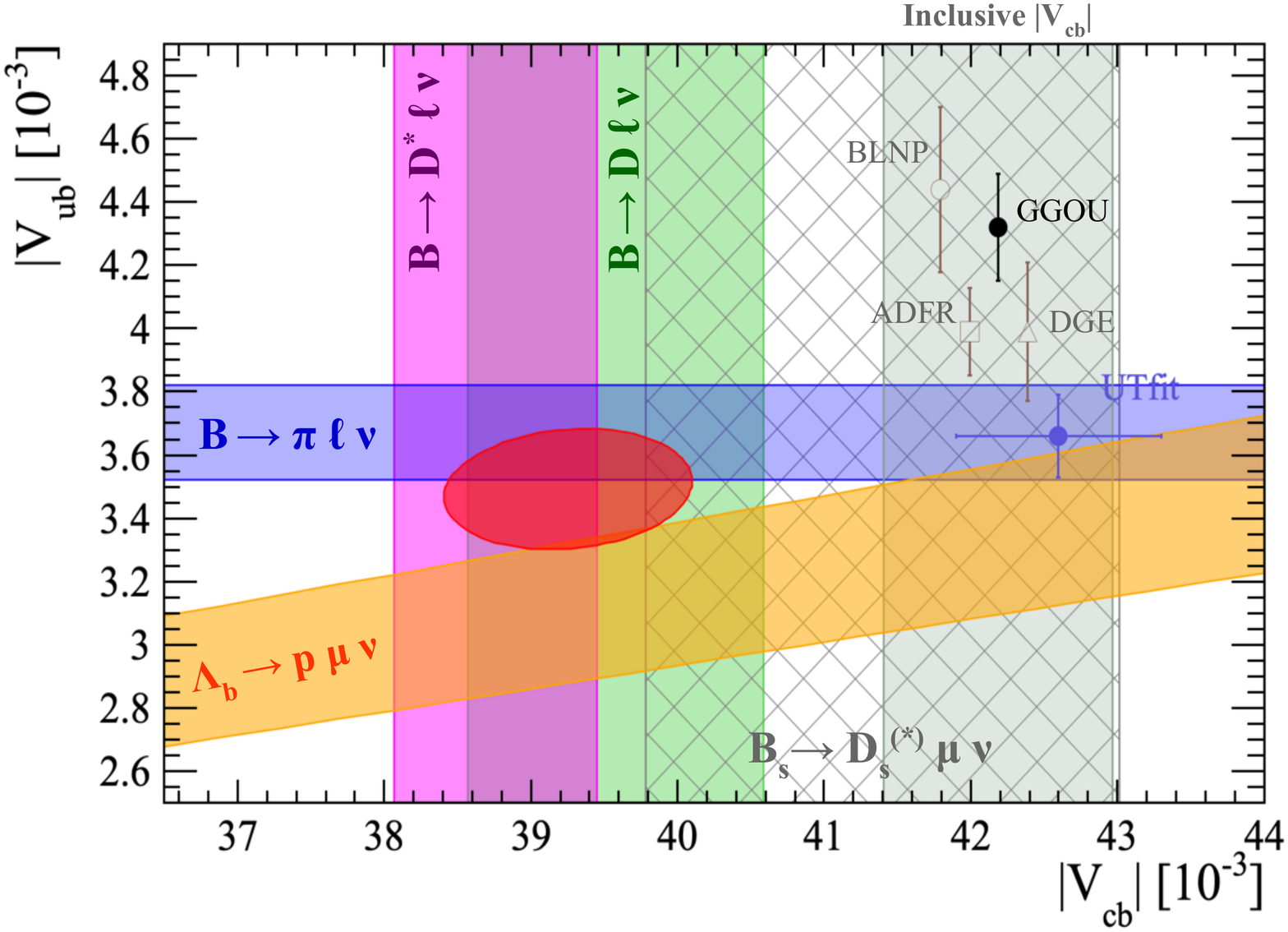}
  \vspace*{-5mm} 
  \caption{The combined $V_{ub}-V_{cb}$ average (red ellipses) obtained by HFLAV including the exclusive measurements of $|V_{ub}|$ from $B\to \pi\ell\nu$ (blue), $|V_{cb}|$ from \btodslnu (magenta) and \btodlnu (green) decays, and $|V_{ub}|/|V_{cb}|$ (orange) from $\Lambda_b\to p\mu\nu$ decay. 
  The band filled with grid pattern corresponds to the LHCb result with $B_s\to D_s^{(*)}\mu\nu_\mu$ decays.
  The grey band refers to inclusive $|V_{cb}|$ in the kinetic scheme. The result on inclusive $|V_{ub}|$ with different calculations are the four points with vertical error bars. The shift along the x-axis of these four points is just arbitrary and has no meaning. 
  The blue point is the result of the indirect predictions of $|V_{ub}|$ and $|V_{cb}|$ obtained by the UTfit collaboration~\cite{UTfit} and based on the global fit to the unitarity triangle.}
  \label{fig:vub_vcb_summary}
\end{figure}

In figure \ref{fig:vub_vcb_summary} we summarize exclusive and inclusive determinations of $|V_{cb}|$ and compare with the analogous determinations of $|V_{ub}|$.
The CKM parameter
$|V_{ub}|$ shares with $|V_{cb}|$
the discrepancy between inclusive and exclusive values, which is labelled,  similarly, the $|V_{ub}|$ puzzle (for concise reviews see e.g. \cite{Ricciardi:2016pmh, Ricciardi:2014iga,Ricciardi:2013cda,Ricciardi:2012pf}).

The most precise estimates of $|V_{cb}|$ 
stem from the \btodslnu channel with inputs from lattice, followed by determinations based on inclusive measurements. Their uncertainties all stay around 1.8\%. In  figure~\ref{fig:vub_vcb_summary} the vertical bands represent the different determinations of $|V_{cb}|$. 
We have separated the bands relative to exclusive $|V_{cb}|$ determination with \btodslnu and \btodlnu decays.
  Both show a discrepancy with the the band relative to the inclusive determination. 
 The bands relative to the exclusive $B\to D$ and $B\to D^\ast$ decays are  the HFLAV averages done with the CLN parameterizations. Also considering the slightly larger uncertainty associated with the BGL fit to \btodslnu decays described before, the discrepancy with 
 the inclusive determinations remains significant. The tension amounts to about 3$\sigma$. The result using $B_s\to D_s^{(*)}\mu\nu_\mu$, still affected by large uncertainties, is 
 compatible with both inclusive and exclusive determinations of $|V_{cb}|$.

It is  also possible to determine $|V_{cb}|$ indirectly, using the CKM unitarity relations together with CP violation and flavour data, excluding direct information on decays. The indirect fits  provided  by the CKMfitter collaboration \cite{CKMfitter} and by the UTfit collaboration~\cite{UTfit} are in agreement between them and  seem to prefer the inclusive value for $|V_{cb}|$,
as shown in figure~\ref{fig:vub_vcb_summary}.


 
 In figure~\ref{fig:vub_vcb_summary} we  also report the  world average values of the CKM parameter $|V_{ub}|$ obtained by the HFLAV collaboration. 
The most precise values for $|V_{ub}|$ are also obtained from semileptonic decays.
The CKM-suppressed decay $B \to \pi \ell\nu_\ell$ is the typical exclusive channel used to extract $|V_{ub}|$, being better controlled both experimentally and theoretically.
We represent also the band constrained by the $|V_{ub}|/|V_{cb}|$ ratio measurement reported in equation \eqref{eq:lhcb_vubvcb}. The LHCb measurement is consistent with the prediction from the indirect determination.

 Most of the theoretical and experimental considerations presented in this review  also apply  to the $|V_{ub}|$ determination.
 The main differences between $|V_{cb}|$ and $|V_{ub}|$ determinations emerge in inclusive decays. Due to the large background to $B \to X_u \ell \nu_\ell$ decays represented by $B\to X_c \ell  \nu_\ell$ decays, the phase space region is strongly limited by the experimental  cuts needed to reduce the background. This  requires to address theoretical issues absent in the inclusive $|V_{cb}|$ determination, since the experimental  cuts enhance the relevance of a region in the phase space, the so-called threshold region, where the applicability of HQE is compromised. In place of a widely accepted theoretical tool as the HQE, several models or schemes have been devised.  They are  all tailored to analyze data in the threshold region,  but differ in their treatment of perturbative corrections and the parameterization of non-perturbative effects. 
In figure~\ref{fig:vub_vcb_summary} we show the results for the four theoretical approaches  included in the HFLAV averages \cite{Amhis:2019ckw}:
ADFR (Aglietti, Di Lodovico, Ferrera, Ricciardi)~\cite{Aglietti:2004fz, Aglietti:2006yb,  Aglietti:2007ik}, BLNP (Bosch, Lange, Neubert, Paz)~\cite{Lange:2005yw, Bosch:2004th, Bosch:2004cb},  DGE (Dressed Gluon Exponentiation)~\cite{Andersen:2005mj} and GGOU (Gambino, Giordano, Ossola, Uraltsev)~\cite{Gambino:2007rp}. The results are based on the same experimental inputs (apart the one from ADFR which does not include the latest result from BaBar \cite{TheBaBar:2016lja}), and are slightly above the exclusive $|V_{ub}|$ value, extracted from both $B\to \pi\ell\nu_\ell$ and $\Lambda_b\to p \mu \nu_\mu$ decays.

\section{Future prospects}
\label{futureprospects}
The pattern of quark and lepton masses and mixings remains one of the most debated and interesting open questions in particle physics, in spite of a plethora of new
experimental results. The precise determination of the CKM matrix elements connects flavour physics with the Higgs sector, since they represent the couplings of the Higgs boson to fermions.
Generations of dedicated experiments have provided us with more and more precise
measurements and exposed a flavor pattern of an highly non-generic
structure, begging  for an underlying organizing principle, which is still unveiled.
Experimental hints for deviations from SM predictions in flavour processes are one of our
best hopes to direct research towards the right energy scale of new physics. As suggested by the 2020 EPPSU update \cite{Strategy:2019vxc}, flavor physics should remain at the forefront of the European particle physics  strategy.
In this wide perspective, the search for very high precision in $|V_{cb}|$ determination  is actively pursued on both experimental and theoretical sides.

In exclusive semileptonic $B$ meson  decays, the  $|V_{cb}|$ determination from \btodslnu decays has the
largest theoretical uncertainty, 
amounting to about 1.4$\%$, as can be seen by comparing the averages
\eqref{eq:vcb_dslnu} and \eqref{eq:vcb_dlnu}. By the same comparison, one observes that 
instead  the experimental error is maximum,  about 2$\%$, for $|V_{cb}|$ values extracted from \btodlnu decays.
%
A theoretical research area with direct impact on future experimental programs is lattice gauge theory, the only systematically improvable method for nonperturbative calculations in QCD.
%
Like in  the  case  of the exclusive $|V_{cb}|$ determination from \btodlnu decays,  determinations from 
 \btodslnu decays  are  expected to improve  significantly    as  soon  as  lattice  calculations  of the form factors at non-zero recoil will become fully available.
The pivotal importance of precise information on the form factors is a clear outcome of the analyses on form factors parameterization in exclusive determinations, discussed in section~\ref{results633}. For example, it has been noted~\cite{Gambino:2019sif} that a possible steeper slope of the form factor ${\cal{F}}(w)$ at zero recoil could lift the value of the exclusive $|V_{cb}|$ determination towards agreement with inclusive determinations.

While lattice unquenched results for the form factors of $ B \to D^{(\ast)}$ semileptonic decays have been available since at least 10 years, lattice analyses for inclusive decays are now moving their first significant steps. On lattice it is not straightforward  to  extract  inclusive  observables,  i.e. quantities that are summed over all multi-particle final states. Major  challenges are that the lattice  calculations are performed in a finite volume and naturally formulated in the Euclidean space, which complicates the analyses of correlation functions for the case of multi-particle states in the kinematic region accessible on the lattice. A large body of work has already gone into developing algorithms and theory to overcome these and similar limitations, with significant implications also on other branches of physics and mathematics (for details see e.g.~\cite{Lehner:2019wvv}).  A different suggestion, specific to inclusive semileptonic $B$ decays, is to analytically continue the amplitude from the experimentally accessible physical kinematic region to an nonphysical region in which the lattice calculation can be performed \cite{Hashimoto:2017wqo}.

 \begin{figure}[t!]
  \centering
  \includegraphics[width=0.95\textwidth]{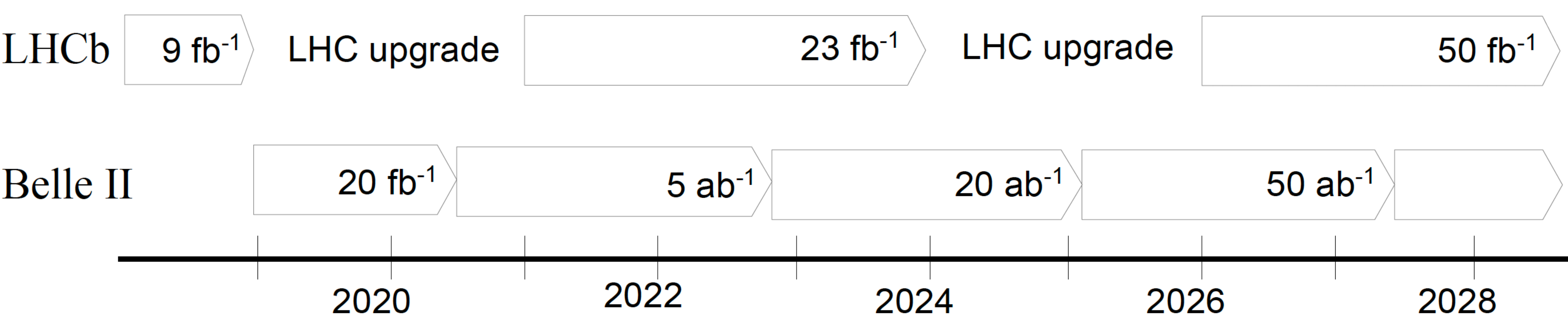}
  \caption{An overview of the expected Belle II and LHCb timelines along with their estimated integrated luminosity at various milestones.}
  \label{fig:timeline}
\end{figure}

The Belle II experiment at KEK started recently to the take data from the renewed $e^+e^-$ KEK-B accelerator (SuperKEKB), designed to reach an instantaneous luminosity 50 times higher than  KEK-B. The final goal of Belle II is to collect $50\, {\rm ab}^{-1}$ by  2027. In figure ~\ref{fig:timeline} the expected timelines for both Belle II and LHCb are reported. The precise study of semileptonic $B$  meson decays is a substantial part of the Belle II program~\cite{Kou:2018nap}. However, the increase in luminosity is not enough by itself, because most of the measurements that we have presented above are limited by the systematics and not by the statistical uncertainties.
For the exclusive $|V_{cb}|$, the most precise measurement comes from the Belle untagged analysis of \btodslnu decays~\cite{Abdesselam:2018nnh} (described in section~\ref{Belleuntaggedmeasurement}), where the systematic uncertainty is 2.5 times the statistical one. The largest contributions to the systematics are due to the tracking and the particle identification, followed by the ones  due to the external inputs, like the branching fractions of $D$ mesons and $f_{00}$.  

It is foreseen that the hadronic $B$ tagged analysis will be the preferred approach to study semileptonic decays at Belle II. 
The tagged analyses of \btodslnu are reaching the precision of the untagged measurements, but are at present affected by large uncertainties due to the calibration of the hadronic $B$ tagging. The reduction of this sources of systematics is paramount to exploit the huge statistics available at Belle II.
With  large statistics available the analyses approaches have to be revisited. For instance, the BaBar tagged measurement~\cite{Dey:2019bgc} described in section~\ref{BaBartaggedmeasurement}, which performed a truly four-dimensional fit,
has reached  precision comparable with the untagged Belle analysis \cite{Abdesselam:2018nnh}, despite the fact that the signal yield was only 1/30 of the Belle signal yield.

The Belle II data taking is ongoing, and the first studies confirm the expected detector performances. Very recently, Belle II collaboration has released an untagged measurement of the branching fractions of the $B^0\to D^{*-}\ell^+{\nu_\ell}$ decays using $8.7~fb^{-1}$ of data \cite{Abudinen:2020yxf}. The result is consistent with the existing measurements. While the uncertainties are not competitive with the ones of the most recent results at $B$-Factories, this measurement validates the full chain of detector operation, calibration and analysis. 


As mentioned in section \ref{Formfactors2}, other interesting exclusive channels are the semileptonic $B_s \to D_s^{(*)} \ell \nu_\ell$ decays. The study of $B_s$ decays at Belle II would require to run SuperKEKB at the energy corresponding to the $\Upsilon(5S)$ mass. At present there are no expected plans for Belle II to collect data at energy higher than the $\Upsilon(4S)$ mass. But the $B_s^0$ are copiously produced at LHC and recently LHCb has exploited these new calculations in the pioneering measurement of $|V_{cb}|$ using the semileptonic $B_s$ decays~\cite{Aaij:2020hsi} (see section \ref{sec:bs}). This measurement is at present limited by the precision of external parameters, but the developed technique can be applied to $B$ meson decays, where their impact is reduced.

The LHCb experiment is undergoing a major upgrade of the detector, which was planned and designed in the 2011 \cite{LHCb-TDR-012} and  should end in 2021, when LHC will restart the activity (see the timeline in figure \ref{fig:timeline}). The upgrade will allow to collect data at higher instantaneous luminosity, so about five $pp$ collisions per bunch crossing are foreseen. To cope with the higher occupancy in the detector, besides the improvements in the various subdetectors, a fully software L0 trigger will be employed (a configuration called {\it triggerless}). The software L0 trigger will add flexibility to the data taking, allowing to reduce the thresholds for muon and hadron trigger decisions and enlarge the physics capabilities.  The analyses of semileptonic decays with tauons and electrons will benefit of the lower trigger thresholds in terms of signal efficiencies.
With this upgraded detector, LHCb is planning to integrate a luminosity of $23$~fb$^{-1}$ by the 2024, and
collect a total sample of $50$~fb$^{-1}$ by the 2028-2029, after LHC will have switched to higher luminosity. 

A promising field of study are $\Lambda_b$ baryons, which represent approximately  20\%  of  all  bottom  hadrons  produced  at  the  LHC. 
The  measurement  of  the  ratio  of $\Lambda^0_b \to p \mu^- \bar \nu_\mu$  and $\Lambda^0_b \to \Lambda^+_c \mu^- \bar \nu_\mu$  decay  rates  at  LHCb, combined  with  a  lattice  QCD  calculation  of  the  $\Lambda_b  \to p$ and  $\Lambda_b  \to \Lambda_c$ form  factors~\cite{Detmold:2015aaa},  has allowed the first determination of $|V_{ub}|/|V_{cb}|$ at an hadron collider~\cite{Aaij:2015bfa}, as described in section~\ref{sec:lbtopmunu}. We have shown the band of results in figure~\ref{fig:vub_vcb_summary}.  
Right now, theory uncertainties are  approximately  5\%, comparable with the experimental uncertainty. 
As the latter is expected to reach about 3\% 
at the integrated luminosity of 23~${\rm fb}^{-1}$ foreseen by 2024 (see figure~\ref{fig:timeline}), 
further theoretical progress is needed, which could come from lattice improvements to the form factors computation. With the huge data available in the next years, there are prospects to extend the measurement of $\Lambda_b\to p\mu\nu$ to a differential measurement in bins of $q^2$. 
The baryon semileptonic decays are sensitive to both the vector and axial-vector currents in the weak effective Hamiltonian, and their high precision measurements can also represent a check of right-handed couplings beyond the SM. 

Progress is also expected for $B$ decays to excited $D$ meson states. 
Form factors must be determined in all modes through precise differential measurements. 
The required accuracy could come from Belle II which has the potential to precisely isolate all four orbitally excited modes and characterize their sub-decay modes, constraining and measuring the branching ratios with higher accuracy~\cite{Kou:2018nap}. LHCb has the capability to study with high precision the kinematics of the decays into narrow states. Furthermore  LHCb can study in detail the production of excited states in semileptonic $B_s^0$ and $\Lambda_b^0$ decays.

%
Lattice studies are in progress with realistic  charm mass,
and  results on  $ B \to D^{\ast \ast } \ell \nu$ form factors are available, still at a preliminary stage, since 2013~\cite{Atoui:2013ksa}.
For recent and more complete reviews on open charmed systems see e.g.~\cite{Chen:2016spr, Yaouanc:2014ypa}.


The semileptonic $B$  decays we have considered are tree-level processes in the SM, which are generally assumed, in all analyses, unaffected by NP contributions. Because of their pivotal role in precise measurements of the CKM matrix elements, it is not without importance to ascertain the validity of this assumption, given also the tensions underlined above.
 There are many analyses addressing this issue (see e.g. 
\cite{Crivellin:2014zpa, Colangelo:2016ymy, Jung:2018lfu, Colangelo:2018cnj}) and several models  which do not seem to support  evidence of NP in decays driven by $b \to c \ell \nu_\ell$ transitions, where $\ell$ is a light lepton. Particular attention deserves $R(D^{(\ast)})$, discussed in section \ref{Exclusivedecaysintoheavyleptons}, whose measured value differs from the SM prediction. A better understanding of this discrepancy could shed light on possible NP and as such it is a priority for Belle II and for the future planned LHCb upgrade.


\ack

It is a pleasure to thank Aoifa Barucha and Francesco Polci for their invitation to give  GDR-InF lectures at the Institut Henri Poincaré  (Paris, France), where part of this work was performed, and for providing a stimulating environment.
G.R. acknowledges partial financial support from MIUR under 
Project No. 2015P5SBHT and from the INFN research initiative ENP.
\section{Bibliography}

\bibliographystyle{iopart-num} 
\bibliography{main,main_ex}

\end{document}